\documentclass[a4paper, amsfonts, amssymb, amsmath, reprint, showkeys, nofootinbib, twoside, superscriptaddress]{revtex4-1}
\usepackage{amsmath,amsfonts,amssymb} %
\usepackage{mathtools} %
\usepackage{wasysym} %
\usepackage{bm} %
\usepackage[bbgreekl]{mathbbol} %
\usepackage{graphicx} %
\usepackage[dvipsnames,table]{xcolor}
\usepackage[acronym]{glossaries} %
\usepackage{braket} %
\usepackage{multirow} 
\usepackage{quantikz}
\usepackage{fancyhdr} %
\usepackage{microtype} %
\usepackage{natmove}
\usepackage{smartref} %
\usepackage[breaklinks, %
colorlinks, linkcolor=Blue, citecolor=Blue, urlcolor=Blue]{hyperref} %
\usepackage[version=3]{mhchem}
\newcommand{\bea}{\begin{eqnarray}}
\newcommand{\eea}{\end{eqnarray}}
\newcommand{\eq}[1]{Eq.~(\ref{#1})} %
\newcommand{\fig}[1]{Fig.~\ref{#1}} %

\newcommand{\Tr}[0]{\text{Tr}}

\newcommand{\imag}{\mathrm{i}}
\newcommand{\AV}[1]{\langle #1 \rangle}
\newcommand{\VA}[1]{\text{Var}(#1)}
\begin{document}

\title{Quantum Measurement for Quantum Chemistry on a Quantum Computer}

\author{Smik Patel}
\affiliation{Chemical Physics Theory Group, Department of Chemistry, University of Toronto, Toronto, Ontario M5S 3H6, Canada}
\affiliation{Department of Physical and Environmental Sciences, University of Toronto Scarborough, Toronto, Ontario M1C 1A4, Canada}
\author{Praveen Jayakumar} 
\affiliation{Chemical Physics Theory Group, Department of Chemistry, University of Toronto, Toronto, Ontario M5S 3H6, Canada}
\affiliation{Department of Physical and Environmental Sciences, University of Toronto Scarborough, Toronto, Ontario M1C 1A4, Canada}
\author{Tzu-Ching Yen} 
\affiliation{Columbia Business School, Columbia University, New York, NY 10027, USA}
\author{Artur F. Izmaylov}
\affiliation{Chemical Physics Theory Group, Department of Chemistry, University of Toronto, Toronto, Ontario M5S 3H6, Canada}
\affiliation{Department of Physical and Environmental Sciences, University of Toronto Scarborough, Toronto, Ontario M1C 1A4, Canada}

\date{\today} 

\newacronym{LCU}{LCU}{linear combination of unitaries}
\newacronym{QPE}{QPE}{quantum phase estimation}
\newacronym{MTD}{MTD}{Majorana tensor decomposition}
\newacronym{SVD}{SVD}{singular value decomposition}
\newacronym{MPS}{MPS}{matrix product state}
\newacronym{SF}{SF}{single factorization}
\newacronym{DF}{DF}{double factorization}
\newacronym{AC}{AC}{anticommuting}
\newacronym{CSA}{CSA}{Cartan sub-algebra}
\newacronym{THC}{THC}{tensor hypercontraction}
\newcommand{\myceil}[1]{\lceil #1 \rceil}
\newcommand{\myfloor}[1]{\lfloor #1 \rfloor}
\newcommand{\BCC}[1]{{\color{blue} {#1}}}

\begin{abstract}
Quantum chemistry is among the most promising applications of quantum computing, offering the potential to solve complex electronic structure problems more efficiently than classical approaches. A critical component of any quantum algorithm is the measurement step, where the desired properties are extracted from a quantum computer. This review focuses on recent advancements in quantum measurement techniques tailored for quantum chemistry, particularly within the second quantized framework suitable for current and near-term quantum hardware.

We provide a comprehensive overview of measurement strategies developed primarily for the Variational Quantum Eigensolver (VQE) and its derivatives. These strategies address the inherent challenges posed by complexity of the electronic Hamiltonian operator. Additionally, we examine methods for estimating excited states and one- and two-electron properties, extending the applicability of quantum algorithms to broader chemical phenomena.

Key aspects of the review include approaches for constructing measurement operators with reduced classical preprocessing and quantum implementation costs, techniques to minimize the number of measurements required for a given accuracy, and error mitigation strategies that leverage symmetries and other properties of the measurement operators. Furthermore, we explore measurement schemes rooted in Quantum Phase Estimation (QPE), which are expected to become viable with the advent of fault-tolerant quantum computing.

This review emphasizes foundational concepts and methodologies rather than numerical benchmarks, serving as a resource for researchers aiming to enhance the efficiency and accuracy of quantum measurements in quantum chemistry.
\end{abstract}


\maketitle

\section{Introduction}

Quantum chemistry is one of the most promising applications of quantum computing \cite{CR_QC:2019,RevModPhys.92.015003}. Several quantum algorithms have been developed to address the electronic structure problem \cite{CR_QA4QC}. Regardless of the specific algorithm, the final step in any quantum calculation involves measuring the property of interest on a quantum computer. Without this step, the results of the calculation cannot be obtained. This review focuses on recent advancements in measurement techniques designed to efficiently extract information from quantum computers.

Our primary focus is on measurement methods tailored to solving the electronic structure problem on current and near-future quantum hardware. The Hamiltonian we will consider is the second-quantized electronic Hamiltonian which has been projected onto a set of $N$ spin orbitals
\begin{equation}
    \hat{H}_e = \sum_{p,q=1}^{N} h_{pq} \hat{a}_p^\dagger \hat{a}_q + \sum_{p,q,r,s=1}^{N} g_{pqrs} \hat{a}_p^\dagger \hat{a}_q \hat{a}_r^\dagger \hat{a}_s,
\end{equation}
where $\hat{a}_p^\dagger$, and $\hat{a}_p$ are fermionic creation and annihilation operators for spin orbital $p$, and $h_{pq}, g_{pqrs}$ are obtained from the one- and two-body integrals. Much of the existing research has concentrated on ground-state energy estimations, which already have potential industrial applications \cite{QB_DARPA}. We also explore techniques for measuring excited states and calculating one- and two-electron properties.

Most measurement schemes have been developed within the framework of the Variational Quantum Eigensolver (VQE) \cite{peruzzoVariationalEigenvalueSolver2014}, where the expectation value of the electronic Hamiltonian is calculated. In VQE, a trial state $\ket{\Phi}$ is optimized based on its energy, $\bra{\Phi}\hat{H}\ket{\Phi}$. However, measuring this expectation value is challenging because the electronic Hamiltonian, $\hat{H}$, contains approximately $\sim N^4$ Pauli product terms, where $N$ is the number of qubits or spin orbitals.

The rise in popularity of VQE stems from the limitations of the Quantum Phase Estimation (QPE) algorithm \cite{SLoyed_QPE,Aspuru_Guzik_2005}, which cannot be implemented on near-term (NISQ) devices due to the high circuit depth required, leading to significant error accumulation \cite{NISQ_Preskill,RevModPhys.94.015004}. While QPE is infeasible for NISQ hardware, its more sophisticated measurement schemes are expected to become viable with advances in gate fidelities and error-correction capabilities. Thus, this review also considers QPE-based measurement schemes, assuming the emergence of fault-tolerant quantum computing in the future.

In both VQE and QPE, a state close to the eigenstate of interest, $\ket{\Phi}$, is prepared. To measure any non-trivial property $\hat{O}$, an additional unitary rotation $\hat{U}$ is applied, producing the modified state $\hat{U}\ket{\Phi}$. Measuring the Pauli single qubit $\hat z$ operators for $\hat{U}\ket{\Phi}$ provides an estimate of $\bra{\Phi}\hat{O}\ket{\Phi}$. Consequently, a central theme of this review is the construction of $\hat{U}$ for obtaining estimates of quantities of interest.

We focus on three key aspects of measurement schemes based on $\hat{U}$:
\begin{enumerate}
\item The classical pre-processing cost of generating $\hat{U}$ for a specific observable.
\item The quantum circuit implementation cost of $\hat{U}$.
\item The number of measurements required to achieve a desired accuracy.
\end{enumerate}
Additionally, for pre-error-corrected algorithms, we consider opportunities for error mitigation leveraging properties of $\hat{U}$, such as conserved symmetries.

This review emphasizes the conceptual principles of various approaches rather than an extensive analysis of numerical results on specific molecular systems. Readers are encouraged to consult the original publications for detailed data. The structure of the review is as follows:
First, we discuss various techniques of formulating measurable fragments (Sec.~\ref{Sec:MFragments}). This angle to 
 the measurement problem originated from the necessity to measure  $\bra{\Phi}\hat H\ket{\Phi}$ in VQE. We discuss how the measurable 
 fragments can be found using various algebraic relations related to 
 solvability of the many-body eigenvalue problem. An alternative approach based on classical shadow tomography is discussed as well. Second, Sec.~\ref{Sec:Embedding} considers techniques that embed the Hamiltonian to a bigger space and thus provide more ways to 
 measure expectation values of the Hamiltonian. Third, one of the main concerns in measurement is the number of measurements and how it can be reduced, which is covered in Sec.~\ref{Sec:MeasurementNReduction}.
 Fourth, Sec.~\ref{Sec:OtherQuantities} discusses approaches for measurement beyond a single electronic state or energy only considerations. Fifth, another crucial aspect of 
 the measurement on near-term quantum devices is the mitigation of errors due to noise, which is covered in Sec.~\ref{Sec:NoiseReduction}. Finally, we discuss measurement schemes inspired by QPE, highlighting their potential for fault-tolerant quantum computing in Sec.~\ref{Sec:Heisenberg}.

\section{Using Measurable Fragments of the Hamiltonian}
\label{Sec:MFragments}

\subsection{The Computational Basis and Ising Hamiltonians}
\label{Subsec:IsingH}

The goal of quantum simulation is to obtain useful quantitative information about physical systems, like eigenvalues of Hermitian operators, or density matrix elements, which can be used to make predictions about physical and chemical properties (e.g., molecular geometry, reaction rates, spectroscopic information, etc). Information is extracted from the quantum computer via quantum measurements. The basis that a quantum computer measures is called the computational basis. For a single qubit, the computational basis states are usually denoted by $\ket{0}$ and $\ket{1}$. A single qubit state is a normalized linear combination of the computational basis states:
\begin{equation}
    \ket{\psi} = c_0\ket{0} + c_1\ket{1},
\end{equation}
where $c_0,c_1 \in \mathbb{C}, |c_0|^2 + |c_1|^2 = 1$. Measurement of the qubit in this state will produce measurement result $z \in \mathbb{F}_2$ with probability $|c_z|^2$, where $\mathbb{F}_2$ denotes the binary field $\{0,1\}$. The computational basis is also the eigenbasis of the Pauli $\hat{z}$ operator, as $\hat{z}\ket{z} = (-1)^{z}\ket{z}, z \in \mathbb{F}_2$.

On $N$ qubits, there are $2^N$ computational basis states
\begin{equation}
    \ket{\vec{z}} := \ket{z_1 z_2 \ldots z_N}, \ z_i \in \mathbb{F}_2.
\end{equation}
A general $N$ qubit state can be written as
\begin{equation}
    \ket{\psi} = \sum_{\vec{z} \in \mathbb{F}_2^N} c_{\vec{z}} \ket{\vec{z}}, \label{quantum_state}
\end{equation}
where $\sum_{\vec{z} \in \mathbb{F}_2^N} |c_{\vec{z}}|^2 = 1$. Measurement of all $N$ qubits produces a bit string $\vec{z}$ with probability $|c_{\vec{z}}|^2$. Such a measurement is equivalent to a simultaneous measurement of the $N$ mutually-commuting single qubit Pauli $z$ operators $\hat{z}_k$, producing, as a measurement result, a binary-string $\vec{z}$, corresponding to measured eigenvalues $(-1)^{z_k}$ for the Pauli $\hat{z}$ operators, as $\hat{z}_k \ket{\vec{z}} = (-1)^{z_k}\ket{\vec{z}}$.

The binary vectors $\vec{z}$ defining the eigenvalues of $\hat{z}_k$ constitute the elementary outputs of the quantum computer. These bit-strings, together with knowledge of the quantum circuit used to prepare the measured states, are used to estimate all other desired quantities. In a basic measurement experiment, one prepares and measures the state $\ket{\psi}$ of \eq{quantum_state} a total of $M$ times, producing result $\vec{z}$ a total of $M_{\vec{z}}$ times. Then, since $|c_{\vec{z}}|^2$ is equal to the probability of obtaining $\vec{z}$ as a measurement result, it can be estimated from the measurement results as $M_{\vec{z}} / M$. As measurements are always done in the $\hat{z}$ basis, the simplest quantities to estimate beyond $|c_{\vec{z}}|^2$ are expectation values of qubit Hamiltonians which only contain $\hat{z}$ operators. Such Hamiltonians, which are called Ising Hamiltonians, take the following form
\begin{align}
    \hat{H}_{\text{ising}} &= d_0 + \sum_{k=1}^{N} d_k \hat{z}_k + \sum_{k>l}^{N} d_{kl} \hat{z}_k \hat{z}_l + \cdots\\
    &= p(\hat{z}_1,\ldots,\hat{z}_N),
\end{align}
where $p$ is a polynomial function. The eigendecomposition allows us to write $\hat{H}_{\text{ising}}$ in terms of its own eigenvalues and eigenvectors as follows
\begin{align}
    \hat{H}_{\text{ising}} &= \sum_{\vec{z} \in \mathbb{F}_2^N} E_{\vec{z}} \ket{\vec{z}} \bra{\vec{z}}\nonumber\\
    E_{\vec{z}} &= p\big((-1)^{z_1},\ldots,(-1)^{z_N}\big).\label{ising_eigendecomp}
\end{align}
In \eq{ising_eigendecomp}, $E_{\vec{z}} \in \mathbb{R}$ is the eigenvalue of $\hat{H}_{\text{ising}}$ corresponding to eigenvector $\ket{\vec{z}}$, obtained by substituting $(-1)^{z_1},\ldots,(-1)^{z_N}$ into the polynomial defining $\hat{H}_{\text{ising}}$. The estimator of $\braket{\psi|\hat{H}_{\text{ising}}|\psi}$ is then
\begin{equation}
\bar{H}_{\text{ising}} = \sum_{\vec{z} \in \mathbb{F}_2^N} \frac{M_{\vec{z}}}{M} E_{\vec{z}}.\label{ising_estimator}
\end{equation}
Note that, although \eq{ising_estimator} contains exponentially many terms, a polynomial number of measurements will produce a polynomial number of non-zero $M_{\vec{z}}$.

\begin{figure*}
    \includegraphics[width=0.9\textwidth]{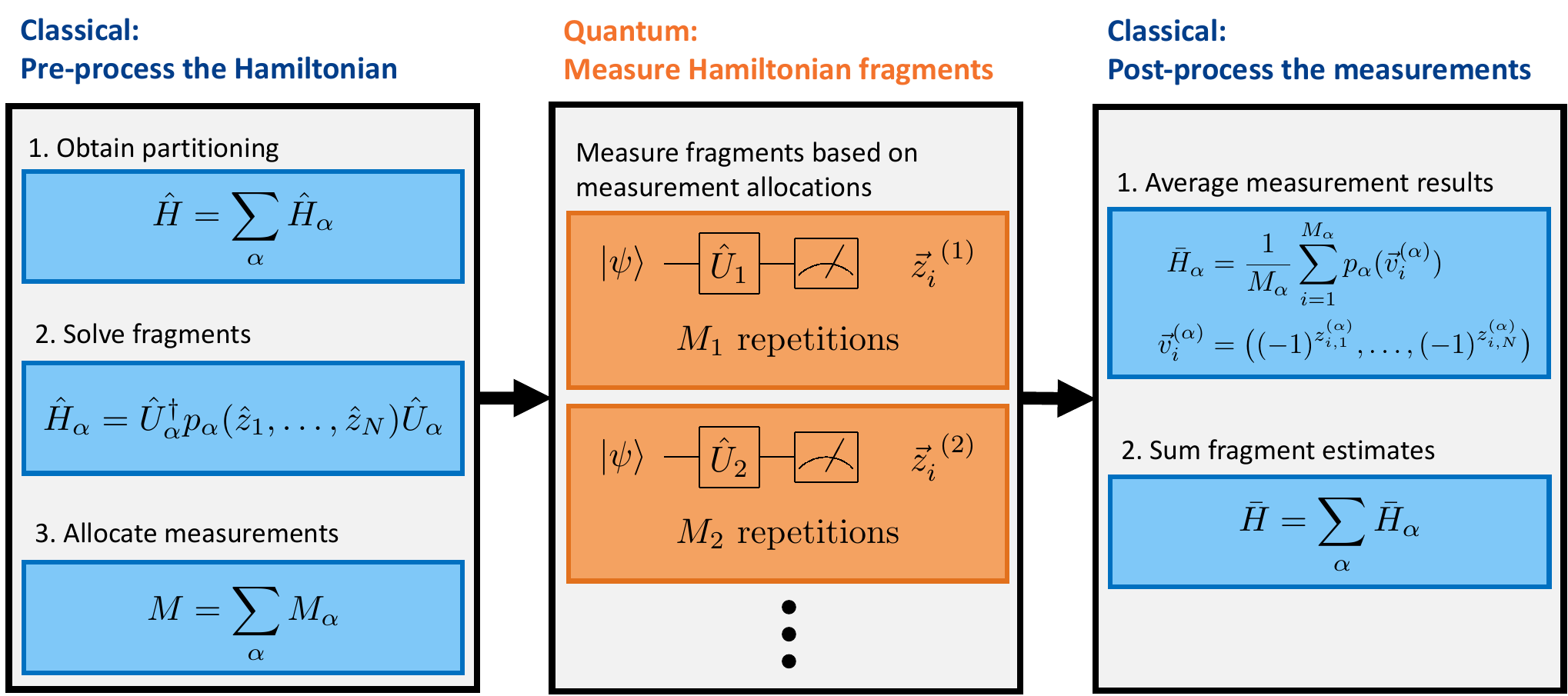}
    \caption{Diagrammatic representation of the framework for Hamiltonian energy estimation using a partitioning of the Hamiltonian to measurable fragments, which can be divided into three components. In the first, the classical computer obtains the Hamiltonian partitioning $\hat{H} = \sum_\alpha \hat{H}_\alpha$, solves the fragments to obtain the measurement circuits $\hat{U}_\alpha$ and the Ising polynomial $p_\alpha(\hat{z}_1,\ldots,\hat{z}_N)$ encoding the energies, and allocates a total of $M$ measurements to the fragments (see Sec. \ref{Sec:MeasurementNReduction} for a discussion of measurement allocations). In the second, the quantum computer is used to prepare the target state $\ket{\psi}$, and measure the fragments. In the third, the classical computer processes the measurement results to obtain an estimate $\bar{H}$ of $\braket{\psi|\hat{H}|\psi}$.}
    \label{partitioning_procedure_fig}
\end{figure*}

The projection operators $\ket{\vec{z}}\bra{\vec{z}}$ onto the computational basis constitute a complete set of orthogonal projection operators, also known as a projection valued measure (PVM), defining an eigenbasis for a general Ising Hamiltonian. Most physical Hamiltonians are not Ising Hamiltonians, and therefore contain terms with Pauli $\hat{x}$ and $\hat{y}$ operators. The computational basis is not, in general, an eigenbasis of such Hamiltonians. However, because Ising Hamiltonians correspond exactly to the set of Hermitian operators which are diagonal in the computational basis, any qubit Hamiltonian $\hat{H}$ can be transformed to an Ising Hamiltonian by a unitary transformation $\hat{U}$:
\begin{equation}
    \hat{H} = \hat{U}^\dagger p(\hat{z}_1,\ldots,\hat{z}_N) \hat{U},\label{h_diag}
\end{equation}
Any $\hat{U}$ satisfying \eq{h_diag} corresponds to a PVM $\{\hat{U}\ket{\vec{z}}\bra{\vec{z}}\hat{U}^\dagger\}$ defining an eigenbasis of $\hat{H}$. Therefore, $\hat{H}$ can be measured by transforming to the eigenbasis of $\hat{H}$ via $\hat{U}$,
\begin{equation}
    \braket{\psi|\hat{H}|\psi} = \braket{\hat{U}\psi|p(\hat{z}_1,\ldots,\hat{z}_N)|\hat{U}\psi},\label{h_measure}
\end{equation}
Equation (\ref{h_measure}) implies that a sufficient condition for measurability of $\hat{H}$ is the ability to find such a unitary transformation $\hat{U}$ diagonalizing $\hat{H}$, which can be expressed as a quantum circuit. However, finding $\hat{U}$ is equivalent to diagonalizing the Hamiltonian $\hat{H}$, which is exponentially hard in the number of qubits for a general $\hat{H}$. This constitutes the measurement problem in quantum simulation. In the remainder of this section, we describe two popular approaches to measuring general Hamiltonians on a quantum computer: Hamiltonian partitioning (Sec. \ref{partitioning_subsec}) and classical shadows (Sec. \ref{classical_shadow_subsec}). Both approaches circumvent the need to find a $\hat{U}$ diagonalizing $\hat{H}$ by instead using unitaries which diagonalize fragments of $\hat{H}$, which are much simpler to describe and implement.

\subsection{Hamiltonian Partitioning into Measurable Fragments\label{partitioning_subsec}}

The target Hamiltonian is often not possible to measure directly on a quantum computer. A simple solution is to identify classes of Hamiltonians which can be directly measured via a polynomial sized quantum circuit. Such Hamiltonians can be used in decompositions of the target Hamiltonian into measurable fragments:
\begin{equation}
    \hat{H} = \sum_\alpha \hat{H}_\alpha,\label{partitioning}
\end{equation}
where each $\hat{H}_\alpha$ can be measured directly on the quantum computer. Linearity of the expectation value implies
\begin{equation}
    \braket{\psi|\hat{H}|\psi} = \sum_\alpha \braket{\psi|\hat{H}_\alpha|\psi}.
\end{equation}
Therefore, this is sufficient to obtain an estimator for the target Hamiltonian expectation value:
\begin{align}
    \bar{H} &= \sum_\alpha \bar{H}_\alpha\nonumber\\
    \bar{H}_\alpha &= \frac{1}{M_\alpha}\sum_{i=1}^{M_\alpha} H_{\alpha,i},\label{expectation_estimators}
\end{align}
where $H_{\alpha, i}$ is the result of the $i^{\text{th}}$ measurement of $\braket{\psi|\hat{H}_\alpha|\psi}$, $\bar{H}_\alpha$ are the empirical means, and $M_\alpha$ is the total number of times that $\braket{\psi|\hat{H}_\alpha|\psi}$ is measured. The full procedure to obtain $\bar{H}$ using a partitioning of $\hat{H}$ to measurable fragments is shown in \fig{partitioning_procedure_fig}. To use Hamiltonian partitioning in practice, we must identify classes of Hamiltonians which can be directly measured on a quantum computer, i.e. easily diagonalized by a unitary whose quantum circuit decomposition is known, and develop classical partitioning algorithms to obtain the decomposition \eq{partitioning}.

\subsubsection{Fragments Formed from Commuting Pauli Operators\label{commuting_pauli_subsec}}

The simplest class of measurable Hamiltonians are those which are already in Ising form: $\hat{H}_{\text{ising}} = p(\hat{z}_1,\ldots,\hat{z}_N)$, which were introduced in Sec. \ref{Subsec:IsingH}. As most physically relevant Hamiltonians, like the electronic Hamiltonian, are not Ising Hamiltonians, additional classes of measurable Hamiltonians are needed. 

A natural measurable generalization of Ising Hamiltonians in the qubit algebra are fully-commuting (FC) Hamiltonians, which are polynomial functions $\hat{H}_{\text{fc}} = p(\hat{C}_1,\ldots,\hat{C}_K)$ of a set of mutually-commuting, independent Pauli operators. \cite{yenMeasuringAllCompatible2020, gokhaleMinimizingStatePreparations2019, jenaPauliPartitioningRespect2019} Independent in this context means that no $\hat{C}_i$ can be written as a product, up to a phase, of the other $\hat{C}_j$, $j\not=i$. Such Hamiltonians can be transformed to all $\hat{z}$ form $\hat{U}_c \hat{H}_{\text{fc}} \hat{U}_c^\dagger = p(\hat{z}_1,\ldots,\hat{z}_K)$ by a Clifford transformation $\hat{U}_c$.  One can find the correct Clifford transformation using a reduction of the problem to a linear system of equations over the binary field $\mathbb{F}_2$, which was introduced in Ref. \cite{calderbankQuantumErrorCorrection1997}. Any Clifford transformation can be implemented using a quantum circuit consisting of $O(N^2 / \log(N))$ controlled-not (CNOT), phase, and Hadamard gates. \cite{aaronsonImprovedSimulationStabilizer2004} On NISQ computers, where the fidelity of the state drops with the circuit depth, general Clifford transformations could prove problematic, motivating the usage of restricted classes of FC Hamiltonians which can be solved by simpler Clifford transformations. See Sec. \ref{Sec:NoiseReduction}
for a more thorough discussion of the errors and error mitigation techniques in quantum measurement. 

A special case of FC Hamiltonians with shorter measurement circuits are qubit-wise commuting (QWC) Hamiltonians, in which, for all Pauli terms $\hat{P} = \otimes_{i=1}^N \hat{\sigma}_i$, $\hat{Q} = \otimes_{i=1}^N \hat{\tau}_i$ in the Hamiltonian, $[\hat{\sigma}_i, \hat{\tau}_i] = 0$ for all $i$. \cite{verteletskyiMeasurementOptimizationVariational2020} Such Hamiltonians sacrifice flexibility compared to FC Hamiltonians, and thus produce smaller fragments. However, since all Pauli terms in a QWC fragment share a tensor product basis (TPB), they can be mapped to Ising form by single-qubit Clifford transformations which contain no CNOT gates. Therefore the measurement circuits are less susceptible to errors caused by decoherence in NISQ architectures. Ref. \cite{dalfavero$k$commutativityMeasurementReduction2024} introduced measurable classes of operators which interpolate between the QWC and FC classes, by allowing for commutativity between Pauli factors of $\hat{P}$ and $\hat{Q}$ which act on sets of qubits, rather than individual qubits.

\begin{figure}
    \includegraphics[width=0.80\columnwidth]{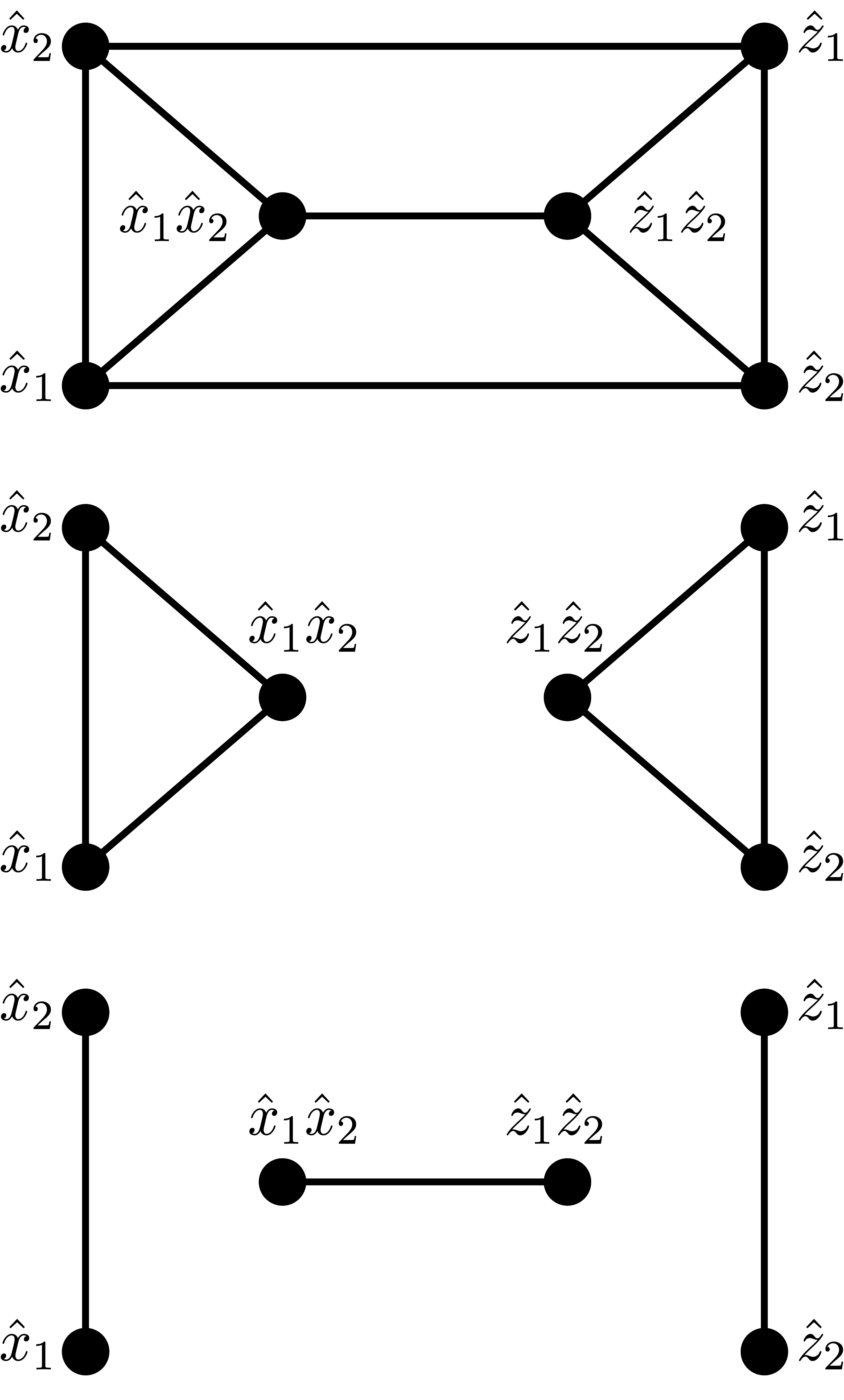}
    \caption{Compatibility graph for the set of Pauli operators $\{\hat{x}_1, \hat{x}_2, \hat{z}_1, \hat{z}_2, \hat{x}_1 \hat{x}_2, \hat{z}_1\hat{z}_2\}$ (upper) and two partitionings of the graph into cliques (middle, lower).}
    \label{FC_graph_fig}
\end{figure}

To obtain a partitioning of the Hamiltonian into FC fragments, one can use an encoding of the network of commutation and anti-commutation relations between the Pauli terms in the Hamiltonian as a graph. The graph, called the ``compatibility'' or ``commutation'' graph, is defined by associating, to each Pauli operator in the target Hamiltonian $\hat{H}$, a vertex, and connecting vertices with an edge if and only if their corresponding Pauli terms commute - see Fig. \ref{FC_graph_fig} for an example. Since any two Pauli operators either commute or anti-commute, the compatibility graph of a Hamiltonian losslessly encodes the commutation relations between the Hamiltonian terms. In this formulation, cliques in the compatibility graph (i.e. complete subgraphs) correspond to groups of Hamiltonian terms which all mutually commute. Therefore, one can obtain a decomposition of $\hat{H}$ to FC fragments by obtaining a partitioning of its compatibility graph into disjoint cliques, which defines a so-called clique cover problem in graph theory. The partitioning into cliques is not unique, as demonstrated in Fig. \ref{FC_graph_fig}, and finding optimal clique covers is in general NP-hard. Thus,  various heuristic polynomial-time algorithms have been developed to obtain approximate solutions. The algorithm used to obtain the clique cover has substantial influence on the subsequent measurement cost. \cite{verteletskyiMeasurementOptimizationVariational2020,yenMeasuringAllCompatible2020,Crawford2021efficientquantum}

One can also search directly in the fermionic representation for Hamiltonian fragments which transform to FC Hamiltonians under a fermion-to-qubit transformation. For what follows, we define the fermionic excitation operator $\hat{E}_{pq}$ which annhilates an electron in spin orbital $q$ and creates an electron in spin orbital $p$:
\begin{equation}
    \hat{E}_{pq} = \hat{a}_p^\dagger \hat{a}_q.
\end{equation}
The simplest class of Hamiltonians which easily transform to FC Hamiltonians are polynomial functions of number operators $\hat{n}_p = \hat{E}_{pp}$ in the orbital basis used to write the second quantized Hamiltonian. Such Hamiltonians map to Ising Hamiltonians under the Jordan-Wigner transformation, as $\hat{n}_p$ maps to $(1 - \hat{z}_p)/2$, implying that one cannot partition the entire Hamiltonian into such fragments unless the Hamiltonian has been diagonalized already. An alternative is to group together Hermitian excitation operators $\hat{E}_{pq} + \hat{E}_{qp}, \hat{E}_{pq} \hat{E}_{rs} + \hat{E}_{sr} \hat{E}_{qp}$ to form fragments which are mapped to FC Hamiltonians by a fermion-to-qubit transformation. This can be done by noting that any individual excitation operator is mapped to an FC Hamiltonian by a fermion-to-qubit transformation, and that Pauli operators obtained from the Jordan-Wigner transformation of any pair of two-body excitations which have completely disjoint indices will commute. Therefore, one can group together $O(N)$ mutually commuting excitations to obtain $O(N^3)$ measurable fragments of $\hat{H}$, as was done in Refs. \cite{gokhaleMeasurementCostBaranyai} and \cite{csakany2023OptimisedBaranyaiPartitioning2023} using Baranyai's theorem from the theory of hypergraphs. \cite{baranyaiEdgecoloringCompleteHypergraphs1979} We note that one can similarly also use heuristic graph partitioning algorithms on the compatibility graph of the excitation operators for finding such fragments, as was done for FC and QWC Hamiltonians. 

\subsubsection{Fragments Formed from Commuting One-Electron Operators \label{commuting_ferm_subsec}}

In the fermionic algebra, another class of solvable Hamiltonians are one-electron (also called free-fermionic) Hamiltonians which take the following form
\begin{equation}
    \hat{H}_{1e} = \sum_{p \geq q}^{N} h_{pq} \hat{E}_{pq},
\end{equation}
where $h_{pq} = h_{qp}$ is a symmetric coefficient matrix. Such Hamiltonians, which are in general constructed from non-commuting fermionic excitation terms $\hat{E}_{pq}$, can be diagonalized by an orbital transformation $\hat{U}$ which is generated by an anti-symmetric one-electron operator
\begin{align}
    \hat{U} \hat{H}_{1e} \hat{U}^\dagger &= \sum_{p=1}^{N} \epsilon_p \hat{n}_p\label{eq:one_e_solvability}\\
    \hat{U} &= \exp\left(\sum_{p>q}^{N} \theta_{pq}(\hat{E}_{pq} - \hat{E}_{qp})\right),\label{eq:orbitaltransform}
\end{align}
where $\epsilon_p$ are eigenvalues of the $h_{pq}$ coefficient matrix. The orbital rotation angles satisfy $\theta_{pq} = [\log(u)]_{pq}$, where $u$ is an $N \times N$ unitary matrix which diagonalizes the $h_{pq}$ coefficients, and $\log$ denotes the matrix logarithm. Therefore, the one-electron Hamiltonian can be written as a linear combination $\hat{H}_{1e} = \sum_{p=1}^{N} \epsilon_p \hat{C}_p$ of independent, mutually-commuting, one-electron symmetries $\hat{C}_p = \hat{U}^\dagger \hat{n}_p \hat{U}$ which can be simultaneously mapped to number operator form by the orbital transformation $\hat{U}$.  Exact implementation of $\hat{U}$ as a quantum circuit requires first computing a decomposition of $\hat{U}$ into a product of at most $N(N-1)/2$ elementary fermionic rotations which take the following form
\begin{equation}
    \hat{G}_{pq}(\phi_{pq}) = \exp\big(\phi_{pq}(\hat{E}_{pq} - \hat{E}_{qp})\big).
\end{equation}
$\hat{U}$ can then be written as
\begin{equation}\label{orbital_givens}
\hat{U} = \prod_{p>q}^{N} \hat{G}_{pq}(\phi_{pq}).
\end{equation}
The $\phi_{pq}$ are, in general, non-linear functions of the $\theta_{rs}$ parameters, but can be obtained efficiently using, for example, the method of Ref. \cite{kivlichanQuantumSimulationElectronic2018}. By ordering the product defining $\hat{U}$ in \eq{orbital_givens} such that $\hat{G}_{pq}(\phi_{pq})$ acting on disjoint qubits can be simultaneously applied, and using SWAP gates to bring non-adjacent $p$ and $q$ together when necessary, one can implement $\hat{U}$ with an $O(N)$ depth circuit consisting of $O(N^2)$ quantum gates. \cite{reckExperimentalRealizationAny1994,weckerSolvingStronglyCorrelated2015,kivlichanQuantumSimulationElectronic2018,jiangQuantumAlgorithmsSimulate2018,arrazola_universal_2022}

One can obtain general fermionic Hamiltonians that are solvable by an orbital transformation by considering polynomial functions of one-electron mutually-commuting operators in a fixed but arbitrary orbital basis: 
\begin{equation}
    \hat{H}_{2e} = \sum_{p \geq q}^{N} \lambda_{pq} \hat{C}_p \hat{C}_q, \quad \hat{C}_p = \hat{U}^\dagger \hat{n}_p \hat{U}\label{ferm_csa}
\end{equation}
where $\hat{U}$ is a real-orbital transformation, and $\lambda_{pq}$ are real-number coefficients. Here, we restrict ourselves to quadratic polynomials since the electronic Hamiltonian is two-body, but higher-degree polynomials of the $\hat{C}_p$ are still solvable by $\hat{U}$. Such Hamiltonians are linear combinations of non-commuting one- and two-body hermitian excitation operators, but can be transformed to a polynomial of number operators $\hat{U}\hat{H}_{2e}\hat{U}^\dagger = \sum_{p\geq q} \lambda_{pq} \hat{n}_p \hat{n}_q$ by the orbital transformation $\hat{U}$. The solvable fragments can be obtained via non-linear optimization of some distance function of the Hamiltonian to the fragments:
\begin{equation}
    f(\lambda_{pq}^{(\alpha)}, \hat{U}^{(\alpha)}) = \left|\left| \hat{H}_e -  \sum_\alpha U_\alpha^\dagger \left(\sum_{p \geq q}\lambda_{pq}^{(\alpha)}  \hat{n}_p \hat{n}_q \right)\hat{U}_\alpha\right|\right|\label{csa_cost}
\end{equation}
Typically, either the $\ell_1$ or $\ell_2$ norm of the Hamiltonian two-body-tensor in a spatial orbital basis is used, but alternative cost functions which also include the one-body-tensor can also be used. This can be cast as an un-constrained optimization by using the parametrization \eq{eq:orbitaltransform} for the orbital rotations $\hat{U}_\alpha$. This non-linear optimization can be very expensive, even for small systems, due to the $O(N^5)$ complexity in evaluating the orbital transformation. \cite{yoshimineConstructionHamiltonianMatrix1973} To mitigate this cost, one can use a ``greedy'' approach and obtain each fragment one at a time. As discussed in Sec. \ref{reduction_greedy}, the greedy approach is also beneficial for reducing the measurement cost associated to the subsequent decomposition. \cite{yen2021cartan} In Ref. \cite{cohnQuantumFilterDiagonalization2021}, it was proposed to further reduce the cost by optimizing the $\hat{U}^{(\alpha)}$ and the $\lambda_{pq}^{(\alpha)}$ separately. In this formulation, the $\lambda_{pq}^{(\alpha)}$ can be globally optimized via least-squares. 

In the above approaches, the optimized coefficient matrix $\lambda_{pq}^{(\alpha)}$ can be a full-rank matrix, and the obtained fragments are referred to as full-rank (FR) fragments. Alternatively, a more efficient approach which circumvents the non-linear optimization altogether requires the usage of a constrained, low-rank (LR) coefficient matrix $\lambda_{pq}^{(\alpha)} = \epsilon_p^{(\alpha)} \epsilon_q^{(\alpha)}$ in the definition of the fragments, such that the fragments can be expressed as a square of a solvable one-body Hamiltonian $\left(\sum_{p=1}^{N} \epsilon_p^{(\alpha)} \hat{C}_p\right)^2$. In contrast to the FR fragments defined in \eq{ferm_csa}, these can be found via a singular-value decomposition (SVD) or Cholesky decomposition of the target Hamiltonian two-body-tensor $g_{pq,rs}$ when regarded as a $N^2 \times N^2$ dimensional matrix, and can thus be obtained much more efficiently than the methods relying on a non-linear optimization of the orbital transformation angles. \cite{pengHighlyEfficientScalable2017,mottaLowRankRepresentations2021,Huggins_Babbush:2021}

\subsubsection{Cartan Sub-Algebra Decomposition\label{csa_subsec}}

One can view the commuting symmetries defining the fermionic fragment types (\eq{ferm_csa}) within a Lie algebraic framework. A set of Hermitian operators $\{\hat{X}_i\}_{i=1}^{d}$ generates a $d$ dimensional compact Lie algebra $\mathcal{A} = \text{span}\{\hat{X}_i\}$  if there exist so-called structure constants $\{c_{ij}^{(k)}\}$ which satisfy \cite{hallLieGroupsLie2015} 
\begin{equation}
    [\hat{X}_i, \hat{X}_j] = \imag\sum_k c_{ij}^{(k)} \hat{X}_k.
\end{equation}
Associated to the compact Lie algebra $\mathcal{A}$ are the Cartan sub-algebras (CSA), which are any maximal sub-algebra $\mathcal{C} \subset \mathcal{A}$ consisting of only mutually-commuting operators. Furthermore, associated to $\mathcal{A}$ is the compact connected Lie group of unitaries of the form $\exp(\imag\hat{X})$ for $\hat{X} \in \mathcal{A}$. As a consequence of the maximal torus theorem for compact Lie groups, if a Hamiltonian $\hat{H}$ is an element of $\mathcal{A}$, then it can be transformed to an element of a CSA with generators $\{\hat{C}_k\}_{k=1}^{r}$, by a unitary $\hat{U}$ in the corresponding Lie group. Then, one can write:
\begin{equation}
    \hat{H} = \sum_{i=1}^{d} c_i \hat{X}_i = \sum_{k=1}^{r} \epsilon_k \hat{U}^\dagger \hat{C}_k \hat{U},
\end{equation}
Therefore, since the operators $\{\hat{U}^\dagger \hat{C}_k \hat{U}\}$ mutually commute, and $\hat{H}$ is constructed from a linear combination of them, it follows that they form $r$ symmetries of $\hat{H}$. 

One can also construct Hamiltonians which are polynomial functions of the Lie algebra generators
\begin{equation}
    \hat{H} = c_0 + \sum_{i=1}^{d} c_i \hat{X}_i + \sum_{i\geq j}^{d} c_{ij} \hat{X}_i \hat{X}_j + \cdots,
\end{equation}
which is not an element of the Lie algebra $\mathcal{A}$, but rather, an element of the universal enveloping algebra (UEA) $\mathcal{E}_{\mathcal{A}}$ defined as follows:
\begin{equation}
    \mathcal{E}_{\mathcal{A}} = \mathbb{F} \oplus \mathcal{A} \oplus \mathcal{A}^2 \oplus \mathcal{A}^3 \oplus \cdots,
\end{equation}
where $\mathbb{F}$ is the field ($\mathbb{R}$ or $\mathbb{C}$ in practice), and $\mathcal{A}^k$ is the $k$-fold tensor product of $\mathcal{A}$. The associated CSAs $\mathcal{E}_{\mathcal{C}}$ are formed from polynomial functions of the mutually-commuting generators of $\mathcal{C}$. Although the UEA is itself a compact Lie algebra, its dimension is typically exponential in the dimension of $\mathcal{A}$, implying that finding the diagonalizing unitary which solves $\hat{H} \in \mathcal{E}_{\mathcal{A}}$ is prohibitively expensive. This is true even if $\hat{H}$ is a low degree polynomial in $\mathcal{E}_{\mathcal{A}}$ (e.g., an element of $\mathcal{A}^2$), as this does not imply that the corresponding diagonalizing unitary will be generated by a low degree polynomial in $\mathcal{E}_{\mathcal{A}}$. 

The solvability of one-electron fermionic Hamiltonians can be seen as a realization of the Lie algebraic formalism to the $u(N)$ algebra generated by $\{\hat{E}_{pq} + \hat{E}_{qp}\} \cup \{\imag(\hat{E}_{pq} - \hat{E}_{qp})\}$, whose anti-symmetric terms $\{\imag(\hat{E}_{pq} - \hat{E}_{qp})\}$ generate an $so(N)$ sub-algebra. For both algebras, the corresponding CSA is generated by the number operators $\{\hat{n}_p\}$ \cite{izmaylovHowDefineQuantum2021, yen2021cartan} Orbital transformations, which have one-electron generators, form the corresponding Lie group elements that map the one-electron Hamiltonian to a commuting form. Within the Lie algebraic framework, the FR and LR fragments are described as quadratic polynomial functions of $u(N)$ CSA generators $\{\hat{U}^\dagger \hat{n}_p \hat{U}\}$. \cite{yen2021cartan} Although such quadratic polynomials belong to the exponentially large UEA of $u(N)$, their restricted form as polynomials of CSA generators allows them to be solved efficiently. 

In the qubit algebra, the analogue are qubit-mean-field (QMF) Hamiltonians \cite{ryabinkinRelationFermionicQubit2018,izmaylovHowDefineQuantum2021} which are polynomial functions $\hat{H}_{\text{QMF}} = p(\hat{S}_1,\ldots,\hat{S}_N)$ of a set of single qubit Hamiltonians
\begin{equation}
    \hat{S}_k = \alpha_k \hat{x}_k + \beta_k \hat{y}_k + \gamma_k \hat{z}_k + \delta_k \hat{1}_k,
\end{equation}
where $k$ denotes the qubit that $\hat{S}_k$ acts non-trivially on. The $\hat{S}_k$ form $N$ mutually-commuting single-qubit symmetry operators of $\hat{H}_{\text{QMF}}$ which can be mapped to Ising form by a rotation of the corresponding qubit
\begin{equation}
    \hat{S}_k = e^{-\imag\theta_k \hat{\eta}_k} \big(\epsilon_k\hat{z}_k +\delta_k \hat{1}_k\big)e^{\imag\theta_k \hat{\eta}_k},
\end{equation}
where $\hat{\eta}_k = \tau_x^{(k)} \hat{x}_k +  \tau_y^{(k)} \hat{y}_k +  \tau_z^{(k)} \hat{z}_k$ is a unit vector on the Bloch sphere of the $k$th qubit, and $\epsilon_k = \sqrt{\alpha_k^2 + \beta_k^2 + \gamma_k^2}$. Therefore, $\hat{H}$ can be mapped to Ising form by a product of these single-qubit rotations. The associated algebra within the Lie algebraic framework defining this class of Hamiltonians is the direct sum $\oplus_{k=1}^{N} u(2)$ of algebras generated by $\{\hat{x}_k, \hat{y}_k, \hat{z}_k, \hat{1}_k\}$ associated to each qubit, which has a $2N$ dimensional CSA generated by $\{\hat{z}_k, \hat{1}_k\}$. QMF fragments are challenging to obtain, both due to the non-linear search procedure one must use, as well as the fact that the non-linear nature of fermion-to-qubit transformations implies that a high degree QMF polynomial is necessary to recover all the Pauli terms present in the target fermionic Hamiltonian. Therefore, QMF Hamiltonians have not been used to generate measurable fragments of the electronic Hamiltonian.

For identification of alternative Lie algebras to construct measurable fragments, one can use the realization of the electronic Hamiltonian in terms of Majorana-fermion operators which take the following form
\begin{align}
    \hat{\gamma}_{2p} &= \hat{a}_p^\dagger + \hat{a}_p\nonumber\\
    \hat{\gamma}_{2p+1} &= \imag(\hat{a}_p^\dagger - \hat{a}_p),\label{maj_from_ferm}
\end{align}
under which the electronic Hamiltonian can be written as
\begin{equation}
    \hat{H} = 2\imag\sum_{ij=1}^{2N} \tilde{h}_{ij} \hat{\gamma}_i \hat{\gamma}_j + \sum_{ijkl=1}^{2N} \tilde{g}_{ijkl} \hat{\gamma}_i \hat{\gamma}_j \hat{\gamma}_k \hat{\gamma}_l.\label{majorana_form}
\end{equation}
Similarly with fermionic operators $\hat{a}_p, \hat{a}_p^\dagger$, non-interacting Majorana-fermionic Hamiltonians, which are quadratic in the Majorana operators
\begin{equation}
    \hat{H}_{1m} = 2\imag\sum_{i>j}^{2N} h_{ij} \hat{\gamma}_i \hat{\gamma}_j
\end{equation}
can also be directly measured on a quantum computer. This follows since the set of Majorana operators $\{\hat{\gamma}_i\}_{i=1}^{2N}$ are mutually anti-commuting, involutory operators, implying they satisfy the Clifford algebra relation
\begin{equation}
    \{\hat{\gamma}_i, \hat{\gamma}_j\} = 2\delta_{ij},
\end{equation}
where $\{\hat{A},\hat{B}\} = \hat{A}\hat{B} + \hat{B}\hat{A}$ denotes the anti-commutator. The commutator closure of the Clifford algebra generators produces an $so(2N+1)$ Lie algebra generated by $\{\hat{\gamma}_i\}_{i=1}^{2N} \cup \{\imag\hat{\gamma}_i \hat{\gamma}_j\}_{i>j}^{2N}$. Furthermore, the purely quadratic elements $\{\imag\hat{\gamma}_i \hat{\gamma}_j\}_{i>j}^{2N}$ generate an $so(2N)$ Lie sub-algebra. In both cases, the CSA is generated by $N$ mutually commuting operators $\{\imag\hat{\gamma}_{2k-1} \hat{\gamma}_{2k}\}_{k=1}^{N}$. Therefore, the Hamiltonian $\hat{H}_{1m}$ can be transformed to the CSA by a so-called fermionic Gaussian unitary (FGU) in $\text{Spin}(2N)$ which takes the following form:
\begin{equation}
    \hat{U}_g = \exp\left(-\sum_{i>j}^{N}\theta_{ij} \hat{\gamma}_i \hat{\gamma}_j\right)\label{fgu_definition}
\end{equation}
such that
\begin{equation}
    \hat{U}_g \hat{H}_{1m} \hat{U}_g^\dagger = 2\imag \sum_{k=1}^{N} \epsilon_k \hat{\gamma}_{2k-1} \hat{\gamma}_{2k}.
\end{equation}
Noting the relationship $\imag\hat{\gamma}_{2k-1} \hat{\gamma}_{2k} = 1 - 2 \hat{n}_k$, which follows from \eq{maj_from_ferm}, this implies that applying $\hat{U}_g$ on the quantum computer is necessary to measure $\hat{H}_{1m}$. The set of fermionic Gaussian unitaries coincides with the class of quantum circuits known as nearest-neighbour matchgate circuits, which are generated by a restricted set of two-qubit Pauli operators acting on adjacent qubits, and can always be implemented with a poly$(N)$ depth circuit. \cite{valiantQuantumComputersThat2001, terhalClassicalSimulationNoninteractingfermion2002, jozsaMatchgatesClassicalSimulation2008, brodEfficientClassicalSimulation2016}

The Clifford algebra - Lie algebra connection allows one to construct a solvable Hamiltonian out of any set of involutory Majorana or Pauli operators $\{\hat{A}_i\}_{i=1}^{K}$ which mutually anti-commute, and therefore generate an $so(K+1)$ Lie algebra. In the anti-commuting grouping (AC grouping) approach, one obtains fragments of $\hat{H}$ of the form $\hat{H}_\text{ac} = \sum_{i=1}^{K} c_i \hat{A}_i$, where $\{A_i\}_{i=1}^{K}$ mutually anti-commute.  \cite{zhaoMeasurementReductionVariational2020,izmaylovUnitaryPartitioningApproach2020} Such a fragment can be transformed to $||\vec{c}|| \hat{A}_1$ by a unitary in the associated Lie group, where $||\vec{c}||$ denotes the $\ell_2$ norm of the coefficients. To find such fragments, one can use the fact that the mutually anti-commuting Pauli terms generate independent sets in the compatibility graph of $\hat{H}$. Equivalently, one can consider the compliment graph to the compatibility graph, called the ``frustration'', ``anti-compatibility'', or ``anti-commutation'' graph, in which two vertices share an edge if and only if their corresponding Pauli operators anti-commute. In the anti-compatibility graph, mutually anti-commuting Pauli operators form cliques, implying that one can obtain the decomposition by solving the clique cover problem on the anti-compatibility graph. In comparison to the FC grouping, the AC grouping produces smaller fragments, as the largest anti-commuting sets on $N$ qubits only contain $2N + 1$ Pauli operators, compared with $2^N$ for fully-commuting sets. \cite{bonet-monroigNearlyOptimalMeasurement2020} The larger size for fully-commuting sets is a consequence of the fact that the property of being commutative is closed under products, whereas anti-commutativity is not closed under products. Nevertheless, there exists measurement techniques which require a decomposition of the target Hamiltonian into a linear combination of unitaries (LCU), for which such a decomposition can be used (see Sec. \ref{hadamard_test_lcu_section} for details).

Recent works have looked at various extensions to the FC and AC solvable Hamiltonian types which are still measurable. Reference \cite{patelExactlySolvableHamiltonian2024} introduced so-called term-wise commuting AC (TWC-AC) Hamiltonians which generalize both the FC and the AC classes. They are constructed as a sum $\hat{H}_\text{twc-ac} = \sum_{i} \hat{H}^{(i)}_{\text{ac}}$ of AC Hamiltonians $\hat{H}^{(i)}_{\text{ac}}$ with the additional property that all Pauli products in one AC term $\hat{H}^{(i)}_{\text{ac}}$ commute with all Pauli products in another AC term $\hat{H}^{(j)}_{\text{ac}}, i \ne j$. The commutativity between AC terms allows the TWC-AC Hamiltonian to be transformed to FC form by a product of Lie group rotations. To find such fragments, one must solve a generalization of the clique cover problem, as the anti-compatibility graphs of TWC-AC Hamiltonians are defined by the property that they are a disjoint union of cliques, rather than a single clique. Another extension of AC Hamiltonians was proposed in Ref. \cite{sawayaNonCliffordDiagonalizationMeasurement2024}. Here, the measurable Hamiltonians are solvable due to a locality property, and take the following form
\begin{align}
    \hat{H}_{k-\text{local}} &= \sum_r \hat{W}_r\\
    \hat{W}_r &= A_{r,1}^{(S_{r,1})} \otimes A_{r,2}^{(S_{r,2})} \otimes \cdots,
\end{align}
where $S_{r,i}$ is the subset of qubits that $\hat{A}_{r,i}^{(S_{r,i})}$ acts on, and $|S_{r,i}| \leq k$. The measurability of $\hat{H}_{k-\text{local}}$ is enforced by the commutativity constraint $[\hat{W}_r, \hat{W}_s] = 0$, which implies that all $\hat{W}_r$ are simultaneously diagonalizable. The tensor product structure of the $\hat{W}_r$ implies that such a unitary can also be written in a simple tensor product structure, and thus can be found and implemented efficiently.

Another generalization of AC Hamiltonians are those which are formed from the full set of $so(K+1)$ Lie algebra generators associated to a set of mutually anti-commuting Pauli operators, which take the following form \cite{patelExactlySolvableHamiltonian2024}
\begin{equation}
    \hat{H}_{so} = \sum_{i=1}^{K} c_i \hat{A}_i + \imag\sum_{j>k}^{K} d_{jk} \hat{A}_j \hat{A}_k.\label{so_quad}
\end{equation} 
Like with AC Hamiltonians, this can also be extended to sums of term-wise commuting Hamiltonians $\hat{H}_{\text{twc-}so} = \sum_{i=1}^L \hat{H}_{so}^{(i)}$ whose Pauli terms generate a $\oplus_{i=1}^{L} so(K_i + 1)$ algebra. While such fragments are strictly more general than those that can be obtained with AC grouping, Ref. \cite{patelExactlySolvableHamiltonian2024} found two issues which limit their applicability in practice. First, it was found that one can only include a subset of $so(K+1)$ algebra generators within a single fragment of a real-symmetric Hamiltonian $\hat{H}$, like the electronic Hamiltonian. This is because, within any $so(3)$ sub-algebra of $so(L+1)$ generated by $\{\hat{A}_i, \hat{A}_j, \imag\hat{A}_i \hat{A}_j\}$, there will be at least one anti-symmetric operator which cannot be present in any decomposition of a real-symmetric Hamiltonian $\hat{H}$ into fragments, unless the fragments include terms which are not present in the Hamiltonian (e.g., ghost Pauli products, \cite{choi2022improving} described in Sec. \ref{reduction_coefficient_splitting}).  Second, one cannot use the standard graph partitioning algorithms to obtain fragments of this form. In Ref. \cite{chapmanCharacterizationSolvableSpin2020}, it was shown that the anti-compatibility graph of any Hamiltonian whose Pauli terms generate an $so(K+1)$ algebra has a well-defined structure - it takes the form of a so-called ``line graph'' \cite{krausz1943demonstration}  - which can be identified in $O(K^2)$ time. \cite{roussopoulosMaxAlgorithmDetermining1973, lehotOptimalAlgorithmDetect1974, degiorgiDynamicAlgorithmLine1995} However, although the anti-compatibility graph losslessly encodes the full network of commutation and anti-commutation relations between Pauli terms, it discards all other functional relations, and therefore does not encode the information that $\hat{A}_i \hat{A}_j$ is a product of $\hat{A}_i$ and $\hat{A}_j$, which is also necessary for the Lie algebra closure. Therefore, additional information beyond what is encoded in the anti-compatibility graph is necessary to identify operators of the form shown in \eq{so_quad}.

Lastly, the electronic Hamiltonian is itself a polynomial function of Lie algebra generators: one-electron $u(N)$ generators in the fermionic representation; $so(2N)$ generators in the Majorana representation; or $\oplus u(2)$ generators in the qubit representation. These can be thought of as ``mean-field'' Lie algebras, since the elements of these Lie algebras are exactly-solvable single-particle Hamiltonians of the respective particle type. As the electronic Hamiltonian encodes particle interactions, it belongs to the exponentially large UEA of the associated Lie algebras, and thus can be solved in principle by a unitary in the associated Lie group. To mitigate the exponential scaling in this approach, one can also obtain modest sized Lie algebras by constructing UEAs of mean-field algebras on a restricted set of fermionic modes/qubits. In the fermionic algebra, this can be accomplished by defining a set of fermionic modes of size $K$, analogous to a complete active space, and considering all products of one-electron excitations that act on the $K$ spin-orbitals. To find fragments of the electronic Hamiltonian which belong to small UEAs, one can consider all subsets of spin-orbitals of fixed size $K$, and group electronic Hamiltonian excitation terms together based on which subset they interact in. This idea can also work in the qubit space, but it is less practical for the electronic Hamiltonian, since fermion-to-qubit maps are non-locality preserving. Therefore, qubit representations of the electronic Hamiltonian have Pauli terms which act on all qubits.

\subsubsection{Incorporation of Pauli Symmetries\label{symmetries_subsec}}

One can construct more general measurable Hamiltonians within the Lie algebraic formalism in the case when a Lie algebra generated by $\{\hat{X}_a\}$ has a set of mutually commuting, independent, measurable symmetries $\{\hat{C}_k\}$, for which $[\hat{C}_k, \hat{C}_l] = [\hat{C}_k, \hat{X}_a] = 0$ for all $k,l,a$. \cite{patelExtensionExactlySolvableHamiltonians2024} When such symmetries exist, one can form a solvable Hamiltonian by constructing linear combinations of the algebra generators
\begin{equation}
    \hat{H} = \sum_{a=1}^d \hat{c}_a \hat{X}_a \label{symmetry_solvable}
\end{equation}
where the coefficients $\hat{c}_a$ of the linear combination are polynomial functions of the symmetries
\begin{equation}
    \hat{c}_a = p_a(\hat{C}_1,\ldots,\hat{C}_K).
\end{equation}
The terms in $\hat{H}$ are not mutually-commuting, and the Lie algebra they generate will, in general, have a dimension that is exponential in the number of symmetries $\hat{C}_k$. Nonetheless, $\hat{H}$ is still measurable since it decomposes as a direct sum over the simultaneous eigenspaces of the symmetries $\hat{C}_k$, and, within each subspace, it acts as an element of the Lie algebra:
\begin{align}
    \hat{H} \hat{\mathcal{P}}_{\vec{v}} &= \hat{H}_{\vec{v}}\hat{\mathcal{P}}_{\vec{v}}\nonumber\\
    \hat{H}_{\vec{v}} &= \sum_{a=1}^{d} p_a(v_1,\ldots,v_K)\hat{X}_a,
\end{align}
where $\vec{v}$ is a vector of eigenvalues of the $\hat{C}_1,\ldots,\hat{C}_K$ and $\hat{\mathcal{P}}_{\vec{v}}$ is the projection operator onto the associated simultaneous eigenspace. Measuring $\hat{H}$ requires three steps: (1) measure the commuting symmetries $\hat{C}_k$ after transforming them to Ising form, obtaining some result $\vec{v}$, (2) calculate, on a classical computer, the circuit which transforms $\hat{H}_{\vec{v}}$ to Ising form, and (3) measure $\hat{H}_{\vec{v}}$ after applying the derived circuit. Note that, although pre-computation of the circuits which diagonalize $\hat{H}_{\vec{v}}$ is possible, the number of distinct circuits scales exponentially in the number of symmetries $\hat{C}_k$, so it is in general infeasible to pre-compute all the diagonalizing circuits. 

In the qubit algebra, Hamiltonians of this form can be constructed using $\hat{X}_a$ which belong to any Lie algebra described in Sec. \ref{csa_subsec}, and $\hat{C}_k$ which are Pauli operator symmetries. \cite{patelExactlySolvableHamiltonian2024} For such a Hamiltonian, the unitary that diagonalizes the symmetries $\hat{C}_k$ will be a Clifford transformation, and the unitary that diagonalizes the projected Hamiltonian $\hat{H}_{\vec{v}}$ will be a rotation in the Lie group generated by the $\hat{X}_a$, followed by a Clifford transformation. For obtaining solvable fragments of the form shown in \eq{symmetry_solvable}, in the most general case where the $\hat{X}_a$ generate a direct sum of $so$ Lie algebras, one can characterize such Hamiltonians in terms of the structure of their anti-compatibility graphs, and use graph partitioning algorithms. For all $a$, the set of Pauli operators in the term $\hat{c}_a \hat{X}_a$ correspond to twin vertices in the anti-compatibility graph $G$ of $\hat{H}$, meaning that they are adjacent to the exact same vertices in $G$, and are not adjacent to each other. \cite{chapmanCharacterizationSolvableSpin2020, patelExactlySolvableHamiltonian2024} Once these twin vertices are identified and removed, one will obtain the graph of the Lie algebra generators $\{\hat{X}_a\}$ (in general, a disjoint union of line graphs, see Sec. \ref{csa_subsec}). The overall procedure of obtaining the anti-compatibility graph of the Lie algebra generators $\{\hat{X}_a\}$ and checking its structure can be done in time that is polynomial in the number of Pauli terms in $\hat{H}$, thus providing a method to solve the graph partitioning problem necessary to obtain the fragments. 

\begin{figure}
    \includegraphics[width=0.95\columnwidth]{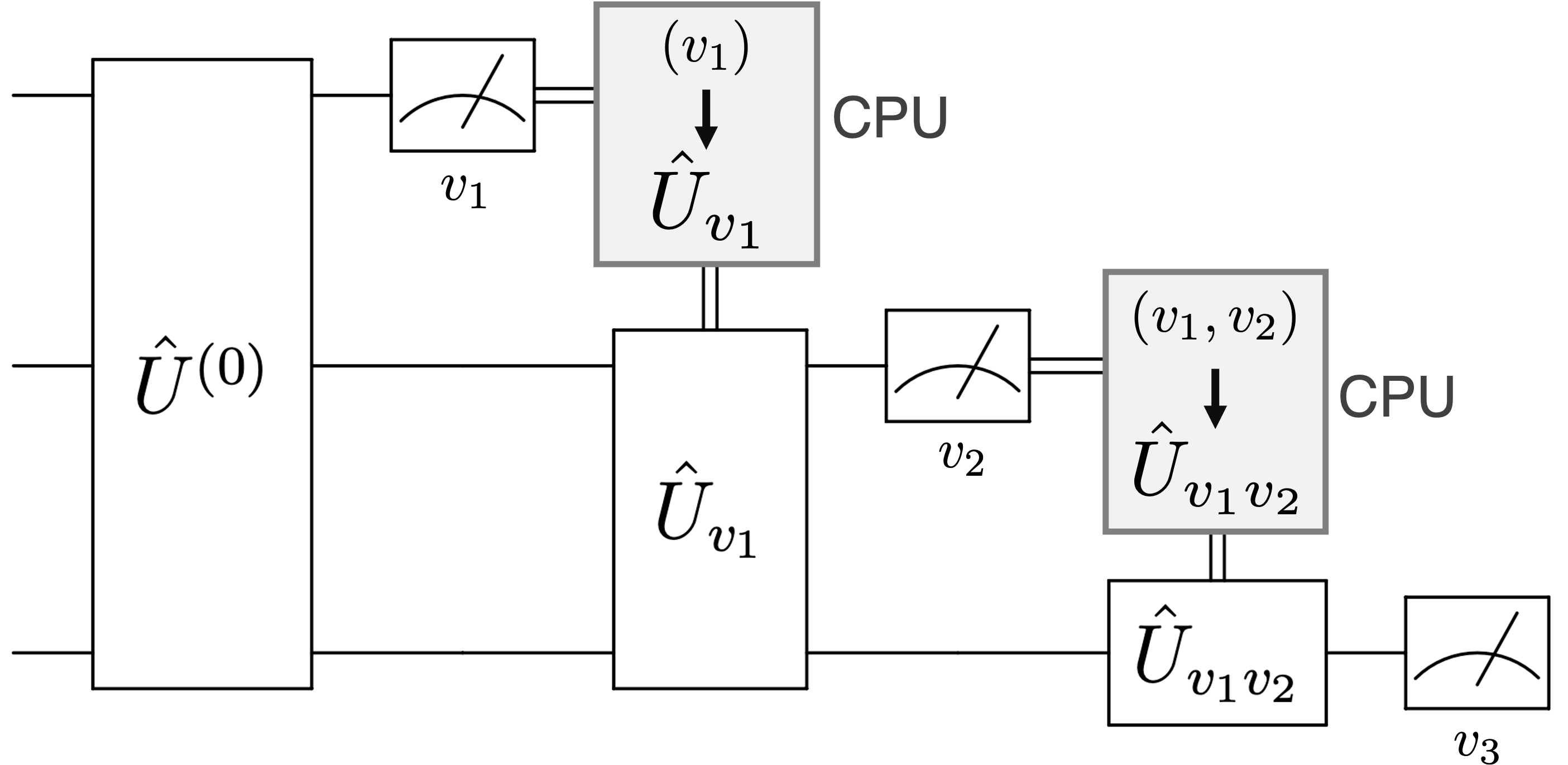}
    \caption{Example of a three qubit feed-forward circuit used to measure a Hamiltonian with single qubit symmetry operator. In this circuit, a unitary $\hat{U}^{(0)}$ is applied before measurement of the first symmetry, which is mapped to $\hat{z}_0$ on the first qubit. The measurement result $v_1$ of the first qubit defines the transformation $\hat{U}_{v_1}$ which transforms the subsequent symmetry to $\hat{z}_1$. Then, the vector of measurement results $(v_1,v_2)$ defines the transformation $\hat{U}_{v_1 v_2}$ which is used to transform the final symmetry to $\hat{z}_2$.}
    \label{feedforward_fig}
\end{figure}

Incorporation of symmetries can also be done in the fermionic or Majorana Lie algebras. In the construction of Ref. \cite{patelExtensionExactlySolvableHamiltonians2024}, the Lie algebra generators are excitation operators $\{ \hat{U}^\dagger \hat{E}_{pq} \hat{U}\}$ on a subset of orbitals $S_1$, and the symmetries are number operators $\{ \hat{U}^\dagger \hat{n}_k \hat{U}\}$ on the remaining orbitals $S_2$. Here, $\hat{U}$ is an orbital rotation that acts on all the orbitals. Then, the measurable fermionic Hamiltonian takes the following form
\begin{equation}
    \hat{H} = \hat{U}^\dagger \Bigg[\sum_{pq \in S_1} \left(\sum_{k \in S_2} h_{pq}^{(k)} \hat{n}_k\right) \hat{E}_{pq} + \sum_{kl \in S_2} \lambda_{kl} \hat{n}_k \hat{n}_l\Bigg] \hat{U}
\end{equation}
For finding such fragments, one can perform a similar non-linear optimization of the two-body-tensor distance shown in \eq{csa_cost} used to obtain the CSA fragments. However, it was found that such fragments, once obtained, yield considerably worse measurement statistics compared with FR and LR fragments.

Another measurement approach which uses Hamiltonians that are not necessarily easily transformable to Ising form by a unitary, but which contain symmetries that can be efficiently measured, was introduced in Ref. \cite{izmaylovRevisingMeasurementProcess2019}. Here, the measurable Hamiltonians were defined inductively as follows. The starting Hamiltonian $\hat{H}$, has a single qubit symmetry operator $\hat{O}_1^{(k_1)}$, on qubit $k_1$, obtained potentially after an initial rotation is applied. Upon measurement of $\hat{O}_1^{(k_1)}$, producing value $v_1 \in \{-1,1\}$, one obtains an effective Hamiltonian $\hat{H}(v_1) = \hat{H} \hat{\mathcal{P}}_1(v_1)$ on the remaining qubits, where $\hat{\mathcal{P}}_1(v_1)$ is the projection operator onto the space where $\hat{O}_1^{(k_1)} = v_1$. Now, if one assumes that for all $v_1$, $\hat{H}(v_1)$ has a single qubit symmetry operator $\hat{O}_2^{(k_2)}(v_1)$, then one could repeat this procedure by measuring $\hat{O}_2^{(k_2)}(v_1)$, and produce a new effective Hamiltonian $\hat{H}(v_1, v_2) = \hat{H} \hat{\mathcal{P}}_1(v_1) \hat{\mathcal{P}}_2(v_2)$. The Hamiltonians considered in Ref. \cite{izmaylovRevisingMeasurementProcess2019} are defined by the condition that all Hamiltonians $\hat{H}(\vec{v})$ generated in this manner,  for some starting Hamiltonian $\hat{H}$ will have a single qubit symmetry operator that can be directly measured. Therefore, such Hamiltonians can be measured using a ``feed-forward'' 
circuit, as shown in Fig. \ref{feedforward_fig}, in which at each step, a single qubit is measured, and the measured eigenvalue of the given single-qubit symmetry operator defines the unitary transformation which diagonalizes the subsequent symmetry.

\subsection{Classical Shadow Methods\label{classical_shadow_subsec}}

\begin{figure*}
    \centering
    \includegraphics[width=0.9\textwidth]{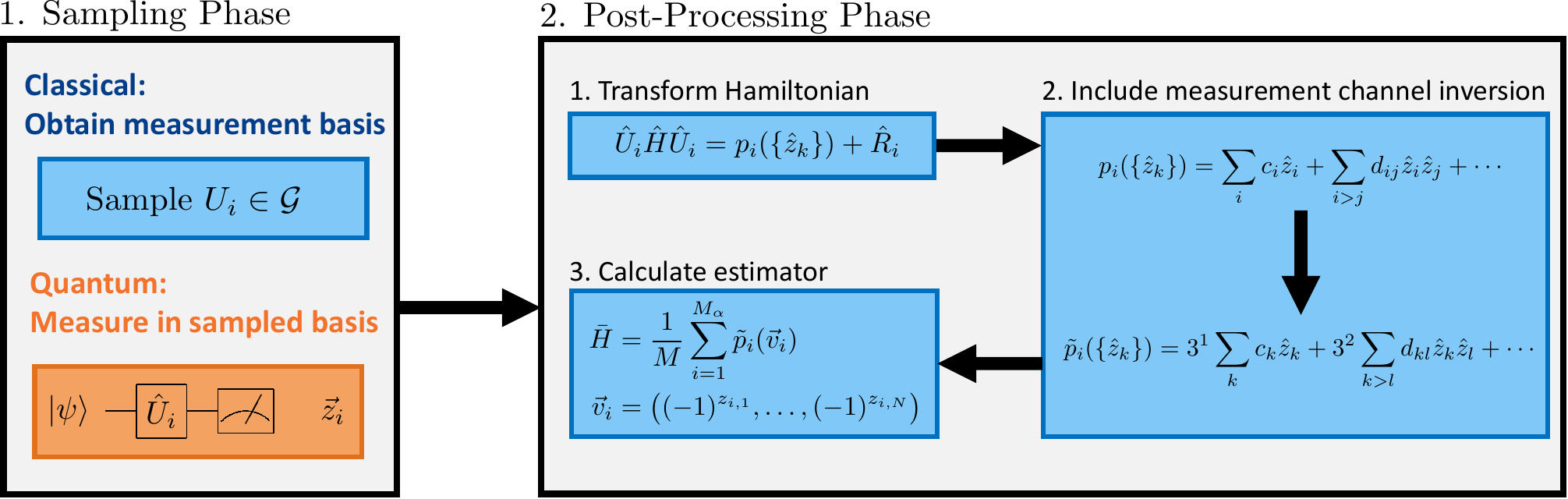}
    \caption{Diagrammatic representation of the framework for Hamiltonian energy estimation using classical shadow tomography for a pure input state $\ket{\psi}$, which can be divided into two components. In sampling phase, the classical computer samples unitary transformations from a group $\mathcal{G}$, which, in the Pauli classical shadows method of Ref. \cite{huang2020predicting}, is taken to be the single-qubit Clifford group, sampled uniformly. For each sampled unitary, the target state $\ket{\psi}$ is measured a single time in the associated basis. In the post-processing phase, the classical computer uses both the unitaries defining the measurement bases along with the measurement results to construct an estimate $\bar{H}$ of $\braket{\psi|\hat{H}|\psi}$. In a given measurement basis encoded by $\hat{U}_i$, only the Ising part of the Hamiltonian $p_i(\{\hat{z}_k\})$ contributes to the measured energy, as the non-Ising ``remainder'' $\hat{R}_i$ satisfies $\braket{\vec{z}|\hat{R}_i|\vec{z}} = 0$ for all computational basis states $\ket{\vec{z}}$. However, to obtain an unbiased estimate of the Hamiltonian expectation value, the coefficients of the Ising part must be modified to emulate the inversion of the measurement channel.}
    \label{fig:shadows_procedure}
\end{figure*}

Classical Shadow Tomography (CST) \cite{Huang_Preskill:2021} approaches evaluation of the expectation value for operator $\hat O$, $\bra{\psi}\hat O\ket{\psi} = {\rm Tr}[\hat \rho \hat O]$, through evaluating the trace function on a classical computer by reconstructing the density $\hat \rho$ from classically tractable parts. These parts are obtained from quantum measurements of $\hat \rho$, but for the purpose of reconstruction, prior to the measurement, $\hat \rho$ is rotated in different unitary frames. Thus, the main technical question is how to reconstruct $\hat \rho$ from the projections performed in multiple unitary frames? This question can be addressed using quantum channel theory. Quantum measurements in the computational basis can be described mathematically as the following quantum channel acting on $\hat \rho$ 
\begin{equation}
    \mathcal M(\hat \rho) = \sum_{\vec z \in \mathbb F_2^{N}} \ket{\vec z}\bra{\vec z} \hat \rho \ket{\vec z}\bra{\vec z}
    \label{eq:dephase}
\end{equation}
where $\mathbb F_2^N$ denotes the set of all $N$ length bit-strings $\vec z = (z_1, \dots, z_{N})$ with $z_i \in \{0, 1\}$. 
To reconstruct the density matrix, one must invert the measurement channel: $\hat \rho = \mathcal M^{-1}(\mathcal M(\hat \rho))$. 
However, the measurement channel in \eq{eq:dephase} 
leads to loss of information by removing the coherences and is therefore not invertible. To construct an invertible channel using
quantum measurements, CST applies unitary transformations $\hat U$ before the measurement and $\hat U^\dagger$ on the measured state during the classical post-processing. $\hat U$ is chosen randomly from a set forming a group $\mathcal G$ that is also 
informationally complete. 
The combined quantum channel is then written as
\bea    \label{eqn:twirl}
    \mathcal M_{\mathcal G}(\hat \rho) &=& \mathbb E_{\mathcal U \in \mathcal G}[\mathcal U^\dagger \circ \mathcal M\circ \mathcal U (\hat \rho)] \\
    &=& \mathbb E_{\mathcal U \in \mathcal G}
    \left[\sum_{\vec z} \mathcal U^\dagger (\ket{\vec z}\bra{\vec z}) \langle \vec z |\mathcal U(\hat \rho)|\vec z\rangle\right] \label{eq:shadows}
\eea
where $\mathcal U(\rho) = \hat U \hat \rho\hat U^\dagger$ denotes the unitary channel defined by the unitary operator $\hat U$ and $\circ$ operator denotes quantum channel composition. 
The notation $\mathbb E_{U \in \mathcal G}[.]$ denotes the average over all group elements in the case of a discrete group, and an integral with the uniform Haar measure over the group elements in the case of the continuous group. Each $\ket{\vec z}$ component for a 
particular $\mathcal U$ in \eq{eq:shadows} is a {\it classical shadow} of the quantum density.   

The channel $\mathcal M_{\mathcal G}(\hat \rho)$ is known as the \textit{twirl} of the measurement channel with the group $\mathcal G$. Intuitively, at the origin of $\mathcal M_{\mathcal G}$ invertibility is preservation of density coherences when the measurement is done in rotated bases. 
More formally, the expectation value of the Hamiltonian can be written
as 
\bea\notag
    \Tr(\hat H \hat \rho) &=& \Tr[\hat H \cdot \mathcal{M}_{\mathcal G}^{-1}\circ \mathcal{M}_{\mathcal G} (\hat \rho)] \\ \notag
    &=& 
    \mathbb E_{\mathcal U \in \mathcal G}
    \Bigg{[}\sum_{\vec z} \Tr\left(\hat H \cdot \mathcal M_{\mathcal G}^{-1}\circ \mathcal U^\dagger (\ket{\vec z}\bra{\vec z})\right) \\
    &\times & \langle \vec z |\mathcal U(\hat \rho)|\vec z\rangle\Bigg{]},
\eea
where $\Tr\left(\hat H \cdot \mathcal M_{\mathcal G}^{-1}\circ \mathcal U^\dagger (\ket{\vec z}\bra{\vec z})\right)$ can be evaluated efficiently on a 
classical computer, and  $\langle \vec z |\mathcal U(\hat \rho)|\vec z\rangle$
is measured. Appendix \ref{app:group} provides more rigorous mathematical justification of invertibility of $\mathcal{M}_{\mathcal G}$ and details of the inversion process. 



Although seemly different from the Hamiltonian partitioning, CST can be seen as an alternative approach to construct a partitioning scheme. 
This can be illustrated if we consider the single-qubit Clifford group as $\mathcal G$. 
Applying a unitary from this group to the density 
and then measuring is equivalent to rotating the Hamiltonian using 
the same unitary and measuring its Ising part; the expectation 
value of the non-Ising part is zero and thus does not contribute - as shown in \fig{fig:shadows_procedure}.  
Estimates obtained in the Hamiltonian fragmentation method is only a linear sum of estimates of expectation values of Ising terms of the predetermined fragments. Whereas the classical shadows estimate of a single observable is a linear sum of estimates of Ising parts in different random frames, now weighted with pre-determined coefficients. The randomization converts the measurement channel to an invertible channel and the coefficients are added to emulate the inverse of the measurement channel (see Appendix \ref{app:group} for details).
For example, following notation introduced in \cite{Hadfield_Mezzacapo:2022}, the uniformly random Pauli classical shadow method suggested by \citet{huang2020predicting} estimates the expectation value of Hamiltonian as \bea 
    \bar H &=& \frac{1}{M} \sum_{i=1}^{M} \bar H^{(i)} \\ 
    \bar H^{(i)} &=& \sum_{P} \alpha_P f(\hat B^{(i)}, \hat P) \mu (\hat B^{(i)}, \text{supp}(\hat P))
    \label{eq:random-pcs-estimator}
\eea 
where $\bar H^{(i)}$ is seen as a single-shot estimate of the Hamiltonian's expectation value, $\alpha_P$ is the coefficients of $\hat P$ in $\hat H$, $\hat B^{(i)}$ is the measurement basis randomly chosen for the $i^{\text{th}}$ measurement with equal probability of $\{\hat x, \hat y, \hat z\}$ on all qubits. The function $\mu (\hat B^{(i)}, \text{supp}(\hat P))$ determines the Pauli products $\hat P$ in the observable that are in Ising form in the basis $\hat B^{(i)}$ and evaluates the Pauli product $\hat P$ to be $\pm 1$ given the measurement results of the $i^{\text{th}}$ measurement, or $0$ if $\hat P$ is incompatible (not in Ising form) with the measurement basis $\hat B^{(i)}$. For $\hat P = \otimes_{i=1}^N \hat \sigma_i $, $f(\hat B^{(i)}, \hat P)$ is the appropriate weighting obtained through the theory of Pauli classical shadow: \bea
    f(\hat B, \hat P) &=& \prod_{j=1}^K f_j (\hat B_j, \hat \sigma_j) \\
    f_j (\hat B_j, \hat \sigma_j) &=& \begin{cases}
        1 & \text{if } \hat \sigma_j = \hat 1 \\
        3 & \text{if } \hat B_j = \hat \sigma_j \neq \hat 1\\
        0 & \text{otherwise.}
    \end{cases} \label{cs_local_clif_estimator}
\eea 
The estimator in \eq{eq:random-pcs-estimator} can be rewritten as follows: \bea 
    \bar H = \sum_{P} \sum_{i=1}^M \alpha_P \frac{f(\hat B^{(i)}, \hat P)}{M} 
    \mu (\hat B^{(i)}, \text{supp}(\hat P)). 
\eea 
Note that by design, $\mu (\hat B^{(i)}, \text{supp}(\hat P))$ is expected to be evaluated for exactly $M / {f(\hat B^{(i)}, \hat P)}$ times, so the inner summation can be seen as simply taking the average of measured values for each Pauli product. 

\section{Increased Number of Qubits}
\label{Sec:Embedding}

Here, we discuss methods to estimate $\langle \hat{H} \rangle$ which require introducing additional qubits. The motivation for introducing additional qubits beyond the number necessary to model the system is to facilitate the implementation of more flexible measurement protocols. All such measurement protocols described in this section involve entangling the system qubits with the additional ancilla qubits. 

\subsection{Hadamard Test \label{hadamard_test_lcu_section}}

\begin{figure}
    \includegraphics[width=0.9\columnwidth]{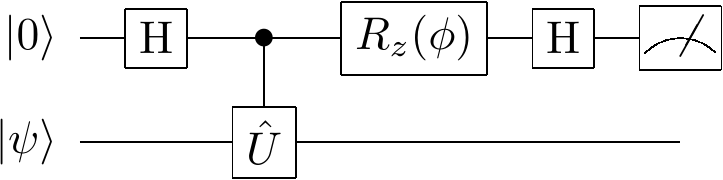}
    \caption{Hadamard test circuits for estimating $\text{Re}\braket{\psi|\hat{U}|\psi}$ ($\phi = 0$) and $\text{Im}\braket{\psi|\hat{U}|\psi}$ ($\phi = \pi/2$). The expectation values are encoded in the probability of measuring $0$ in the ancilla qubit. Here, $R_z(\phi) = e^{-\imag \phi\hat z / 2}$ is the Pauli-$\hat{z}$ rotation operator, and $\mathrm{H}$ denotes the Hadamard gate.}
    \label{hadamard_circuit_fig}
\end{figure}

An approach to measurement of a general Hamiltonian which requires a decomposition into fragments, but uses ancilla qubits, relies on repeated applications of the Hadamard test. The Hadamard test produces an estimate of the expectation value of a unitary operator $\hat{U}$, given access to a quantum circuit implementing $\hat{U}$, and a single ancilla qubit. The expectation value is encoded in the probability of measuring $0$ in the ancilla qubit, for the quantum circuit shown in Fig. \ref{hadamard_circuit_fig}. Since the expectation value is encoded as a probability, it can be estimated to error $\epsilon$ via $O(1/\epsilon^2)$ repetitions of the Hadamard test circuit.

The Hadamard test can be applied to obtain the expectation value of an operator $\hat{O}$, given a decomposition of $\hat{O}$ to a linear combination of unitaries (LCU)
\begin{equation}
    \hat{O} = \sum_j \alpha_j \hat{U}_j, \quad \hat{U}_j \hat{U}_j^\dagger = \hat{1}.
\end{equation}
Then, given a quantum state $\ket{\psi}$, linearity of the expectation value implies
\begin{equation}
    \braket{\psi|\hat{O}|\psi} = \sum_j \alpha_j \braket{\psi|\hat{U}_j|\psi},
\end{equation}
implying that the expectation value of $\hat{O}$ can be calculated via repeated Hadamard tests applied to all unitaries in an LCU decomposition of $\hat{O}$. 

For the electronic Hamiltonian $\hat{H}$, the question of obtaining an LCU decomposition is well-studied. Note that most investigations of LCU decompositions of the electronic Hamiltonian were done in the context of developing more efficient block-encodings of $\hat{H}$ for estimation of electronic energies via quantum phase estimation (discussed in more detail in Sec. \ref{Sec:Heisenberg}), but the same LCUs can be used for measurement via the Hadamard test. The simplest procedure to obtain an LCU decomposition is to write $\hat{H}$ in terms of Pauli products - which are unitary operators - via a fermion-to-qubit map like the Jordan-Wigner \cite{jordanUeberPaulischeAequivalenzverbot1928} or Bravyi-Kitaev transformations. \cite{Bravyi_BKTransf,seeleyBravyiKitaevTransformationQuantum2012} Furthermore, as discussed in Sec. \ref{csa_subsec}, linear combinations of anti-commuting Pauli operators are proportional to unitaries. Therefore, the AC grouping approach, which finds such linear combinations within $\hat{H}$, can produce a more compact LCU representation compared with using Pauli products. In the fermionic algebra, simple reflection operators can be formed as $U^\dagger \hat{r}_p \hat{U}$, where
\begin{equation}
    \hat{r}_p = 1 - 2 \hat{n}_p. \label{fermionic_reflection}
\end{equation}
Based on \eq{fermionic_reflection}, Ref. \cite{loaizaReducingMolecularElectronic2023a} proposed a two step procedure for obtaining an LCU decomposition of a two-electron operator $\hat{H}_{2e}$: (1) obtain a decomposition of $\hat{H}_{2e}$ into fermionic fragments of the form shown in \eq{ferm_csa}, and (2): use \eq{fermionic_reflection} to write each fragment as an LCU. \cite{loaizaReducingMolecularElectronic2023a} Another approach is the tensor hypercontraction (THC) method, \cite{leeEvenMoreEfficient2021} in which a decomposition of the two-body-tensor is used to produce the following decomposition of the two-electron operator
\begin{equation}
    \hat{H}_{2e} = \sum_{\mu\nu=1}^{M} \xi_{\mu\nu} \hat{U}_\mu^\dagger (\hat{n}_{1\downarrow} + \hat{n}_{1\uparrow}) \hat{U}_\mu \hat{U}_\nu^\dagger (\hat{n}_{1\downarrow} + \hat{n}_{1\uparrow}) \hat{U}_\nu,\label{thc_decomp}
\end{equation}
where $\hat{n}_{1\uparrow}$ and $\hat{n}_{1\downarrow}$ are the number operators for the first spatial orbital with different spin projections. From \eq{thc_decomp}, the conversion \eq{fermionic_reflection} is used to obtain the LCU. Alternatively, the class of methods known as Majorana tensor decomposition, introduced in Ref. \cite{loaiza2024LCU3}, obtains LCU decompositions via a factorization of the coefficient tensors of $\hat{H}$ when written in the Majorana form of \eq{majorana_form}, and forms a direct generalization of other fermionic LCU approaches. 

For implementation of the Hadamard test, the only requirement is a quantum circuit for implementing $\hat{U}_j$, and an ancilla qubit. Such a quantum circuit can be transformed into a quantum circuit implementing controlled $\hat{U}_j$ by controlling every gate in the circuit decomposition of $\hat{U}_j$. However, such controlled circuits can be prohibitively expensive to implement on near-term noisy hardware. Despite this, it is true that, for many of the LCUs that have been developed, the $\hat{U}_j$ are also exactly solvable, and therefore can be measured directly without an ancilla qubit. (Note that direct measurement of a unitary $\hat{U}_j$ without an ancilla qubit requires implementing a different unitary that maps $\hat{U}_j$ to Ising form.) Therefore, the Hadamard test approach closely resembles the measurement technique based on Hamiltonian partitioning to exactly solvable fragments, but with the additional constraints of unitarity of the fragments, requiring an ancilla qubit, and potentially requiring deeper measurement circuits. As such, this technique is not typically considered for estimation of the expectation value of the Hamiltonian. However, it has potential applications for measuring other quantities, such as matrix elements, to be discussed in Sec. \ref{Sec:OtherQuantities}.

\subsection{Joint Bell Measurements}

The joint Bell measurement approach of Ref. \cite{caoAcceleratedVariationalQuantum2024} provides a method to estimate the expectation value $\braket{\psi|\hat{H}|\psi}$ of the target qubit Hamiltonian $\hat{H} = \sum_k c_k \hat{P}_k$, where $\hat{P}_k$ are Pauli operators, using additional qubits as a resource. In this approach, one uses the fact that, for the full set of $4^N$ Pauli operators $\{\hat{P}_k\}_{k=1}^{4^N}$ on $N$ qubits, one can form a fully-commuting set by considering the set of doubled Pauli operators $\{\hat{P}_k \otimes \hat{P}_k\}_{k=1}^{4^N}$ on $2N$ qubits. Therefore, for a general $2N$ qubit state $\ket{\Psi}$, one can estimate all $\braket{\Psi|\hat{P}_k \otimes \hat{P}_k|\Psi}$ simultaneously after applying a Clifford transformation which maps all doubled Pauli operators to Ising form. This is termed a joint Bell measurement, as the Clifford transformation is defined by pairing qubits between the $N$ qubit subsystems and transforming the pair of qubits from the computational basis to the Bell basis. Taking $\ket{\Psi} = \ket{\psi} \otimes \ket{\psi}$ allows one to estimate the absolute value of all Pauli terms in $\hat{H}$ simultaneously, as
\begin{equation}
    \bra{\Psi} \hat{P}_k \otimes \hat{P}_k \ket{\Psi} = |\braket{\psi|\hat{P}_k|\psi}|^2,
\end{equation}
implying that joint Bell measurements can produce an estimate of the quantity $\sum_k c_k |\langle \hat{P}_k \rangle |$ without computing a decomposition of $\hat{H}$ into measurable fragments. In the joint Bell measurement VQE approach, an assumption is made that the sign of $\langle \hat{P}_k \rangle$ changes infrequently during the VQE optimization. Therefore, one can use the sign from a previous iteration, calculated via a direct estimation of $\langle \hat{P}_k \rangle$ using standard measurement techniques (e.g., partitioning into measurable fragments), in subsequent iterations, and only recomputing the sign after many iterations have been done. An interesting question is whether different choices of the operator mapping to an extended space, and of the extended state $\ket{\Psi}$ can be used to obtain more information beyond the absolute values of all $\hat{P}_k$ simultaneously.

\subsection{Informationally Complete Generalized Measurements}\label{Subsec:IC_POVM}

The formalism of positive operator-valued measures (POVMs) provides a powerful framework for describing measurement procedures on quantum computers. POVMs are a generalization of the simpler PVMs, described in Sec. \ref{Subsec:IsingH}. The motivation for introducing POVMs stems from the fact that not all quantum measurements can be described by PVMs. For example, a more general framework is needed to describe measurements whose outcomes correspond to non-orthogonal states, which cannot form an eigenbasis of a Hermitian operator. In the most general case, quantum measurements are described by a set of measurement operators $\hat{M}_m$ that satisfy $\sum_m \hat{M}_m^\dagger \hat{M}_m = \hat{1}$. \cite{Nielsen_Chuang_2010} The probability of obtaining the measurement result $m$ corresponding to $\hat{M}_m$ is given by $\Pr(m) = \braket{\psi|\hat{M}_m^\dagger \hat{M}_m|\psi}$, and the post-measurement state is proportional to $\hat{M}_m\ket{\psi}$. In the case that one is only interested in the probabilities, and not the post-measurement state, one only needs to consider the set of operators $\hat{\Pi}_m = \hat{M}_m^\dagger \hat{M}_m$, which have non-negative eigenvalues. This motivates the definition of a POVM, which is a set of operators $\{\hat{\Pi}_m\}$ that are positive, $\braket{\psi|\hat{\Pi}_m|\psi} \geq 0$ for all $\ket{\psi}$, and sum to the identity, $\sum_m \hat{\Pi}_m = \hat{1}$. The individual operators $\hat{\Pi}_m$ are called POVM effects. 

The approaches to quantum measurement based on Hamiltonian fragments can be interpreted as special cases of POVM measurements. The motivation for using a Hamiltonian partitioning is that the PVM which implements a measurement of the target Hamiltonian eigenbasis is prohibitively challenging to find and implement. As a consequence, the Hamiltonian is decomposed into fragments $\hat{H} = \sum_\alpha \hat{H}_\alpha$ which can be directly measured via unitaries $\hat{U}_\alpha$. The eigendecomposition of the fragments, which reveals the associated PVMs for measuring in the fragment eigenbases, is 
\begin{equation}
    \hat{H}_\alpha = \sum_{\vec{z} \in \mathbb{F}_2^N} e_{\vec{z}, \alpha} \hat{U}_\alpha \ket{\vec{z}}\bra{\vec{z}} \hat{U}_\alpha^\dagger,
\end{equation}
where $e_{\vec{z}, \alpha}$ are the eigenvalues of $\hat{H}_\alpha$. From the collection of PVMs $\mathcal{P}_\alpha = \{\hat{U}_\alpha \ket{\vec{z}}\bra{\vec{z}} \hat{U}_\alpha^\dagger\}$, and any set of numbers $w_\alpha \geq 0$ such that $\sum_\alpha w_\alpha = 1$, it follows that the set $\bigcup_\alpha w_\alpha \mathcal{P}_\alpha$ is a POVM. Provided that all $w_\alpha$ are non-zero, this produces a decomposition of the target Hamiltonian into a linear combination of POVM effects
\begin{equation}
    \hat{H} = \sum_\alpha \sum_{\vec{z} \in \mathbb{F}_2^N} \left(\frac{e_{\vec{z}, \alpha}}{w_\alpha}\right) w_\alpha \hat{U}_\alpha \ket{\vec{z}}\bra{\vec{z}}\hat{U}_\alpha^\dagger,\label{fragment_based_povm_decomp}
\end{equation}
which can be used as a substitute for the eigendecomposition of $\hat{H}$ for expectation value estimation via measurements. Formally, implementation of the POVM measurement is accomplished by (1) randomly sampling a fragment $\alpha$ based on the probability distribution $w_\alpha$, and (2) measuring the corresponding PVM, which is done by applying $\hat{U}_\alpha$ as a quantum circuit and then measuring in the computational basis. In an actual quantum computation, where $\langle \hat{H} \rangle$ is estimated via quantum measurements of the fragment eigenbases, the PVMs are not randomly sampled. Instead, the number of measurements allocated to each fragment is specified beforehand via a deterministic procedure designed for minimizing the error in the expectation value estimate, described in Sec. \ref{reduction_greedy}. Nonetheless, in this more deterministic framework, the $w_\alpha$ can be interpreted as the proportion of total measurements allocated to fragment $\hat{H}_\alpha$, and therefore have an effect on the error in the estimate. Classical shadow tomography is also easily described as an instance of observable estimation via POVM measurements. In classical shadows, the computational basis is measured after implementation of a unitary $\hat{U}$ which is uniformly sampled from a group $\mathcal{G}$. This constitutes an implementation of a POVM measurement with POVM effects $\bigcup_{\hat{U} \in \mathcal{G}}\{1/|\mathcal{G}| \ \hat{U} \ket{\vec{z}}\bra{\vec{z}}\hat{U}^\dagger\}$. Like with Hamiltonian partitioning, one can obtain improved estimates by instead using a different POVM measurement which samples the associated PVMs non-uniformly, with the goal of minimizing the error in the expectation value estimate. This idea forms the basis of biasing and derandomization techniques which are described in Sec. \ref{reduction_shadows}.

Measurement techniques based on using Hamiltonian fragments are not the only application of POVM measurements. Reference \cite{GarciaPerezPOVMMeasurement} proposed the use of so-called informationally-complete (IC) POVMs, which are POVMs that span the space of observables, for expectation value estimation. IC-POVMs $\{\hat{\Pi}_m\}$ are useful since the target Hamiltonian can always be expressed as a linear combination of the associated POVM effects
\begin{equation}
    \hat{H} = \sum_m \omega_m \hat{\Pi}_m. \label{povm_decomp_general}
\end{equation}
Like \eq{fragment_based_povm_decomp},  \eq{povm_decomp_general} can be interpreted as a generalization of the eigendecomposition of $\hat{H}$, in which the $\omega_m$ generalize the eigenvalues of $\hat{H}$, and the $\hat{\Pi}_m$ generalize the eigenspace projectors. Noting that $\Pr(m) = \langle \hat{\Pi}_m \rangle$, it follows that one can estimate the Hamiltonian expectation value $\langle \hat{H} \rangle = \sum_m \omega_m \langle \hat{\Pi}_m \rangle$ from repeated measurements of the IC-POVM as
\begin{equation}
    \bar{H} = \frac{1}{S} \sum_{s=1}^{S} \omega_{m_s}, \label{povm_est}
\end{equation}
where $S$ is the number of measurement shots, and $m_s$ is the result of the $s^{\rm th}$ measurement. In the IC-POVM framework, one does not need to express $\hat{H}$ in the IC-POVM basis directly, as the application of \eq{povm_est} only requires that the IC-POVM can be measured efficiently, and that the $\omega_m$ can be classically calculated efficiently. The procedure to measure a general IC-POVM is non-trivial, as a quantum computer can only measure in the computational basis. Therefore, the implementation of a POVM measurement must be accomplished via a measurement of one, or multiple, PVMs that themselves can be measured efficiently. In general, it is always possible to implement the measurement of an IC-POVM $\{\hat{\Pi}_m\}_{m=1}^{4^N}$ on $N$ qubits via the measurement of a single PVM on $2N$ qubits. Such a PVM $\{\hat{U} \ket{m}\bra{m}\hat{U}^\dagger\}$ on $2N$ qubits can be used to measure the IC-POVM if, for all $N$ qubit states $\ket{\psi}$, we have
\begin{equation}
    \braket{\psi|\hat{\Pi}_m|\psi} = \bra{\psi}\bra{\phi_0} \hat{U}\ket{m}\bra{m}\hat{U}^\dagger \ket{\psi}\ket{\phi_0},\label{general_povm_measurement}
\end{equation}
where $\ket{\phi_0}$ is a fixed reference state on the ancilla qubits, taken, without loss of generality, to be the all zero state.

The informational completeness of the POVMs used in the approach based on \eq{general_povm_measurement} implies that, in principle, this technique works for any Hamiltonian and any choice of IC-POVM. Nonetheless, for a fixed target Hamiltonian $\hat{H}$, the specific choice of IC-POVM influences the computational cost for estimating $\langle \hat{H} \rangle$, both due to the additional circuit for implementing $\hat{U}$, and due to the measurement overhead which depends on the choice of IC-POVM. To address the circuit overhead, Ref. \cite{GarciaPerezPOVMMeasurement} proposed to use $N$ qubit IC-POVMs which are defined in terms of a set of ``local'' one-qubit IC-POVMs, which can be measured in parallel using one ancilla per qubit. IC-POVMs on a single qubit consist of 4 POVM effects $\{\hat{\Pi}_{i\alpha}, i,\alpha \in \{0,1\}\}$ that span the 4-dimensional space of Hermitian operators that act on the single-qubit Hilbert space, and can be measured via an association to a two qubit PVM  $\{ \hat{U} \ket{i\alpha}\bra{i\alpha}\hat{U}^\dagger\}$ which satisfies the criteria \eq{general_povm_measurement}. The association of POVMs to PVMs also implies that we can parametrize a set of single qubit IC-POVMs via two-qubit unitary operators. Consider a two-qubit unitary operator $\hat{U}$, which we write in the computational basis as follows
\begin{equation}
    \hat{U} = \sum_{ij\alpha \beta} u^{i \alpha}_{j \beta} |i\alpha\rangle\langle j\beta|.
\end{equation}
The POVM associated to implementing $\hat{U}$ and measuring in the computational basis can be recovered as $\hat{\Pi}_{i\alpha} = |\pi_{i\alpha}\rangle\langle\pi_{i\alpha}|$, where
\begin{equation}
|\pi_{i\alpha}\rangle = \sum_{j} (u_{j 0}^{i\alpha})^{\ast} |j\rangle.\label{unitary_defn_povm}
\end{equation}
Using unitaries on a larger space is not the only way to parametrize a set of IC-POVMs. An alternative approach for generating a single qubit IC-POVM was introduced in Ref. \cite{glosAdaptivePOVMImplementations2022}, in which PVM measurements are combined with classical randomness in a way analogous to how POVM measurements describe fragment based measurement procedures. Essentially, a POVM with $K$ effects is implemented from a collection $\bigcup_\mu \{\hat{P}_0^{(\mu)},\hat{P}_1^{(\mu)}\}$ of projective measurements by (1): randomly selecting the $\{\hat{P}_0^{(\mu)},\hat{P}_1^{(\mu)}\}$ projective measurement with probability $p_\mu$, (2): measuring the associated projective measurement, obtaining result $i \in \{0,1\}$, and (3): randomly selecting the outcome $1 \leq m \leq 4$ of the IC-POVM by sampling a probability distribution $\Pr(m|\mu, i)$. 

In general, for an $N$ qubit system, and a set of local IC-POVMs $\mathcal{M}_k = \{\hat{\Pi}_{m_k}^{(k)} : 1 \leq m_k  \leq 4 \}$, for $1 \leq k \leq N$, one can form an IC-POVM on the $N$ qubit space using the $4^N$ operators formed from tensor products as follows
\begin{equation}
    \hat{\Pi}_{\mathbf{m}} =  \hat{\Pi}_{m_{1}}^{(1)} \otimes \cdots \otimes \hat{\Pi}_{m_{N}}^{(N)}.
\end{equation}
To express a Pauli product $\hat{P}_k = \otimes_i \hat{\sigma}_i^{(k)}$ in terms of the IC-POVM effects, one merely needs to express each $\hat{\sigma}_i^{(k)}$ in terms of the local IC-POVM basis elements $\mathcal{M}_k$, which can be done by solving a 4 dimensional linear system of equations. Therefore, since the coefficient $d_{k, \mathbf{m}}$ of each $\hat{P}_k$ in the $\hat{\Pi}_{\mathbf{m}}$ basis can be calculated efficiently, and there are polynomially many Pauli products $\hat{P}_k$ in the target Hamiltonian, each $\omega_{\mathbf{m}} = \sum_k c_k d_{k, \mathbf{m}}$ can be calculated efficiently, implying that the POVM measurement strategy can be used. The locality of the IC-POVMs is essential for two reasons: (1) to calculate the coefficients $\omega_{\mathbf{m}}$ efficiently, and (2) to ensure that the associated $2N$ qubit unitary $\hat{U}$ which implements the measurement is implementable via a short measurement circuit.

\section{Reducing the Number of Measurements} 
\label{Sec:MeasurementNReduction}

\subsection{Grouping Techniques\label{reduction_grouping}}

Equipped with numerous measurement schemes, one needs a suitable metric of merits to identify the superior approach.
Although early works \cite{verteletskyiMeasurementOptimizationVariational2020,izmaylovUnitaryPartitioningApproach2020,yenMeasuringAllCompatible2020} focused on the number of fragments that one can partition the qubit Hamiltonian into, 
the appropriate metric to evaluate the number of measurements needed to accurately estimate the Hamiltonian expectation value is the variance of the estimator.
In what follows we justify why variance captures the quality of a measurement scheme in the regime of many measurements, and discuss various heuristic approaches to minimize the resulting estimator variance of a measurement scheme. 

Suppose one has a partitioning of the qubit Hamiltonian $\hat H = \sum_\alpha \hat H_\alpha$, where each $\hat H_\alpha$ can be transformed into Ising polynomials using simple unitaries. 
A natural corresponding measurement scheme for $\bra{\psi} \hat H \ket{\psi}$ would be to sample the values of individual fragments for $M_\alpha$ times, compute their means, and sum over the means to obtain the estimator $\bar H$ for the expectation value $\bra{\psi} \hat H \ket{\psi}$. To analyze the statistical property of this estimator, we use the definition of the estimators $\bar{H}$ and $\bar{H}_\alpha$ for $\braket{\psi|\hat{H}|\psi}$ and $\braket{\psi|\hat{H}_\alpha|\psi}$ respectively given in \eq{expectation_estimators}. Note that by the measurement postulates, the random variable $\hat H_{\alpha}$ has mean $\AV{\hat H_\alpha} = \bra{\psi} \hat H_\alpha \ket{\psi}$ and finite variance $\VA{\hat H_\alpha} =\bra{\psi} \hat H_\alpha^2 \ket{\psi} - \bra{\psi} \hat H_\alpha \ket{\psi}^2$. 
Then, if we consider $\bar H_\alpha$ as a new random variable, the central limit theorem states that the distribution of $\bar H_\alpha$ quickly approaches the normal distribution with mean $\mu_\alpha$ and variance $\VA{\hat H_\alpha}/M_\alpha$ as $M_\alpha$ increases. 
Then $\bar H$, as the sum of independent estimators $\bar H_\alpha$, also follows the normal distribution with mean $\bra{\psi} \hat H \ket{\psi} = \sum_\alpha \mu_\alpha$ and variance \bea
    \text{Var}[\bar H] = \sum_\alpha \text{Var}[\bar H_\alpha] = \sum_\alpha \frac{\VA{\hat H_\alpha}}{M_\alpha}. 
\eea 
Therefore, the variance of $\bar H$ fully characterizes the resulting distribution of a measurement scheme in the limit of sufficient measurements.

The variance $\text{Var}[\bar H_\alpha]$ allows us to write probabilistic bounds on the deviation of $\bar H_\alpha$ from 
$\AV{\hat H_\alpha}$. Chebyshev's inequality, \cite{bienayme1853,chebyshev1867} applicable for any random variable with a finite variance, states that 
\bea
    \text{Pr}\left(\left|\bar H_\alpha - \AV{\hat H_\alpha}\right| \geq k \text{Var}^{1/2}[\bar H_\alpha]\right) \leq \frac{1}{k^2}
    \label{eq:chebyshev}
\eea 
for any $k > 0$. On the other hand, using the fact that measurement results of $\hat{H}_\alpha$ are bounded variables, Hoeffding's inequality \cite{hoeffdingProbabilityInequalitiesSums1963} provides a stronger exponentially decaying bounds 
\bea
    \text{Pr}\left(\left|\bar H_\alpha - \AV{\hat H_\alpha}\right| \geq t \right) \leq 2e^{-2M_\alpha t^2 / (\Delta E_\alpha)^2} ,
    \label{eq:hoeffding_t}
\eea 
where $\Delta E_\alpha = E_\text{max}^{(\alpha)} -E_\text{min}^{(\alpha)}$, and $E_\text{min/max}^{(\alpha)}$ are the lowest/largest eigenvalues of $\hat H_\alpha$.
One can rewrite this bound in variance of $\hat H_\alpha$ 

\begin{align}
    &\text{Pr}\left(\left|\bar H_\alpha - \AV{\hat H_\alpha}\right| \geq 
    k \text{Var}^{1/2}[\bar H_\alpha]\right) \nonumber\\
    \leq  
    &2\exp{\left[-\frac{k^2\VA{\hat H_\alpha}}{2\VA{\hat H_\alpha}_{\max}}\right]}, 
    \label{eq:hoeffding_k}
\end{align}
where $\VA{\hat H_\alpha}_{\max} = (\Delta E_\alpha)^2/4$ is the maximum variance of $\hat H_\alpha$. 
Note that the dependence on $k$ in \eq{eq:hoeffding_k} is exponential, whereas the dependence of $k$ in \eq{eq:chebyshev} is only quadratic. 
The former provides a stronger guarantee using the information that $\AV{\hat H_\alpha}$ is a bounded variable. 
The main advantage of the Chebyshev and Hoeffding bounds 
is that they are applicable even if $\hat H_\alpha$ are not 
measured many times.
However, we know that with an increasing number of measurements $\bar{H}_\alpha$ distribution becomes normal, hence,
\bea\notag
    \text{Pr}\left(\left|\bar H_\alpha - \mu_\alpha \right| 
    \geq k \text{Var}^{1/2}[\bar H_\alpha]\right) 
    \leq \frac{\sqrt{2}}{k\pi} e^{-k^2 / 2},
\eea  
which gives the tightest bound that decays as $O( e^{-k^2 / 2}/k)$. 
In all three bounds, the deviation of the estimator average from the true mean is controlled by the variance.

Having established variances as the major metric of merit for a measurement scheme, we next discuss efforts 
for minimization of the fragment variances. 
All methods fall in two categories: 1) greedy approaches creating fragments in the norm descending order and 2) covariance-based approaches grouping terms while optimizing fragment variances based on terms covariances.

\subsection{Greedy Approaches\label{reduction_greedy}}

The idea behind greedy approaches is: an uneven distribution of fragments generally decreases the measurement cost of the partitioning scheme. 
To see this, suppose that one measures $\hat H$ according to a partitioning $\hat H = \sum_\alpha \hat H_\alpha$, where each fragment $\hat H_\alpha$ has the corresponding $\VA{\hat H_\alpha}$, and measures each fragment for $M_\alpha$ times with the maximum measurement budget $M = \sum_\alpha M_\alpha$. 
Suppose further that one has perfect information on the values of $\VA{\hat H_\alpha}$. 
Then, one can solve for the optimal $M_\alpha$ as a function of $\VA{\hat H_\alpha}$ and $M$, and find that the optional choice results in a minimum variance that scales linearly with $
    \left(\sum_\alpha \sqrt{\VA{\hat H_\alpha}}\right)^2
$\cite{Crawford2021efficientquantum}.
Evidently, due to the square roots, an uneven set of fragments' variances would generally have a lower overall measurement cost than that of an evenly distributed set.

Furthermore, fragments with larger coefficient 1-norms in their Pauli product representation 
generally have larger variance. This heuristic idea can be based on the following sequence of upper bounds:  
\bea
\VA{\hat H_\alpha}^{1/2} \le \frac{(E_{\max}^{(\alpha)}-E_{\min}^{(\alpha)})}{2}
\le \sum_n |c_n^{(\alpha)}|, 
\eea
where $c_n^{(\alpha)}$'s are coefficients 
of $\hat H_\alpha$ when expressed as a linear combination of Pauli products. The first upper bound can be shown by maximizing the variance over all wavefunctions, while the second upper bound has been shown in Ref.~\cite{loaiza2023LCU1}. Thus, having an uneven set of the fragments' coefficient 1-norms can help reduce measurement cost. 
This idea motivates greedy approaches, which use various heuristics to find uneven partitionings in coefficient 1-norms. 

Proposed by \citet{Crawford2021efficientquantum}, Sorted Insertion is the first greedy algorithm that employs this idea for qubit-based partitioning approaches. In Sorted Insertion, one first sorts all the Pauli terms by the magnitudes of their coefficients in the descending order, and then groups all the Pauli terms that can be measured together. As a result, the first few measurable groups generally contain Pauli terms with larger magnitudes and naturally larger variances. As demonstrated by \citet{Crawford2021efficientquantum}, following the greedy approach resulted in a two- to four-fold reduction in measurement cost compared to grouping methods that focus on finding the grouping with the lowest number of fragments. The greedy idea has also been implemented for decompositions obtained in the fermionic algebra. As described in Sec. \ref{commuting_ferm_subsec}, one can produce a partitioning of the electronic Hamiltonian two-electron part into low-rank exactly solvable fragments via a singular value decomposition (SVD) of the two-body-tensor $g_{pqrs}$. \cite{Huggins_Babbush:2021} In this approach, the fragments $\hat H_\alpha$ will naturally decay in norm due to the property of the SVD algorithm. Alternatively Ref. \cite{yen2021cartan} then proposed a generalization which obtains full-rank fragments (\eq{ferm_csa}) via a non-linear optimization of the fragment parameters. In this approach, one finds the fragments in an iterative fashion, one at a time; at each step, a fragment is found by minimizing the norm of the remaining coefficients of the Hamiltonian. This procedure similarly results in fragments with significantly uneven distribution of coefficients and fragment variances. 

\subsection{Covariance Based Minimization\label{reduction_coefficient_splitting}}

Earlier grouping techniques assigned each Pauli product to a single fragment $\hat H_\alpha$. 
However, one can utilize the fact that each Pauli product can be compatible with multiple fragments to achieve further reduction in measurement cost. 
For example, consider the following Hamiltonian and its partitioning: \bea
    \hat H 
    = \hat H_1 + \hat H_2 \\
    \hat H_1 = \hat P_1 + a \hat P_2,\ \ \hat H_2 &=& (1 - a) \hat P_2 + \hat P_3
\eea where $a \in \mathbb R$, $[\hat P_1, \hat P_2] = [\hat P_2, \hat P_3] = 0$, but $[\hat P_1, \hat P_3] \neq 0$.
Suppose we measure $\hat H_1$ and $\hat H_2$ for $M_1$ and $M_2$ times respectively with the state $\ket{\psi}$. 
We can write the measurement variance as a function of $M_1, M_2$ and $a$: 
\bea
    \text{Var}[\bar H] 
    &=& \frac{\VA{\hat H_1}}{M_1} + \frac{\VA{H_2}}{M_2} \nonumber\\
    &=& [\VA{\hat P_1} + a^2 \VA{\hat P_2} + 2a{\rm Cov}_{1, 2}]/M_1 \nonumber
        \\ &+& [\VA{\hat P_3} + (1-a)^2 \VA{\hat P_2} \nonumber \\
        &+& 2(1 - a){\rm Cov}_{2, 3}]/M_2 \label{eq:simple_cs_var}
\eea 
where 
$\VA{\hat P_i} = 1- \bra{\psi}\hat P_i\ket{\psi}^2$ are the variances of $\hat P_i$ 
and ${\rm Cov}_{i, j} = \bra{\psi}\hat P_i \hat P_j\ket{\psi} - \bra{\psi}\hat P_i\ket{\psi}\bra{\psi}\hat P_j\ket{\psi}$'s are the covariances between $\hat P_i$ and $\hat P_j$. 
Note that the optimal choice of $a$ depends non-trivially on the $M_\alpha$, the variance of the shared term, and the covariances between grouped terms.
For example, if one ignores the covariances ${\rm Cov}_{1, 2}$ and ${\rm Cov}_{2, 3}$ and set $M_1 = M_2$, 
the squared dependencies on $a$ in both terms would suggest than an even distribution of coefficient ($a=0.5$) is optimal.
In contrast, following the greedy approach would result in an suboptimal choice of $a = 0$ or $a = 1$. 
This demonstrates why a coefficient splitting approach \cite{Crawford2021efficientquantum, yen2023deterministic} often provides additional gain beyond the greedy approach. 
However, this gain requires a careful choice of the numbers of measurements ($M_1$, $M_2$) and the split coefficient ($a$) that take into account the the exact values of ${\rm Cov}_{1, 2}$, ${\rm Cov}_{2, 3}$. 

Generally, suppose that $\hat H = \sum_i w_i \hat P_i$ is partitioned into $M$ fragments \bea
    \hat H &=& \sum_\alpha \hat H_\alpha \\
    \hat H_\alpha &=& \sum_i w_{i, \alpha} \hat P_i 
\eea where $\sum_\alpha w_{i, \alpha} = w_i\ \forall i$ and $w_{i, \alpha} = 0$ for all $\hat P_i$ incompatible with other Pauli products in $\hat H_\alpha$. 
Similarly to \eq{eq:simple_cs_var}, the variance of the corresponding energy expectation value estimator $\bar H$ can be written as a function of the fixed covariances and the free parameters $M_\alpha$ and $w_{i, \alpha}$ 
\bea\nonumber
    \text{Var}[\bar H] &=& \sum_{\alpha} \frac{1}{M_\alpha}\Bigg{(}\sum_i w_{i, \alpha}^2 \VA{\hat P_i} \\
    &+& \sum_{ij} 2w_{i, \alpha}w_{j, \alpha} {\rm Cov}_{i, j}\Bigg{)}.~
    \label{eq:full_cs_var}
\eea 
Hence the minimization of $\text{Var}[\bar H]$ reduces to a joint optimization of $M_\alpha$ and $w_{i, \alpha}$ constrained by $M = \sum_\alpha M_\alpha$ and $w_{i} = \sum_{\alpha} w_{i, \alpha}\ \forall i$. 

The idea of moving parts of the Hamiltonian to minimize the variance is first proposed in \citet{yen2023deterministic}, which achieves up to three-fold reduction in measurement cost relative to that of Sorted Insertion. 
The follow-up work by \citet{choi2022improving} further introduces ``ghost" Pauli terms that are not present in the original Hamiltonian but serve to reduce the overall measurement cost when introduced to the fragments. 
However, solving the optimal assignment of $w_\alpha$ is difficult even if one assumes access to the exact covariances of the Pauli terms.
An efficient heuristic for solving this problem is proposed by \citet{wu2023overlapped}. This scheme assigns the split coefficients $w_{i, \alpha}$ of each Pauli products $\hat P_i$ proportional to the number of times $\hat P_i$ is measured in each fragment, i.e. if $\hat P_i$ is measured in both $\hat H_1$ and $\hat H_2$, then $w_{i, 1} = M_1 w /(M_1 + M_2)$ and 
$w_{i, 2} = M_2 w /(M_1 + M_2)$. 
This leaves the choice of $M_\alpha$ the only parameters to optimize in the measurement scheme and greatly reduces the complexity of solving for the optimal parameters. 
\citet{wu2023overlapped} optimizes $M_\alpha$ with a surrogate cost function for variance that only considers the sizes of the coefficients. 
In contrast, \citet{yen2023deterministic} utilizes this heuristic to optimize $M_\alpha$ in \eq{eq:full_cs_var} with an iterative scheme.

The idea of moving parts of the Hamiltonian that do not break diagonalizability of $\hat H_\alpha$ can also be used for fermionic algebra-based partitioning schemes, in which the two-electron fragments are in the form of \eq{ferm_csa}. 
The idempotency of the fermionic number operator $\hat n_p^2 = \hat n_p$ leads to the idempotency of one-electron 
operators built as 
\bea
\hat U_\alpha^\dagger\hat n_p\hat U_\alpha = \hat C_p^{(\alpha)} = \left[\hat C_p^{(\alpha)}\right]^2.
\eea
This permits us to redefine the two-electron fragments $\hat{H}_{2e}^{(\alpha)}$ by removing idempotent one-electron terms as
\begin{align}
        [\hat H_{2e}^{(\alpha)}]' &= \hat H_{2e}^{(\alpha)} - \sum_{p}\epsilon_p^{(\alpha)} \hat C_p^{(\alpha)}\\
        &= \sum_{\alpha, \beta} (\lambda_{pq}^{(\alpha)} - \epsilon_p^{(\alpha)}\delta_{pq})\hat C_p^{(\alpha)}\hat C_q^{(\alpha)}.
\end{align}
where $\epsilon_p^{(\alpha)} \in \mathbb R$. Such a modification preserves the diagonalizability of the modified fragment $[\hat H_{2e}^{(\alpha)}]'$ by $\hat{U}_\alpha$, but can change the corresponding variance of the estimator. To ensure the fragments sum to the full Hamiltonian, the removed terms are recombined with the existing one-electron fragment $\hat H_{1e} = \sum_p \lambda_p \hat U_0^\dagger \hat n_p\hat U_0$. A combined one-electron fragment remains diagonalizable since it is still a one-electron operator, and therefore diagonalizable by an orbital transformation $\hat{U}_0'$
\begin{equation}
    \hat H_{1e} + \sum_\alpha \sum_p  \epsilon_p^{(\alpha)} \hat C_p^{(\alpha)}  = 
    \hat U_0' \left[ \sum_p \lambda_p' \hat n_p \right]\hat U_0'^\dagger.
\end{equation}
By minimizing variances of modified fragments with respect to the parameters $\{\epsilon_p^{(\alpha)}\}$, \citet{choi2023fluid} showed significant reduction to the overall measurement cost, achieving the state-of-the-art number of expected measurement cost on common molecular systems.

   
Evidently, a major problem with these approaches is that variances and co-variances needed for the optimization 
are not available without their evaluation with the wavefunction. Thus, prior works have relied on approximate estimates of these quantities.
The approximations came from classical quantum-chemistry methods \cite{yen2021cartan, yen2023deterministic, choi2023fluid} or accumulated statistics of quantum measurements \cite{shlosberg2023adaptiveestimation}. 
The classical quantum-chemistry methods provide tractable reference wavefunctions (e.g., the Hartree-Fock state) that are similar to the ground state of the Hamiltonian of interest. For approaches based on partitioning Pauli products, \citet{shlosberg2023adaptiveestimation} proposed estimating the covariance using the measurement results directly. The approach utilizes the fact that when measuring two Pauli products $\hat P_1$, $\hat P_2$ in the same fragment, one also obtains a sample of $\braket{\hat P_1 \hat P_2}$ that can then be used to estimate the covariance. 

\subsection{Classical Shadow Methods\label{reduction_shadows}}
 
The difference of classical shadow methods from others is a random selection of the measurement bases from a predetermined group. In this case, the number of measurements for a single observable is upper bounded by the shadow norm. The shadow norm is evaluated by maximizing the expected value of the square of the estimator over all possible input states. The square of the shadow norm of $\hat O = \sum_P\alpha_P \hat P$ has the form
\begin{align}
   ||\hat O||_{sh}^2 = \max_{\mu} \mathbb E_{\hat B}&\mathbb E_{\mu(\hat B)} \Bigg[\sum_{P, Q}\alpha_P\alpha_Q f(\hat B, \hat P)f(\hat B, \hat Q)\nonumber\\
   &\cdot \mu(\hat B, \mathrm{supp}(\hat P))\mu(\hat B, \mathrm{supp}(\hat Q))\Bigg],
\end{align}
where $\mathbb E_{\mu(\hat B)}[.]$ denotes the expectation value of the estimator when measured in the $\hat B$ basis, and $\mathbb E_{\hat B}[.]$ denotes the expected value averaged uniformly over all choices of basis $\hat B$. The maximization of $\mu$ can be thought of as a maximization over the measured density $\hat \rho$. For the Pauli classical shadows, the shadow norm of a single Pauli observable has the form $||\hat P||_{sh} = 3^l$, where $l =  |\mathrm{supp}(\hat P)|$ denotes the locality of the Pauli word. For a general observable, the shadow norm scales as $||\hat O||_{sh} = 4^l||\hat O||_{\infty}$.
 
The scaling of the shadow norm with locality is unfavourable and sampling over all bases is evidently wasteful for estimation of a single observable, since, given the constituents of an observable of interest, measuring under some bases offers more information about the energy expectation value than others. 
It turns out that neither randomness nor uniformity is necessary or efficient. 
``Biasing" techniques \cite{Hadfield:2021, Hadfield_Mezzacapo:2022, hillmich2021decision} increase the probability of selecting unitaries based on the sizes of the coefficients of their correspondingly compatible Pauli products, which is akin to the greedy approach for obtaining a Hamiltonian partitioning. 
For example, in \citet{Hadfield_Mezzacapo:2022}, the measurement basis on $j^{\text{th}}$ qubit $\hat B^{(i)}_j$ is randomly chosen following a probability distribution $\beta_j(\hat B_j)$ for each local measurement basis $\{\hat x_j, \hat y_j, \hat z_j\}$. 
The corresponding energy estimator is \bea
    \bar H = \sum_{P} \alpha_P \sum_{i=1}^M \frac{f(\hat B^{(i)}, \hat P, \beta)}{M} 
    \mu (\hat B^{(i)}, \text{supp}(\hat P))
    \label{eq:LBCS-estimator}
\eea with the corresponding weights
\bea
    f(\hat B, \hat P, \beta) &=& \prod_{j=1}^K f_j (\hat B_j, \hat \sigma_j, \beta_j) \\
    f_j (\hat B_j, \hat \sigma_j) &=& \begin{cases}
        1 & \text{if } \hat \sigma_j = \hat 1 \\
        \beta_j(\hat B_j)^{-1} & \text{if } \hat B_j = \hat \sigma_j \neq \hat 1 \\
        0 & \text{otherwise}
    \end{cases}
\eea 
for $\hat P = \otimes_{j=1}^N \hat \sigma_j$.
Analogously, the inner summation in \eq{eq:LBCS-estimator} can be seen as taking a simple average, since $\mu (\hat B^{(i)}, \text{supp}(\hat P))$ is evaluated $M/f(\hat B^{(i)}, \hat P, \beta)$ times on average.

On the other hand, the ``derandomization" technique proposed by \citet{Huang_Preskill:2021} uses the same greedy heuristic to choose a predetermined set of measurement bases and achieved a more efficient method than the initial random version.
However, while these methods can be shown to fall under the same measurement framework and employ similar heuristics, since the unitaries studied therein are restricted to single-qubit rotations, the Pauli products that can be measured together using these methods have to commute qubit-wise. 
This severely limits the efficiency of these methods when compared to partitioning methods based on fully commuting groups. 

Another recent method, called ``ShadowGrouping'', also exploits the greedy heuristic. \cite{greschGuaranteedEfficientEnergy2025} In this approach, measurement bases are obtained one at a time, via a ranking of Pauli operators that takes into account the magnitude of the associated coefficient, as well as how many times that Pauli operator has already been measured. A utility of this approach is that it works for both qubit-wise commutativity and full-commutativity.

Symmetries of the state or the estimated observables can also be used to reduce the sampling cost of classical shadows. One can choose a unitary ensemble and a measurement basis that is informationally-complete within the target irreducible subspaces of the symmetry. Reference \cite{sauvage2024classical} discusses several methods for constructing symmetric classical shadows for general symmetry groups, and demonstrates an exponential reduction of sampling cost when compared to local and global Clifford classical shadows for estimating observables or states that are invariant under qubit permutation.

\subsection{Hadamard Test and LCU Decompositions\label{reduction_hadamard}}

Methods to estimate expectation values and matrix elements of $\hat{H}$ based on an application of Hadamard tests (see Sec. \ref{hadamard_test_lcu_section}) to each term in an LCU decomposition of $\hat{H}$ 
\begin{equation}
    \hat{H} = \sum_j \alpha_j \hat{U}_j,
\end{equation}
require $O(\lambda^2/\epsilon^2)$ measurements to achieve precision $\epsilon$, where
\begin{equation}
    \lambda = \sum_j |\alpha_j|
\end{equation}
is the 1-norm of the LCU decomposition. \cite{leeSamplingErrorAnalysis2023} This motivates the development of LCU decompositions with reduced 1-norms for the purpose of reducing the overall measurement cost. Methods for 1-norm reduction have been developed in the context of optimizing LCU decompositions for block-encoding of $\hat{H}$ in quantum phase estimation, whose cost also grows with $\lambda$. \cite{LCU_childs} For qubit Hamiltonians, it was shown in Ref. \cite{loaizaReducingMolecularElectronic2023a} that the naive LCU decomposition based on Pauli operators can be improved by grouping together $K$ mutually anti-commuting Pauli operators, as such a linear combination preserves unitarity. However, the unitary that transforms the resulting Hamiltonian to Ising form will contain $K - 1$ Pauli rotations in general, implying a much deeper measurement circuit that will effect the overall quantum circuit depth.  Furthermore, it was shown in Refs. \cite{loaizaReducingMolecularElectronic2023a} and \cite{koridonOrbitalTransformationsReduce2021} that the 1-norm for the Pauli LCU and the anti-commuting grouping LCU can be reduced further by optimizing the orbital basis. For fermionic LCUs based on obtaining a full-rank or low-rank decomposition of the two-electron part of the electronic Hamiltonian (see Sec. \ref{hadamard_test_lcu_section}), Ref. \cite{oumarouAcceleratingQuantumComputations2024} proposed introducing a additional term in the cost function \eq{csa_cost} which penalizes increases in the 1-norm of the Hamiltonian decomposition.

Another method to reduce the 1-norm is based on its relation to the spectral range $\Delta E = E_\text{max} - E_\text{min}$ of the target Hamiltonian. It was shown in Ref. \cite{loaizaReducingMolecularElectronic2023a} that the 1-norm of any LCU decomposition of $\hat{H}$ is lower bounded by half of the spectral range of $\hat{H}$: $\lambda \geq \Delta E / 2$. Therefore, one can improve the efficiency of measurement approaches which scale in the LCU 1-norm by developing methods to engineer modifications of the target electronic Hamiltonian with reduced spectral range, for which the target eigenstate and eigenenergy remain unchanged. The most successful approach to reduce the spectral range of the target Hamiltonian is to directly modify the Hamiltonian with a block invariant symmetry shift (BLISS) operator $\hat{K}$, producing a new Hamiltonian $\hat{H} - \hat{K}$ with reduced spectral range and 1-norm. \cite{loaiza2023LCU2} The operator $\hat{K}$ satisfies two properties: 1) $\hat{K}$ fixes a subspace defined by symmetries $\{\hat{S}_k = s_k\}_{k=1}^{K}$ of $\hat{H}$, which contains the target eigenstate; 2) $\hat{H} - \hat{K}$ has reduced spectral range compared with $\hat{H}$. The operator $\hat{K}$ is chosen to minimize the spectral range of $\hat{H} - \hat{K}$, whose lower bound is the spectral range of $\hat{H}$ when restricted to the subspace in which all $\hat{S}_k = s_k$, which is unmodified by the BLISS operator $\hat{K}$. \cite{cortesAssessingQueryComplexity2024} For the electronic Hamiltonian, the total number operator $\hat{N}_e = \sum_p \hat{n}_p$ is used as the symmetry that defines the BLISS operator, producing the two-electron operator
\begin{equation}
    \hat{K}(\vec{\xi}) = \left(\xi_0 + \sum_{pq} \xi_{pq} \hat{E}_{pq}\right)(\hat{N}_e - N_e) \label{BLISS_op}
\end{equation}
which annihilates any state with $N_e$ electrons. Although $\hat{S}_z$ and $\hat{S}^2$ also commute with the Hamiltonian, it was found that inclusion of such spin-symmetries in the BLISS operator does not yield improvements in spectral range or LCU 1-norms of the electronic Hamiltonian. \cite{loaiza2023LCU2} For obtaining the optimal BLISS operator, various approaches, either based on non-linear gradient based optimization of the BLISS parameters, \cite{loaiza2023LCU2,roccaReducingRuntimeFaultTolerant2024, dekaSimultaneouslyOptimizingSymmetry2024,caesuraFasterQuantumChemistry2025} or via linear programming, \cite{patelGlobalMinimizationElectronic2025} have been developed. Reference \cite{leeReductionFiniteSampling2024} found that the application of a BLISS operator to $\hat{H}$ before computing an LCU decomposition produces between 1 and 2 orders of magnitude improvement in overall cost of computing matrix elements via measurements for small molecules. 

\subsection{IC-POVM Measurements}\label{Subsec:IC_POVM_opt}

In the IC-POVM approach to measurement (see Sec. \ref{Subsec:IC_POVM}), all choices of IC-POVMs can be used in principle for measuring the Hamiltonian expectation value. However, the accuracy of the estimate of the expectation value $\langle \hat{H} \rangle = \sum \omega_{\mathbf{m}} \langle \hat{\Pi}_{\mathbf{m}}\rangle$ depends on the choice of IC-POVM, since $\langle \hat{H} \rangle$ is a function of the probabilities $\Pr(\mathbf{m}) = \langle \hat{\Pi}_\mathbf{m}\rangle$ and weights $\omega_{\mathbf{m}}$. In the adaptive measurement scheme of Ref. \cite{GarciaPerezPOVMMeasurement}, both the locality and the informational completeness properties of the POVMs that are measured is used to simultaneously optimize the choice of POVM while the expectation value $\langle \hat{H} \rangle$ is being estimated. This is accomplished by estimating the gradient of the variance of the estimate as a function of the POVM parameters. This is possible since informational completeness of the POVM being measured can be exploited to estimate the variances of other POVMs, and the locality of the POVMs is necessary for this to be achievable efficiently. For the initial choice of POVM to start the optimization, Ref. \cite{GarciaPerezPOVMMeasurement} proposed the use of so-called symmetric IC-POVMs, which are a set $\{1/2 \ket{\pi_i}\bra{\pi_i} : 1 \leq i \leq 4\}$, where the $\ket{\pi_i}$ are single qubit states which form a tetrahedron on the Bloch sphere.

\section{Measuring Multiple States, Operators, or Matrix Elements\label{Sec:OtherQuantities}}

Previous sections dealt with methods to estimate $\langle \hat{H} \rangle$ via measurements on a quantum computer. Here, we discuss the generalization of this problem to the simultaneous estimation of multiple expectation values. This has many applications, including the treatment of excited states, estimating reduced density matrices, and estimating energy gradients in the variational quantum eigensolver, all of which are also discussed in this section. 

\subsection{One Operator, Multiple States}
\label{Subsec:multiple_states}
Many important chemical processes, such as photochemistry, catalysis, and spectroscopy, require modelling multiple eigenstates of the electronic Hamiltonian; typically this consists of the ground state as well as a subset of low-level excited states. Various excited-state VQE approaches, based on preparing excited states on the quantum computer and estimating their expectation values via repeated quantum measurements, have been developed. \cite{peruzzoVariationalEigenvalueSolver2014,higgottVariationalQuantumComputation2019,parrishQuantumComputationElectronic2019,santagatiWitnessingEigenstatesQuantum2018} A challenge for excited-state VQE algorithms is the increased measurement cost associated to estimating the expectation values of multiple eigenstates $\{\ket{\phi_k}\}$ of the target electronic Hamiltonian $\hat{H}$. Nonetheless, many of the measurement techniques and cost reduction methods discussed thus far are also applicable to this situation. The mesurement cost is quantified as the total number of measurements $M(\epsilon)$ required to obtain an estimate of all energies $\braket{\phi_k|\hat{H}|\phi_k}$ with $\epsilon$ accuracy. Here it is useful to make a distinction between measurement methods which are state-independent (e.g., greedy approaches for Hamiltonian partitioning, or derandomization for classical shadow tomography), and methods which use information about the target state to reduce the measurement cost (e.g., covariance based approaches like coefficient splitting). The former methods can be applied essentially without modification when estimating the expectation values of multiple states. Moreover, as described in Sec. \ref{reduction_greedy}, it is still true that an uneven distribution of fragment variances will generally yield a decrease in the measurement cost of the associated partitioning. Here, since there are multiple states, there will be a distribution of variances for each state being measured. Greedy approaches, which produce fragments with an uneven distribution in norm (1-norm of Pauli coefficients in the qubit representation, or $\ell^2$ norm of the two-body-tensor in the fermionic representation), will generally produce an uneven distribution of fragment variances for all target eigenstates simultaneously, motivating the usage of greedy partitioning methods even when multiple states are being measured. For covariance based approaches, one must optimize the Hamiltonian partitioning to minimize the total measurement cost $M(\epsilon)$. As $M(\epsilon)$ is exponentially hard to calculate exactly, it was proposed in Ref. \cite{choiMeasurementOptimizationTechniques2023} to estimate $M(\epsilon)$ using the measurement cost associated to the equal mixture of classically obtained approximations $\ket{C_k}$ of each $\ket{\phi_k}$
\begin{equation}
    M_\rho(\epsilon) = \frac{1}{\epsilon^2}\sum_\alpha \frac{\Tr(\hat{H}_\alpha^2 \hat{\rho}) - \Tr(\hat{H}_\alpha \hat{\rho})^2}{m_\alpha}, \label{multi_state_cost}
\end{equation}
where $m_\alpha$ is the number of measurements allocated to $\hat{H}_\alpha$, and $\hat{\rho}$ is the associated density matrix 
\begin{equation}
    \hat{\rho} = \frac{1}{N_s} \sum_{k=1}^{N_s}\ket{C_k} \bra{C_k}.
\end{equation}
Reference \cite{choiMeasurementOptimizationTechniques2023} optimized the Hamiltonian partitioning for the excited-state VQE algorithm of Ref. \cite{parrishQuantumComputationElectronic2019} based on using \eq{multi_state_cost} as a cost function, and demonstrated a reduction the true measurement cost $M(\epsilon)$ of up to two orders of magnitude, compared with only using greedy approaches that do not use information about the target eigenstates.

\subsection{One State, Multiple Operators}
\label{Subsec:multiple_operators}
Often, one is interested in the expectation value of multiple operators for a fixed state. In this scenario, the nature of the operators being measured can influence which measurement approach has the lowest measurement cost. We devote this section to discussing two important special cases of this problem: the estimation of fermionic $k$-body reduced density matrices, and energy gradients in variational quantum algorithms.

\subsubsection{Reduced Density Matrices}

The fermionic $k$-body reduced density matrix ($k$-RDM) of a multi-fermionic wavefunction encodes $k$-body interactions between fermions, and takes the following form
\begin{equation}
    D_{q_1\dots q_k}^{p_1\dots p_k} = \Tr(\hat a_{p_1}^\dagger \cdots \hat a_{p_k}^\dagger \hat a_{q_k}\cdots \hat a_{q_1}\hat \rho).
\end{equation}
Elements of the $k$-RDM can be used to obtain estimates of observables composed of fermionic excitation operators. In particular, the $2$-RDM is used to estimate the expectation value of the electronic Hamiltonian and other physical properties, such as pair-correlation functions and order parameters. \cite{Mazziotti2012} It is infeasible in practice to use Pauli classical shadows for the task of estimating RDM elements, as Pauli classical shadows require that the RDM elements be expressed as Pauli products. This is problematic since fermion-to-qubit transformations like the Jordan-Wigner transformation do not preserve the locality of fermionic or Majorana-fermionic operators when expressed in terms of qubit operators. The cost of estimation with Pauli classical shadows, determined by the shadow norm scales exponentially with the non-locality of the Pauli products, rendering this approach prohibitively costly for estimating fermionic RDMs. Nonetheless, CST methods based on groups of fermionic unitaries can still be used, as the corresponding shadow norm does not have this exponential scaling. Reference \cite{RubinMiyake} proposed and benchmarked two families of fermionic unitaries to obtain classical shadows for estimating $k$-RDMs. The first family consists of the set of unitaries which are simultaneously fermionic Gaussian unitaries (FGU), defined in Eq. \eqref{fgu_definition}, and Clifford unitaries, for which the number of samples to estimate each entry of the RDM up to accuracy $\epsilon$ was shown to scale as $O\left[{N \choose k} k^{3/2}\log N/\epsilon^2\right]$. The second family consists of the set of unitaries which are simultaneously orbital rotations, defined in Eq. \eqref{eq:orbitaltransform}, and Clifford unitaries, augmented with an additional layer of single qubit Clifford unitaries, required to make the group informationally complete. In this case, the upper bound of the number of samples was shown to scale as $O\left[{N^k \log N}/{\epsilon^2}\right]$. For both classes of unitaries, the expectation is polynomial in $N$ for fixed $k$. Numerical analysis \cite{RubinMiyake} demonstrates that the orbital rotation based classical shadows had a lower sampling cost for this task. 

The sampling cost reduction observed for both families of unitaries can be seen as an indirect consequence of the symmetry of the measured observables: the degree of fermionic operators in the $k$-RDM remains invariant under these unitary transformations. Often the expectation values of the RDMs are estimated for states of fixed electron number $N_e < N$, and it is sufficient for the scheme to be informationally-complete in a particular electron number subspace. With this observation, \citet{low2022classical} proposed classical shadows with the family of particle number conserving unitaries (without restricting to Clifford unitaries) that demonstrate a sampling scaling that is polynomial in $N_e$ and not $N$.

Reference \cite{AvdicFewerMeasurements2024} proposed to reduce the sampling cost associated to estimating the $2$-RDM by using the information obtained via classical shadows to define an classical optimization problem which obtains the RDM. Shadows obtained by a real orbital rotation $\hat U$ followed by measurement in the computational basis yields estimates the following expectation values 
\begin{align}
    S_{pq} &= \Tr( \hat a_p^\dagger \hat a_q^\dagger \hat a_q\hat a_p \hat U\hat \rho \hat U^\dagger) \nonumber\\ 
    &= \sum_{ijkl} u_{pi}u_{qj}u_{ql}u_{pk} D_{ij}^{kl}\label{eqn:shadow_rdm}
\end{align}
where $D$ is the $2$-RDM, and $u = \exp(\theta)$ is the $N \times N$ orbital rotation matrix defined by exponentiating the matrix $\theta$ of orbital rotation angles as shown in \eq{eq:orbitaltransform}. The problem of estimating 2-RDM from the classical shadows can be recast into finding $D$ that satisfies the constraints set by the shadows in \eq{eqn:shadow_rdm}. Further constraints to the $2$-RDM structure can be obtained from the $N$-representability conditions, which imply various additional properties of $D$, for example: positive-semidefiniteness and a fixed trace of $N(N-1)$, when expressed as an $N^2 \times N^2$ matrix $D_{ij,kl}$. By converting the constrained search problem into a convex optimization to minimize the energy expectation value function, Ref. \cite{AvdicFewerMeasurements2024} showed reduction in the sampling cost of the classical shadows for 2-RDM estimation. Additionally, relaxing the shadow constraints to find $D$ that approximately satisfies Eq. \eqref{eqn:shadow_rdm} along with the $N$-representability conditions helped mitigate noise errors and inaccuracies in the estimated energy.

\subsubsection{Energy Gradients in the Variational Quantum Eigensolver}

Another place where the expectation value of multiple operators is needed is the estimation of energy gradients for VQE. Here, the form of the unitary ansatz plays a role in how the gradient is evaluated. A typical construction of the ansatz, as proposed in qubit coupled cluster (QCC) \cite{ryabinkin2018qubit} and the adaptive derivative-assembled psuedo-Trotter ansatz VQE (ADAPT-VQE) methods, \cite{grimsleyAdaptiveVariationalAlgorithm2019} begins with a set of simple Hermitian operators $\{\hat{G}_k\}$ (the ``operator pool'') and an easy to prepare initial state $\ket{\psi_0}$ (e.g., the Hartree-Fock state). Then, the ansatz takes the form of a product of exponential functions of operators $\hat{G}_k$ acting on $\ket{\psi_0}$:
\begin{equation}
    \ket{\psi(\vec{\theta})} = \prod_{j} e^{\imag \theta_j\hat{G}_{k_j}}\ket{\psi_0} = \hat{U}(\vec{\theta})\ket{\psi_0}.\label{unitary_ansatz}
\end{equation}
Note that there can be repetition of operators from $\{\hat{G}_k\}$ in the unitary ansatz. The unitary $\hat{U}(\vec{\theta})$ is built one exponential operator at a time, until the energy estimate converges. The criterion for deciding which operator $\hat{G}_k$ to include in the $n^{\text{th}}$ iteration of building $\hat{U}(\vec{\theta})$ is based on which operator has the largest gradient $\partial E / \partial \theta$ of the energy functional
\begin{equation}
    E(\vec{\theta}) = \braket{\psi|\hat{U}^\dagger(\vec{\theta}) \hat{H} \hat{U}(\vec{\theta})|\psi}.
\end{equation}
To estimate the gradient, it must be written in terms of expectation values of operators which can be measured on the quantum computer. For this, there are two approaches that have been considered: (1) parameter-shift rules, and (2) commutator approaches, which we discuss here. 

The parameter shift rule (PSR) was originally proposed for unitaries generated by operators with two eigenvalues (e.g., Pauli operators), \cite{liHybridQuantumClassicalApproach2017,mitaraiQuantumCircuitLearning2018,schuldEvaluatingAnalyticGradients2019} and later generalized to unitaries with arbitrary generators. \cite{izmaylovAnalyticGradientsVariational2021,wierichsGeneralParametershiftRules2022} The generalized PSR gives the energy gradient in terms of a linear combination of expectation values:
\begin{equation}
    \frac{\partial E}{\partial \theta} = \sum_{l=1}^{N_t} c_l \braket{\psi|U^\dagger_2 e^{\imag (\theta + s_l)\hat{G}} \hat{U}^\dagger_1 \hat{H} \hat{U}_1 e^{-\imag (\theta + s_l)\hat{G}}\hat{U}_2|\psi} \label{gpsr}
\end{equation}
where $\hat{U}_1, \hat{U}_2$ denote the parts of $\hat{U}(\vec{\theta})$ to the left and right of $e^{-\imag \theta \hat{G}}$.  Therefore, using the generalized PSR allows one to use the same parametrized quantum circuit which measures the expectation value of $\hat{H}$ to estimate the expectation value of the gradient, with the only difference being the requirement to use different values of a single parameter. The decomposition of $\partial E / \partial \theta$ into a linear combination of expectation values is not unique, and the values of $c_l, s_l$ depend in general on the eigenvalues of $\hat{G}$. The total number of terms, $N_t$, which contributes to the cost of using the generalized PSR, depends on the number of distinct eigenvalues $L$ in the spectrum of $\hat{G}$. In the best case scenario, Ref. \cite{izmaylovAnalyticGradientsVariational2021} proposed a procedure to obtain the gradient \eq{gpsr} using an LCU decomposition of $\hat{G}$ in which $N_t = O(\log(L))$. Alternative approaches based on a polynomial decomposition $e^{\imag \theta \hat{G}} = p(\hat{G})$ \cite{izmaylovAnalyticGradientsVariational2021,wierichsGeneralParametershiftRules2022} produce a longer decomposition where $N_t = O(\text{poly}(L))$. Nevertheless, for two commonly used special cases (1): when $\hat{G}$ is a Pauli operator, and (2): when $\hat{G}$ is a fermionic excitation and the state $\ket{\psi}$ is real, only two expectation values are needed to estimate the gradient. \cite{anand2022quantum} A deficiency in generalized PSR approaches to estimating gradients is that, in the regime where the magnitude of $\partial E / \partial \theta$ is substantially smaller than the magnitudes of the corresponding expectation values, the relative error in the estimate of $\partial E / \partial \theta$ can become prohibitively large.

In commutator based approaches, the gradient $\partial E / \partial \theta$ associated to generator $\hat{G}_k$ is evaluated via its relation to the expectation value of $\imag[\hat{G}_k, \hat{H}]$
\begin{equation}
    \frac{\partial}{\partial\theta} \braket{\psi|e^{-\imag\theta \hat{G}_k} \hat{H} e^{\imag \theta \hat{G}_k}|\psi}\Big|_{\theta=0} = \imag\braket{\psi|[\hat{G}_k, \hat{H}]|\psi}.\label{pool_gradients}
\end{equation}
Therefore, the expectation value of the commutator $[\hat{G}_k, \hat{H}]$ must be estimated via measurements for all $\hat{G}_k$ in the operator pool, which can induce a significant overhead in the overall measurement cost for a single VQE step. Also, even though the commutator expression removes the problem of assembling a potentially small gradient value from large expectation values of the Hamiltonian [cf. \eq{gpsr}], note that the commutator expression can be efficiently used only for the generator next to the Hamiltonian.  

The simplest approach to reduce the measurement cost in commutator based approaches is to reduce the size of the operator pool. In fermionic, \cite{grimsleyAdaptiveVariationalAlgorithm2019} qubit, \cite{tangQubitADAPTVQEAdaptiveAlgorithm2021} and qubit-excitation-based \cite{yordanovQubitexcitationbasedAdaptiveVariational2021} ADAPT-VQE, the operator pool contains operators which are derived from fermionic single and double excitation operators, and therefore contains $O(N^4)$ elements. Since the electronic Hamiltonian has $O(N^4)$ operators as well, the total number of distinct operators whose expectation values must be estimated becomes $O(N^8)$. However, Ref. \cite{shkolnikovAvoidingSymmetryRoadblocks2023} demonstrated using Lie algebraic arguments that the smallest operator pools which can guarantee, in principle, an exact representation of any quantum state using the ansatz of the form shown in \eq{unitary_ansatz}, only need $2N - 2$ elements. Furthermore, since the electronic Hamiltonian has Pauli symmetries, it was also shown in Ref. \cite{shkolnikovAvoidingSymmetryRoadblocks2023} that the minimal sized operator pools necessary to produce the ground state can have even fewer than $2N - 2$ elements. Numerical studies of these symmetry-conserving minimal complete pools confirmed that, in practice, they are still able to produce a unitary ansatz which surpasses chemical accuracy in the energy. Alternatively, in Ref. \cite{ryabinkin2020iterative}, an algorithm is developed which uses information about the Hamiltonian terms to construct a polynomial sized operator pool of Pauli operators which are all guaranteed to have non-zero gradient.

Another approach to reduce the measurement cost for gradients in VQE is to collect Pauli operators into measurable groups. Reference \cite{anastasiouHowReallyMeasure2023} demonstrated that the $O(N^4)$ Pauli operators in the qubit ADAPT-VQE operator pool can be collected into $O(N)$ fully-commuting groups, reducing the total number of distinct expectation values from $O(N^8)$ to $O(N^5)$. Yet, this result
does not provide the total number of measurements one needs to do to obtain the gradients with fixed accuracy.

\subsection{Matrix Elements and Overlaps}
\label{Subsec:matrix_elements_overlaps}

The primary barrier for solving the electronic structure problem on a classical computer is the exponential scaling of the Fock space dimension in the system size. For this reason, subspace methods are a popular choice to obtain approximate ground and excited states on both classical and quantum computers. These methods project the target Hamiltonian $\hat{H}$ onto a low-dimensional subspace $W$ of the full Fock space. The subspace $W$ is spanned by a set of states $\{\ket{\phi_\mu}\}$, which in general are not orthogonal. Then, one can approximate the ground and excited state energies by solving a generalized eigenvalue problem for the Hamiltonian projected onto $W$:
\begin{equation}
    \boldsymbol{H}\vec{c} = \boldsymbol{S}E\vec{c} \label{gevp_definition}
\end{equation}
where 
\begin{align}
    \boldsymbol{H}_{\mu\nu} &= \braket{\phi_\mu|\hat{H}|\phi_\nu},\nonumber\\
    \boldsymbol{S}_{\mu\nu} &= \braket{\phi_\mu|\phi_\nu}\label{qsd_equations}
\end{align}
The ground state solution of \eq{gevp_definition} produces the best approximation $\sum_\mu c_\mu \ket{\phi_\mu}$ in $W$ to the ground state of $\hat{H}$. In quantum subspace methods (QSMs), one uses the quantum computer to prepare the states $\ket{\phi_\mu}$ which define the subspace $W$, and estimate the matrix elements $\boldsymbol{H}_{\mu\nu}$ and $\boldsymbol{S}_{\mu\nu}$ via repeated quantum measurements. The choice of measurement technique to use typically depends on how the states $\ket{\phi_\mu}$ are defined, which in turn is dependent on the choice of QSM used - see Ref. \cite{mottaSubspaceMethodsElectronic2023} for a review of various QSMs applied to the electronic structure problem. In the following discussion of these measurement techniques, we will write the states defining $W$ in terms of operators $\{\hat{O}_\mu\}$ acting on a fixed reference state $\ket{\psi_0}$ that can be prepared on the quantum computer: $\ket{\phi_\mu} = \hat{O}_\mu \ket{\psi_0}$. 

Reference \cite{choiMeasurementOptimizationTechniques2023} developed and benchmarked various measurement schemes applied to the QSM approach of Ref. \cite{mccleanHybridQuantumclassicalHierarchy2017}, in which the operators defining the subspace are single excitation operators: $\hat{O}_\mu \in \{\hat{E}_{pq}\}$. A simple measurement approach here is to write the associated matrix elements directly as expectation values of the reference state $\ket{\psi_0}$:
\begin{align}
    \boldsymbol{H}_{\mu\nu} &= \braket{\psi_0|\hat{O}_\mu^\dagger \hat{H} \hat{O}_\nu|\psi_0}\nonumber\\
    \boldsymbol{S}_{\mu\nu}  &= \braket{\psi_0|\hat{O}_\mu^\dagger \hat{O}_\nu|\psi_0}.\label{qse_expectations}
\end{align}
This reformulation allows one to use methods for estimating expectation values for estimating matrix elements and overlaps. Let $\hat{A}_n \in \{\hat{O}_\mu^\dagger \hat{H} \hat{O}_\nu\} \cup \{\hat{O}_\mu^\dagger \hat{O}_\nu\}$ enumerate the full set of operators who expectation values must be estimated. Then, one can apply the Hamiltonian partitioning to all operators $\hat{A}_n$ simultaneously, 
\begin{equation}
    \hat{A}_n = \sum_\alpha \hat{U}_\alpha^\dagger p_n^{(\alpha)}(\hat{z}_1,\ldots,\hat{z}_N) \hat{U}_\alpha. \label{simultaneous_partitioning}
\end{equation}
With this form of the decomposition, fragments associated to multiple operators can be mapped to Ising form by the same unitary, so that a single measurement can provide information about multiple expectation values. In the qubit algebra, the sorted insertion method can be directly applied to the set of Pauli operators in the set of $\hat{A}_n$. This is done by sorting the Pauli operators based on the sum of the magnitudes of their coefficients in all $\hat{A}_n$ before obtaining the grouping. In the fermionic algebra, one can search for fragments of the form shown in \eq{ferm_csa} for all $\hat{A}_n$ simultaneously, constraining the orbital rotation $\hat{U}$ to be the same for all $n$. 

Reference \cite{choiMeasurementOptimizationTechniques2023} also considered applying CST to measurement of \eq{qse_expectations}. The simplest CST methods do not use any information about the operators or the state when defining measurement bases, and thus can be applied without modification to the estimation of all $\langle \hat{A}_n \rangle$. Particularly promising here is the derandomization method, \cite{Huang_Preskill:2021} which exploits the magnitude of coefficients of the Pauli terms in the $\hat{A}_n$ to choose the $\hat{U}_\alpha$ in a way that reduces the total number of measurements required to estimate all $\langle \hat{A}_n \rangle$ simultaneously. The analysis of Ref. \cite{choiMeasurementOptimizationTechniques2023} found that classical shadow techniques can outperform partitioning techniques when many operators are measured simultaneously.

The straightforward grouping method of \eq{simultaneous_partitioning} assumes that the operators $\hat{A}_n$ all have a representation as a linear combination of Pauli products with a number of terms that is polynomial in the number of terms in $\hat{H}$. This is the case in the QSM approach of Ref. \cite{mccleanHybridQuantumclassicalHierarchy2017} since the $\hat{O}_\mu$ are single-excitation operators, but is not necessarily the case in other QSM proposals. There are also methods which can be applied under the assumption that the states which define the subspace can be directly prepared on the quantum computer via a unitary transformation: $\ket{\phi_\mu} = \hat{U}_\mu \ket{\psi_0}$, of an initial reference state $\ket{\psi_0}$. This assumption is satisfied in subspace methods where the states are defined by time-evolution operators \cite{parrishQuantumFilterDiagonalization2019, klymkoRealTimeEvolutionUltracompact2022,mottaDeterminingEigenstatesThermal2020} generated by the target Hamiltonian applied to an approximate ground state, or when the subspace is defined by parametrized quantum circuits. \cite{stairMultireferenceQuantumKrylov2020} One method to estimate matrix elements of quantum states which can be prepared on the quantum computer is based on the Hadamard test approach of Sec. \ref{hadamard_test_lcu_section}, which requires an LCU decomposition of $\hat{H}$
\begin{equation}
    \hat{H} = \sum_k c_k \hat{V}_k.
\end{equation}
Then, since a product of unitaries is still unitary, the matrix elements and overlaps can be written in terms of expectation values of unitary operators as 
\begin{align}
    \boldsymbol{H}_{\mu\nu} &= \sum_k c_k \braket{\psi_0|\hat{U}_\mu^\dagger \hat{V}_k \hat{U}_\nu|\psi_0}\\
    \boldsymbol{S}_{\mu\nu} &= \braket{\psi_0|\hat{U}_\mu^\dagger \hat{U}_\nu|\psi_0}
\end{align}
giving the form required for Hadamard test approach to be used. Reference \cite{cortesQuantumKrylovSubspace2022a} proposed an alternative method to estimate the expectation values of these unitaries which is applicable in the case that the states $\ket{\phi_\mu}$ and unitaries $\hat{V}_k$ have simple symmetries, such as the number-of-electrons operator $\hat{N}_e = \sum_p \hat{n}_p$. In this approach, one calculates the matrix elements via measurements of a superposition of two states with different symmetry values, without having to use an ancilla qubit. 
\begin{figure}
    \includegraphics[width=0.8\columnwidth]{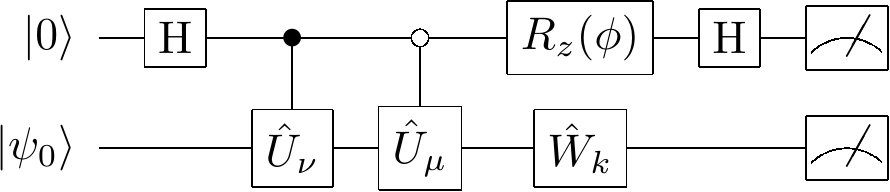}
    \caption{Extended swap test circuits for estimating $\text{Re}\braket{\psi|\hat{H}_k|\psi}$ ($\phi = 0$) and $\text{Im}\braket{\psi|\hat{H}_k|\psi}$ ($\phi = \pi/2$). Here, $R_z(\phi) = e^{-\imag \phi \hat z / 2}$ is the Pauli-$\hat{z}$ rotation operator, $\mathrm{H}$ denotes the Hadamard gate, $\hat{W}_k$ is the unitary that maps $\hat{H}_k$ to Ising form, and $\hat{U}_{\nu}$, $\hat{U}_{\mu}$ are used to prepare the state $\ket{\varphi_{\mu\nu}}$.}
    \label{extended_swap_test_fig}
\end{figure}


Alternatively, in the method known as the extended swap test, \cite{parrishQuantumFilterDiagonalization2019}, one evaluates the matrix element via a standard partitioning-based measurement procedure applied to an operator $\tilde{H}$ on a larger space, defined as follows
\begin{equation}
    \tilde{H} = (\hat x + \imag\hat y) \otimes \hat{H}.
\end{equation}
Given a decomposition of $\hat{H}$ into measurable fragments:
\begin{equation}
    \hat{H} = \sum_k \hat{H}_k = \sum_k \hat{W}_k^\dagger \hat{D}_k \hat{W}_k,
\end{equation}
and the following state in the larger space 
\begin{equation}
    \ket{\varphi_{\mu\nu}} = \frac{1}{\sqrt{2}}(\ket{0} \hat{U}_\mu\ket{\psi_0} + \ket{1}\hat{U}_\nu\ket{\psi_0}),
\end{equation}
we have
\begin{align}
\boldsymbol{H}_{\mu\nu} &= \sum_k \braket{\varphi_{\mu\nu}|(\hat x + \imag\hat y)\otimes \hat{H}_k|\varphi_{\mu\nu}}\\
\boldsymbol{S}_{\mu\nu} &= \braket{\varphi_{\mu\nu}|(\hat x + \imag\hat y)\otimes \hat{1}|\varphi_{\mu\nu}}.
\end{align}
Therefore, $\boldsymbol{H}_{\mu\nu}$ can be calculated via repeated measurements of the exactly solvable Hermitian $\{\hat x \otimes \hat{H}_k\}$ and anti-Hermitian $\{\imag\hat y \otimes \hat{H}_k\}$ fragments of $\tilde{H}$ - see \fig{extended_swap_test_fig} for the full quantum circuit. In practice, since $\hat{H}$ is Hermitian, it follows that both the expectation values $\boldsymbol{H}_{\mu\nu}$ and overlaps $\boldsymbol{S}_{\mu\nu}$ are real. In this case, one only needs to calculate the real expectation values $\hat x \otimes \hat{H}_k$, and not the contributions to the imaginary part associated to the expectation values of the anti-Hermitian fragments $\imag \hat y \otimes \hat{H}_k$. \cite{parrishQuantumFilterDiagonalization2019}

Both the Hadamard test approach and the swap test approach have, associated with them, a finite sampling error, which one can optimize as a function of the LCU in the Hadamard test approach, or choice of measurable fragments in the extended swap test approach. Since the extended swap test approach is identical to the method of measuring Hamiltonian fragments for an effective Hamiltonian on a larger space, all methods for the reduction of measurement cost of Hamiltonian partitioning approaches, discussed in Sections \ref{reduction_greedy} and \ref{reduction_coefficient_splitting} can be applied essentially without modification here. Similarly, all methods to reduce the cost of measurement via Hadamard tests applied to an LCU decomposition of the target Hamiltonian, as described in Sec. \ref{reduction_hadamard} can also be applied without modification here. Reference. \cite{leeReductionFiniteSampling2024} found that, after combining multiple measurement reduction techniques, the method based on extended swap tests requires fewer number of measurements, compared with the method based on the Hadamard test.

Another approach to reduce the measurement cost involves choosing the states such that a fewer number of total measurements are needed to generate the full set of matrix elements $\boldsymbol{H}_{\mu\nu}$ and $\boldsymbol{S}_{\mu\nu}$. In the classically-boosted VQE method of Ref. \cite{radinClassicallyBoostedVariationalQuantum2021}, some of the states which define the subspace are taken to be classically tractable, meaning that the associated matrix elements for those states can be calculated efficiently on a classical computer. Reference \cite{radinClassicallyBoostedVariationalQuantum2021} showed a one to three order of magnitude reduction in the total measurement cost for small molecules by allowing a single state to be the classically tractable Hartree-Fock state.

\section{Noise-Error Reduction}
\label{Sec:NoiseReduction}

In addition to errors in expectation values arising from finite sampling there are errors related to imperfection of quantum hardware. 
The noise related errors are worse than the finite sampling errors because they can introduce not only broader distributions but bias. 
Here we review techniques that mitigate the errors for computational methods intended for non-fault tolerant quantum computing. This topic 
is vast and there are already good reviews written on it \cite{cai2023quantum}. The focus of our discussion will be 
methods that depend on unitaries that are required for 
measurements (e.g., $\hat U_\alpha$ diagonalizing measurable 
fragments) and that use physical meaning of operators that are measured
 (e.g., $p(\{\hat z_k\})$ corresponding to measurable fragments).
Noise related errors can occur during initial state preparation, unitary evolution, and measurement. 

\subsection{Measurement Circuit Modifications}

Here we consider techniques that reduce the circuit part required for the measurement. Due to non-unit fidelity of gates, 
more operations leads to higher errors. On the other hand, 
the measurement part of the circuit allows to measure fragments that 
can have lower quantum variances, $\VA{\hat H_\alpha}$. Therefore,
generally, one wants to strike a balance between the two limits: shallower measurement circuits with larger quantum variances or deeper measurement circuits with larger noise error contributions.  

\subsubsection{Optimization Based on Various Commutativity Definitions}

Lower gate counts in QWC measurement circuits result in lower estimation bias but high variance. In contrast, FC measurement circuits, which involve larger gate counts, produce estimates with higher bias but lower quantum variance. 
Estimator's bias and variance depend non-trivially on the operators, $\{\hat H_\alpha\}$ and error model. \cite{bansingh_fidelity_2022} For certain ranges of noise model parameters and sample counts, a significant reduction in bias can be achieved at the cost of a small increase in variance and vice-versa. Motivated by this observation, several recent works have proposed hybrid methods between FC and QWC to minimize the estimator mean square error (MSE). 

The compatibility criteria for grouping two Pauli products into QWC fragments is stringent: the Pauli operators at each qubit position must commute. By relaxing this criterion to allow commutativity over larger subsets of qubit positions,  larger fragments with lower variances can be constructed, that are diagonalized by measurement unitaries with more gates. \cite{Hamamura2020,Escudero2023} While these larger fragments results in lower variances, the estimates had higher bias compared to QWC fragments, resulting in larger MSE. However, the reduced variance facilitates faster convergence in VQE routines employing this estimation method.
Similarly, \citet{burns2024galichybridmultiqubitwisepauli} proposed a method to constrain the FC Pauli product compatibility criteria to create smaller fragments, diagonalized by unitaries with a controllable number of gates. This approach tailors the gate count to ensure that the resulting MSE remains below a predefined precision threshold.

\subsubsection{Optimization Accounting for Device Connectivity}

The connectivity between qubits in quantum hardware is often limited by the physical implementation of the system. The set of qubit pairs that support two-qubit gates is usually a strict subset of all possible pairs. To implement a two-qubit gate between disconnected qubits, additional SWAP gates are required to bring the target qubit states to connected qubits. Each SWAP gate comprises three CNOT gates, introducing noise overhead.

This overhead can be entirely avoided in QWC measurement circuits, as their unitaries involve only single-qubit gates. In methods like FC, SWAP gate overhead can be reduced by tailoring measurement schemes to the hardware's connectivity. For instance, \citet{miller2022hardwaretailoreddiagonalizationcircuits} used the equivalence between graph state circuits and Clifford circuits to group commuting Pauli product groups that are diagonalized by measurement circuits consisting of two-qubit gates restricted to connected qubit pairs. This approach demonstrated lower variance compared to QWC fragments and achieving lower overall errors for a fixed number of measurements. However, the search for optimal fragments within this framework remains inefficient and poses a classical computing 
bottleneck. 

\subsubsection{Problem Specific Reductions}

Exploiting problem-specific structure can lead to gate-efficient measurement circuits and lower estimation errors. 
For example, in the 2D Hubbard Hamiltonian, a five-fragment approach has been proposed that leverages the locality of nearest neighbour interaction terms in the Hamiltonian. \cite{jiangQuantumAlgorithmsSimulate2018} These fragments have measurement unitaries of constant depth, requiring at most a single layer of CNOT gates. A critical enabler of this low-depth circuit is the use of the Jordan-Wigner transformation for mapping fermionic states and operators to qubits. As noted in Section \ref{commuting_ferm_subsec}, general orbital rotations are decomposed into a product of simpler rotations $\hat G_{pq}(\phi_{pq})$ defined in Eq. \eqref{orbital_givens}. For efficiently implementing $\hat G_{pq}(\phi_{pq})$, the Jordan-Wigner transformation plays a key role. For adjacent qubits ($q = p\pm 1$), the Jordan-Wigner transform of fermionic excitations $(\hat E_{pq} - \hat E_{qp})$ results in Pauli products of the form $i(\hat x_p\hat y_{q} - \hat y_p\hat x_q)/2$ which involve only two qubits. This locality is exploited to construct depth- and gate-efficient circuits for general orbital rotations in fermionic CSA based fragmentation approaches. \cite{arrazola_universal_2022} Thus, the choice of fermion-to-qubit mappings significantly influences circuit depth, gate counts, and the resulting estimation errors.

\subsection{Measurement Modification and Post-Processing}

Based on physical properties of quantities that are measured, such as symmetries, or approximate but classically simulable ways of obtaining the same quantities, one can mitigate errors in measurement. Here we present a few techniques based on these ideas. One additional post-processing technique we cover is noise twirling, it is very analogous to the idea of classical shadow tomography. 

\subsubsection{Symmetry Verification}

The molecular Hamiltonian $\hat H$ has symmetries which form a group of operators, $\mathbb S$ that commute with $\hat H$. For instance, the number operator commutes with $\hat H$ and generates a $U(1) = \{\exp(i\theta \hat N_e)| \theta \in [0, 2\pi)\}$ continuous symmetry group of $\hat H$. The electron number parity operator, $\hat p = (-1)^{\hat N_e}$ commutes with the Hamiltonian and $\{\hat 1, \hat p\}$ forms a symmetry group as $\hat p^2 = \hat 1$. The irreducible representations of $\mathbb S$ divide the Hilbert space into irreducible subspaces, one of which contains the ground state $\ket{\psi}$ of $\hat H$. Denote this subspace by $\mathcal H_0 := \mathrm{span}(\{\hat S \ket{\psi}| \hat S\in \mathbb S\})$. Suppose the noise free prepared state is the ground state of $\hat H$. By projecting the noisy prepared state onto $\mathcal H_0$, one can remove errors that move the ground state out of $\mathcal H_0$ and potentially reduce estimation bias. 
This method to reducing estimation errors is known as symmetry verification (SV) and can be viewed as restoring symmetries of the state. Let $\hat \rho'$ be the density matrix of the noisy prepared state and $\hat{\mathcal P}$ is the projector onto $\mathcal H_0$, then the projected state is $\hat{\mathcal P}\hat \rho' \hat{\mathcal P}/\mathrm{Tr}(\hat \rho' \hat{\mathcal P})$. The expectation value of $\hat H$ on the symmetry projected state is
\begin{equation}\label{symmetry_projected_est}
    \langle \hat H \rangle_{\rm SV} = \frac{\Tr(\hat \rho' \hat{\mathcal P}\hat H)}{\Tr(\hat \rho' \hat{\mathcal P})}.
\end{equation}
To estimate $\langle \hat H\rangle_{\rm SV}$ one needs expectation values of $\hat{\mathcal P} \hat H$ and $\hat{\mathcal P}$. Fragmentation and de-randomized classical shadow approaches have been considered \cite{ruiz2024restoring}, and have been found to be comparable in sampling cost. A drawback to these approaches is the increase in the number of observables needed to be measured can lead to an increase in sampling requirement. Additionally, since the strength of noise in the two estimation procedures need not be the same, the ratio of estimates of expectation values of $\hat{\mathcal P} \hat H$ and $\hat{\mathcal P}$ is usually a biased estimator of $\langle \hat H \rangle_{\rm SV}$.

\begin{figure}
    \centering
    \includegraphics[width=\columnwidth]{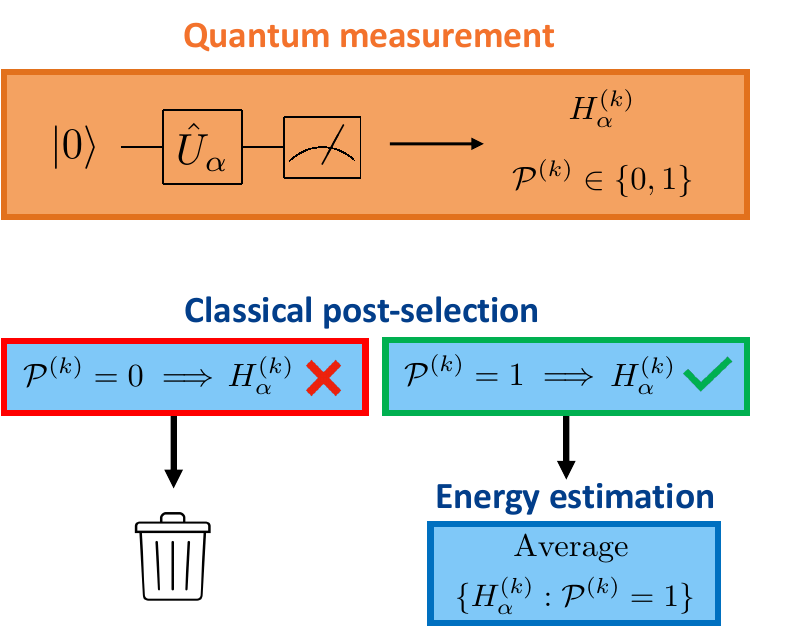}
    \caption{Diagrammatic representation of symmetry verification by post-selection to mitigate errors. The Hamiltonian and symmetry projector are measured simultaneously, and the Hamiltonian measurement $\hat H_\alpha^{(k)}$ is retained only if $\mathcal P^{(k)} = 1$. The Hamiltonian expectation value is estimated using only the retained measurements.}
    \label{fig:symmetry_verification}
\end{figure}

Alternatively, for a Hamiltonian $\hat{H} = \sum_\alpha \hat{H}_\alpha$ that is decomposed to measurable fragments, one often can obtain diagonalizing unitaries $\hat U_\alpha$ for the fragments $\hat{H}_\alpha$ which also diagonalize the projector $\hat{\mathcal P}$. In such a case, a symmetry projected estimate of the fragment $\hat{H}_\alpha$ can be obtained as
\begin{equation}
    \bar{H}_{\alpha,\mathcal P} = \frac{\sum_{k=1}^{M_\alpha} H_\alpha^{(k)} \mathcal P^{(k)}}{\sum_{k=1}^{M_\alpha} \mathcal P^{(k)}}
\end{equation}
where $H_\alpha^{(k)}, \mathcal P^{(k)}$ are measurement results for the fragment $\hat H_\alpha$ and the projector $\hat {\mathcal P}$ respectively, and $M_\alpha$ is the number of measurements allocated to estimating the expectation value of $\hat H_\alpha$. The sum of $\{\bar{H}_{\alpha,\mathcal P}\}$ is an unbiased estimate of $\langle \hat H \rangle_{\rm SV}$. \cite{Stanisic2022Observing} Here $\mathcal P^{(k)} \in \{0, 1\}$ behaves as an indicator function. As a consequence, this method is equivalent to simultaneously measuring the fragments and the symmetry projection operator, and discarding the measurements for which the symmetry projection is zero, implying the following formula for the estimator
\begin{equation}
    \bar{H}_{\alpha, \mathcal{P}} = \frac{1}{N_1} \sum_{k : \mathcal{P}^{(k)} = 1} H_{\alpha}^{(k)},
\end{equation}
where $N_1$ is the number of measurements with correct symmetry values. Since this approach to symmetry verification selects measurements that respect symmetry relations, it is commonly referred to as post-selection - see \fig{fig:symmetry_verification} for an illustration. 

There are two fragmentation methods where post-selection is possible: FC fragments and fermionic CSA based fragments. In FC fragments, we can choose Clifford transformations $\{\hat U_\alpha\}$ to diagonalize the symmetry operators expressed as Pauli products. \cite{yenMeasuringAllCompatible2020} The parity operator, $(-1)^{\hat N_e}$ can be written as $\prod_{i=1}^N \hat z_i$ in the Jordan-Wigner map. In fermionic CSA approaches, the orbital rotations $\{\hat U_\alpha\}$ in Eq. \eqref{eq:orbitaltransform} commute with $\hat N_e$. Since under the Jordan-Wigner transformation, $\hat N_e \rightarrow \sum_{i=1}^N (1 - \hat z_i)/2$ is diagonal, the expectation value of $\hat N_e$ can be estimated along with the expectation value of the fragments $\{\hat H_\alpha\}$. Post-selection has been employed in hardware experiments and have shown more than $\times 2$ improvements in the precision.\cite{Stanisic2022Observing} 

SV of symmetries that can be presented as a single Pauli product can also be done by measuring their eigenvalue using the Hadamard test circuit. Based on the measurement result of the ancilla, we retain or discard samples. This is identical to the standard stabilizer check circuit found in quantum error correction literature. \cite{Nielsen_Chuang_2010, xiong2022circuit} While this method has shown improvements in precision, introducing extra ancilla and controlled gates will lead to larger errors in estimates. 

Note that projection to the symmetry subspace does not remove undetectable parts of the errors that keep the state within the symmetry subspace. Hence, in the presence of noise, $\langle \hat H\rangle_{\rm SV}$ is not an unbiased estimator of the error free expectation value of $\hat H$. 

To reduce this remaining bias, \citet{cai2021expansion} introduced a generalization of SV known as symmetry expansion, where the projector in Eq. \eqref{symmetry_projected_est} is replaced by a weighted sum of the elements of the symmetry group, $\sum_{\hat S \in \mathbb S}w_{S}\hat S/\sum_{\hat S \in \mathbb S}w_S$. The weights $\{w_S\}$ are chosen such that some of the previously detectable errors are now undetectable, but the bias due to these errors is of the opposite sign of the previously undetectable errors, leading to partial cancellation of bias. 

\subsubsection{Error Mitigation by Training}

\begin{figure}
    \centering
    \includegraphics[width=\columnwidth]{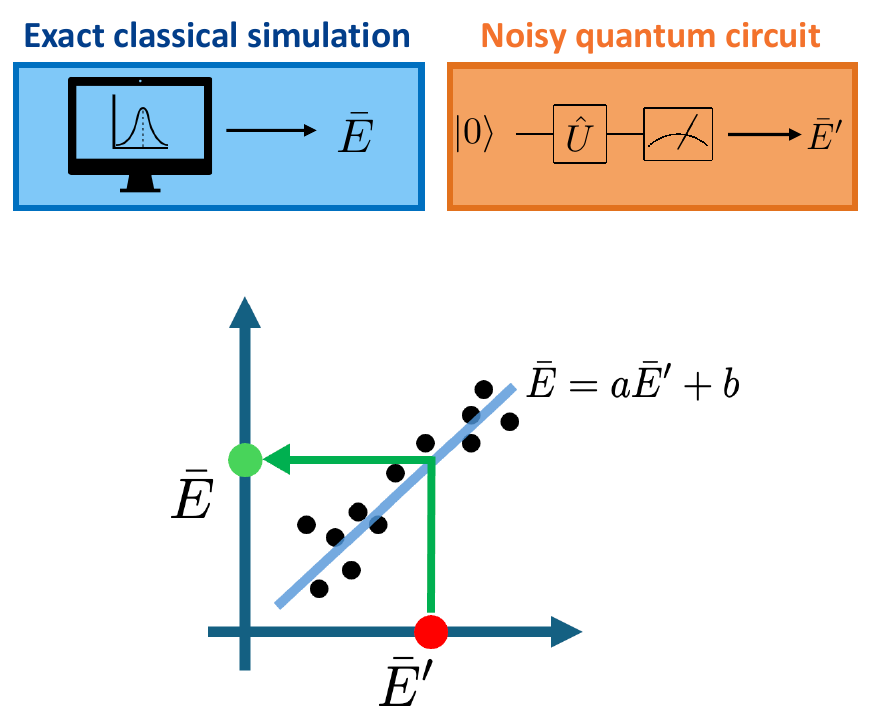}
    \caption{Diagrammatic representation of error mitigation by training. Pairs of noisy and noise-free estimates $\{\bar E', \bar E\}$ are obtained from noisy quantum simulations and classical simulations respectively. These estimates are then used to train a regression model, which in turn is used to predict error-free estimates $\bar E$ (green dot) from noisy estimates $\bar E'$ (red dot).}
    \label{fig:error_mitigation_training}
\end{figure}

In error mitigation by training (EMT), regression models are trained to predict error-free estimates from erroneous estimates of expectation values. Various models have been tested, \cite{liao2023machine} but often a simple linear function, $\bar E= a\bar E' + b$ relating the erroneous estimate $\bar E'$ to the error free expectation value $\bar E$ suffice. The model is trained on pairs of error-free and erroneous expectation values obtained from classical simulations and experiments, respectively. Figure \ref{fig:error_mitigation_training} illustrates this procedure. This approach requires that the expectation values in the training dataset be classically simulable. For effective performance, the training data points should also be close to the target expectation value.

One can predict the noise-free estimate of the expectation value $\braket{\psi|\hat H_\alpha |\psi}$ of a diagonalizable fragment $\hat H_\alpha = \hat U_\alpha^\dagger p(\hat z_1,\dots, \hat z_N) \hat U_\alpha$ on a classical computer. Recognizing that Clifford circuits are classically simulable, \citet{Czarnik2021errormitigation} proposed generating a training set of states $\{\ket{\phi_i}\}$ using unitaries primarily composed of Clifford gates and a fixed number 
of non-Clifford gates. These states approximate $\ket{\hat U_\alpha \psi}$ with respect to a chosen distance metric. The expectation values $\{\braket{\phi_i |p(\hat z_1,\dots, \hat z_N)|\phi_i}\}$ and their corresponding noisy estimates are then used to train a linear regression model. This method, known as Clifford data regression (CDR), works effectively when the circuit contains a small number of non-Clifford gates. Implementations of CDR have demonstrated one to two orders of magnitude reduction in errors. CDR is particularly well-suited for estimation with QWC and FC fragments, as their diagonalization circuits consist of Clifford gates, avoiding the introduction of additional non-Clifford gates.

For parameterized circuits, training data can also be generated within the vicinity of the target expectation value by varying the circuit parameters. This idea was used in Ref. \cite{montanaro2021tflo} to train regression models for predicting expectation values of fermionic operators. In this approach, the state preparation circuit is approximated as a product of fermionic linear optics (FLO) gates. FLO gates have the form $\exp(i\hat h)$, where Hermitian generator $\hat h$ is defined as $\hat h = \sum_{ij} h_{ij}\hat a_i^\dagger \hat a_j + \mu_{ij}\hat a_i^\dagger \hat a_j^\dagger - \mu_{ij}^* \hat a_i\hat a_j$ with parameters $\{h_{ij}, \mu_{ij}\}$. Circuits composed of FLO gates are classically simulable, enabling efficient computation of expectation values for fermionic observables. This variant of EMT, known as training with fermionic linear optics (TFLO), has demonstrated up to $\times 1.5\sim 2$ reduction in errors and has been validated in experiments. \cite{Stanisic2022Observing} Fermionic CSA fragmentation is particularly well suited for TFLO, as the orbital rotations used to diagonalize fragments are specific instances of FLO gates. Additionally, TFLO can be extended to other parameterized, classically simulable circuit classes, such as fermionic Gaussian unitaries (FGU).

\citet{Lolur2023REM} suggested yet another alternative for a training model, they shifted the noisy estimate of the Hamiltonian obtained using a VQE ansatz at target parameter values by the error in the noisy estimate obtained at a classically simulable state such as Hartree Fock state. This method approximates the noise induced bias in the estimate to be the same across the circuit parameter values. Despite it's simplicity, most of the mitigated estimates demonstrated errors in the order of chemical accuracy (1 kcal/mol).  

 \subsubsection{Noise Twirling}

Noise can be considered using quantum channel theory. This theory provides conditions when the channel can be inverted. The inversion of the noise channel would be a way to mitigate the associated error. Unfortunately, bare noise channels are not invertible, however, one can invert noise channels partially if they will be modified by twirling. This is similar to the method of classical shadows described in Section \ref{classical_shadow_subsec}, where twirling is used to create an invertible channel from the measurement channel $\mathcal M$.  

Using the super-operator notation defined in Appendix \ref{app:group}, if the noise channel is denoted by $\Lambda$,  
then the noisy implementation of unitary $ \mathcal V$ can be expressed as $\mathcal V' = \Lambda \circ \mathcal V$.

To twirl the noise $\Lambda$ in $\mathcal V'$ one randomly chooses $\mathcal U\in \mathcal G$ from the twirling group $\mathcal G$ and applies the unitaries $\mathcal V^\dagger\circ \mathcal U \circ \mathcal V$ and $\mathcal U^\dagger$ before and after $\mathcal V'$, respectively. This results in
 \begin{equation}
     \mathcal V'(\hat \rho) \rightarrow \mathbb E_{\mathcal U  \in \mathcal G} \left[ \mathcal U^\dagger \circ \Lambda \circ \mathcal U\circ \mathcal V (\hat \rho)\right] = \Lambda_{\mathcal G} \circ \mathcal V (\hat \rho)
 \end{equation}
 The unitary channel $\mathcal V$ remains unaffected and the twirled noise has a simpler form: $\Lambda_{\mathcal G}$ is expressed as a linear sum of channels:
 \begin{equation}
     \Lambda_{\mathcal G}(\hat \rho) = \sum_\lambda c_\lambda \mathcal P_{\mathcal G}^\lambda(\hat \rho).
 \end{equation}
 where coefficients $\{c_\lambda\}$ depend on the original noise channel, while channels $\{\mathcal P_{\mathcal G}^\lambda\}$ depend only on the twirling group $\mathcal G$. Consequently, the number of parameters needed to describe the twirled noise channel is typically much smaller than $O(8^N)$ parameters required to describe a general noise channel. For instance, in randomized benchmarking, device noise strength is characterized by applying random Clifford unitaries and their inverses to the initial state $\ket{0}^{\otimes N}$, followed by measuring the probability of the state returning to $\ket{0}^{\otimes N}$. Here, the Clifford group $\mathcal G = Cl(N)$ twirls the noise, reducing the channel to a global depolarizing noise channel described by a single parameter (see Appendix \ref{app:globalcliffords}). The depolarizing strength is related to the decay rate, dependent on the device and the circuit depth. By repeating the experiment for several circuit depths, the decay rates characteristic of the device can be extracted.\cite{Helsen2022RB}



 Moreover, twirling the measurement channel can be used to twirl the noise in unitary gates. For a sequence of noisy operations, $\mathcal V' = \Lambda^{(1)}\circ \mathcal V^{(1)}\circ \Lambda^{(2)}\circ \mathcal V^{(2)} \circ \cdots$, the noise channels can be aggregated into an effective noise channel $\Lambda_{\mathrm{eff}}$ towards the end of the circuit, such that
 \begin{equation}
     \mathcal V'\equiv \Lambda_{\mathrm{eff}}\circ \mathcal V^{(1)}\circ \mathcal V^{(2)}\circ \cdots
 \end{equation}
 When twirling is applied to the measurement channel $\mathcal M$, as in classical shadows, it effectively twirls the noisy measurement channel $\mathcal M' = \mathcal M \circ \Lambda_{\mathrm{eff}}$. This modifies the coefficients of the shadows used in estimating the expectation value of $\hat H$ as  $\Tr(\hat H\cdot \left[\mathcal M_{\mathcal G}'\right]^{-1} \circ \mathcal U^\dagger(\ket{\vec z}\bra{\vec z}))$. By constructing estimators to obtain these coefficients, \citet{Chen_robust_2021} proposed noise-resilient \textit{robust} classical shadows that managed to almost completely eliminate the effect of coherent, gate independent noises. Reference \cite{Zhao2024group} generalized these estimators 
in terms of noisy estimates and ideal expectation values of known symmetry operators of the state to construct a symmetry adapted error mitigated classical shadows scheme.


\section{Heisenberg Scaling Methods}
\label{Sec:Heisenberg}

Here we discuss measurement approaches that take their origins from the QPE method. 
The textbook QPE algorithm \cite{Nielsen_Chuang_2010} is an implementation of the von Neumann measurement on a quantum computer. 
To see this, recall that in the von Neumann measurement 
one starts with a tensor product state 
\bea
\ket{q_0}\ket{\Phi} &=& \ket{q_0}\sum_n c_n \ket{\Psi_n}, 
\eea
where $\ket{q_0}$ is a localized pointer state at position $q_0$ and $\ket{\Phi}$ is 
a linear superposition of eigenstates of the system Hamiltonian, 
$\hat H \ket{\Psi_n} = E_n \ket{\Psi_n}$. 
One can construct $\hat U = \exp{[it\hat p\otimes\hat H]}$ 
where $\hat p$ is the pointer momentum operator that 
acts on $\ket{q_0}$ as a shift when exponentiated, $\exp[i\Delta\hat p]\ket{q_0} = \ket{q_0+\Delta}$. 
When $\hat U$ is applied to the initial tensor product state, 
\bea
\hat U \ket{q_0}\ket{\Phi} = \sum_n c_n \ket{q_0+t E_n}\ket{\Psi_n}, 
\eea
one obtains a superposition where $\braket{q_0+t E_m|q_0+t E_n} = \delta_{nm}$ if $t$ is large enough. 
Measuring the pointer state will collapse the entire state to 
$\ket{q_0+t E_n}\ket{\Psi_n}$ with probability $|c_n|^2$. To recognize that the QPE algorithm follows the same scheme, one needs to do two extra steps. 
First, one substitutes the continuous 
spectrum of the pointer states by the discrete spectrum of qubit states. In QPE, the pointer states are product states of ancilla qubits, $\ket{\bar{0}}$. 
Second, due to the discretization of the pointer states, QPE  
substitutes the momentum operator by its Fourier transformed version 
\bea
e^{i\Delta \hat p} f(x) &=& 
\hat U_{FT}^\dagger \left[ \hat U_{FT} e^{ i\Delta \hat p} \hat U_{FT}^\dagger\right] \hat U_{FT} f(x) \\
\hat U_{FT} e^{ i\Delta \hat p} \hat U_{FT}^\dagger &=& e^{2\pi i y \Delta} \\
\hat U_{FT} f(x) &=& \int dx e^{2\pi i yx } f(x) = F(y).
\eea
The advantage of the Fourier transformed version of the momentum operator, $e^{2\pi i y \Delta}$, is that 
it does not require differentiation and is easier to implement as gates via controlled $\exp[it\hat H]$ operations. 
Finally, it is worth mentioning that the 
first set of Hadamard gates applied to the ancilla product state $\ket{\bar{0}}$ 
in QPE is the Fourier transformation. Thus,
the textbook QPE algorithm is a discretized version of the von 
Neumann measurement, where one directly measures the eigenvalues 
of $\hat H$ encoded in ancilla qubit pointer states. 
Assuming a good initial state overlap with the state of interest, $|\bra{\Phi}\Psi_n\rangle|$ one can get the value of the quantity of interest in $O(1/|\bra{\Phi}\Psi_n\rangle|)$ number of measurements.   

Inspired by the efficiency of the measurement part of the QPE algorithm, one can approach the measurement of the expectation value of operator 
$\hat O$ as a phase estimation. The idea is to use apply QPE to a unitary $\hat{U}$ whose phase is related to $\bra{\psi}\hat O\ket{\psi}$. The downside of the QPE method is a deep circuit that scales as 1/$\epsilon$ with  required accuracy $\epsilon$ on the phase value. Thus, the main direction in recent developments is to reduce the circuit depth by trading circuit depth to the number of measurements and to leverage possible simplifications in the structure of $\hat O$, since there is some freedom in its choice. 
This idea gave rise to several methods that are commonly referred to as 
1/$\epsilon$ methods or Heisenberg scaling methods. The origin of this name 
is related to the QPE scaling of circuit depth as 1/$\epsilon$ with required accuracy.
Note that formally, if we are given an eigenstate of the Hamiltonian we will only need 
to measure its energy once using QPE, so 1/$\epsilon$ scaling is not for the number of 
measurements but rather the number of times to call controlled $\hat{U}$ unitary.

There are two questions to address: 1) how to build a unitary operator whose eigenvalues $e^{i\phi}$ are related to the expectation 
value of interest $\bra{\psi}\hat O\ket{\psi}=f(\phi)$; 
2) how to estimate the phase $\phi$ robustly with minimal quantum resources 
(e.g., fewer qubits and shorter circuits) and in the presence of noise.

\subsection{Amplitude Amplification and Estimation}

Many 1/$\epsilon$ methods use the basic idea of the Amplitude Amplification (AA) and Amplitude Estimation (AE) algorithms. These ideas are also key elements of the Grover search algorithm \cite{Nielsen_Chuang_2010,Brassard_2002}. 
Let us review them here.

Any initial state $\ket{\Psi}$ can be split to a normalized component of interest $\ket{\phi}$ and its orthonormal counterpart $\ket{\phi^{\perp}}$
\bea\label{eq:1st_p}
\ket{\Psi} = a\ket{\phi} +\sqrt{1-a^2}\ket{\phi^{\perp}},
\eea
where $a$ is a coefficient. The AA algorithm aims to increase 
$|a|$ while the AE counterpart is used to estimate $a$. 

The unitary transformation that consists of two 
non-commuting reflections, $\hat W = -\hat R_{\Psi}\hat R_{\phi}$, 
$\hat R_{\Psi} = 1-2\ket{\Psi}\bra{\Psi}$ and $\hat R_{\phi} = 1-2\ket{\phi}\bra{\phi}$, 
can both amplify the $\ket{\phi}$ amplitude (AA) and be used to determine its value (AE). To see this, it is easy to check that the linear space spanned by
$\{\ket{\phi},\ket{\phi^{\perp}}\}$ forms a two-dimensional invariant subspace for $\hat W$. 
Thus, one can obtain eigenvalues of $\hat W$ from the matrix
\bea
\mathbf{W} &=& 
\begin{pmatrix}
    \bra{\phi}W\ket{\phi} & \bra{\phi}W\ket{\phi^\perp} \\
    \bra{\phi^\perp}W\ket{\phi} & \bra{\phi^\perp}W\ket{\phi^\perp}
\end{pmatrix} \\
&=&
\begin{pmatrix}
    \cos(\theta) & -\sin(\theta) \\ 
    \sin(\theta) & \cos(\theta)
\end{pmatrix}, \label{eq:Wm}
\eea
where to obtain matrix elements we used a parametrization for 
\bea
\ket{\Psi} = \sin(\theta/2)\ket{\phi} +\cos(\theta/2)\ket{\phi^{\perp}}
\eea
with an angle $\theta$ so that $a= \sin(\theta/2)$ in \eq{eq:1st_p}.
The action of $\hat W$ can be seen as the rotation by angle $\theta$ in the plane of 
$\ket{\phi}$ and $\ket{\phi^{\perp}}$, hence,
\bea\notag
\hat W^k \ket{\Psi} = \sin[\theta(2k+1)/2]\ket{\phi} 
+\cos[\theta(2k+1)/2]\ket{\phi^{\perp}}.
\eea
Clearly this leads to amplification of small initial $\sin(\theta/2)$ coefficients, unless $\theta$ is approaching $\pi/2$. 
Also, using QPE for $W$ operator instead of $e^{i\hat Ht}$ one can obtain its 
eigenvalues $e^{\pm i\theta}$ and phase $\theta$, 
which leads to estimate of the amplitude of $\ket{\phi}$. 



\subsection{Connecting Expectation Values with Amplitude Estimation}

As we have seen in Sec.~\ref{hadamard_test_lcu_section}, the Hamiltonian expectation value can be presented as a 
linear combination of expectation values for unitary operators that come from the 
LCU decomposition of the Hamiltonian. Here, we will show how 
the AE algorithm can be used to obtain the expectation values of unitary operators.
\cite{Knill:PRA2007,aQPE,RAE_PRX21}  
First, we will illustrate this on the simplest case of a Pauli product, $\hat P$. 
The Pauli products are not only unitary but also hermitian, which simplifies their 
eigen-decomposition
\bea
\hat P &=& \hat{\mathcal{P}}_{+} - \hat{\mathcal{P}}_{-} \\
\hat{\mathcal{P}}_{\pm} &=& \sum_i\ket{i_\pm}\bra{i_\pm},
\eea
where $\ket{i_{\pm}}$ are eigenstates, and $\hat{\mathcal{P}}_{\pm}$ are eigen-projectors on eigenspaces corresponding to $\pm 1$ eigenvalues. Since in what follows we will not use the trace-zero property of $\hat P$, this consideration can be used for any hermitian unitary operator assuming that there is a known circuit to implement such an operator.  

Any wavefunction $\ket{\Psi}$ can be written as 
\bea
\ket{\Psi} &=& \sin(\theta/2)\ket{\Psi_-} + \cos(\theta/2)\ket{\Psi_+},
\eea
where $\ket{\Psi_{\pm}} = \hat{\mathcal{P}}\ket{\Psi}/||\hat{\mathcal{P}}\ket{\Psi}||$.
We can establish the connection between the expectation value and angle $\theta$, $\bra{\Psi}\hat P\ket{\Psi} = 
\cos^2(\theta/2)-\sin^2(\theta/2) = \cos(\theta)$. 

To make use of the AE algorithm let us introduce the walker operator, 
$\hat W = \hat R_{\Psi} \hat P$, that has a 2D eigen-subspace spanned by $\{\ket{\Psi_\pm}\}$ or 
$\{\ket{\Psi},\hat P\ket{\Psi}\}$. In the basis of $\{\ket{\Psi_\pm}\}$ 
$\hat W$ has the matrix given by \eq{eq:Wm} and 
eigenvalues $e^{\pm i\theta}$.  
Thus using QPE one can extract $\theta$, which is connected to 
$\bra{\Psi}\hat P\ket{\Psi}$.

Second, a general unitary was initially considered in Ref.~\cite{Knill:PRA2007}. To connect $\bra{\Psi}\hat U\ket{\Psi}$ with the AE algorithm one can build 
the walker operator using, $\hat R_{\Psi}$ and 
$\hat R_{U\Psi} = 1 - 2 \hat U\ket{\Psi}\bra{\Psi}\hat U^\dagger$. It is easy to check 
that $\hat W = \hat R_{\Psi} \hat R_{U\Psi}$ has a 2D eigen-subspace spanned by 
$\{\ket{\Psi},\hat U\ket{\Psi}\}$. For convenience of consideration of the regular 
eigenvalue problem for $\hat W$ rather than a generalized one, one needs to orthogonalize the basis vectors of the 2D eigen-subspace to $\{\ket{\Psi},\ket{\Psi^\perp}\}$, where 
\bea
\ket{\Psi^\perp} = \frac{1}{\sqrt{1-|\langle \hat U\rangle|^2 }}\left[\hat U\ket{\Psi} -  \langle \hat U\rangle \ket{\Psi}\right].
\eea
Then the matrix of $\hat W$ represented in the $\{\ket{\Psi},\ket{\Psi^\perp}\}$ space
is 
\bea
\mathbf{W} = \begin{pmatrix}
    2|\bar{U}|^2-1 & 2\bar{U}\sqrt{1-|\bar{U}|^2} \\ 
    -2\bar{U}^*\sqrt{1-|\bar{U}|^2} & 2|\bar{U}|^2-1
\end{pmatrix}, \label{eq:Wm2}
\eea
where $\bar{U} = \bra{\Psi} \hat U \ket{\Psi}$, and 
the eigenvalues of $\mathbf{W}$ are $e^{\pm i\theta}$ with $\cos(\theta) = 2|\bar{U}|^2-1$. Thus, this case is more complicated than the  Pauli product expectation value, since $\theta$ only gives access to $|\bar{U}|$. To obtain the phase of $\bar{U}$ one can use the same procedure but with the state with extra qubit $\ket{\Psi, +_a}$ 
for two modified unitaries: $\hat U_1 = c-\hat U$ and $\hat U_2 = e^{i\hat z_a \pi/4} \hat U_1$.\cite{Knill:PRA2007} These unitaries deliver the rest of the information for determining $\bar{U}$
\bea
|\bra{\Psi, +_a} \hat U_1 \ket{\Psi, +_a}| &=& |1+\bar{U}|/2 \\
|\bra{\Psi, +_a} \hat U_2 \ket{\Psi, +_a}| &=& |1-i\bar{U}|/2.
\eea
In cases when it is known that $\bar{U}$ is real one can use only $\hat U_1$ procedure.\cite{O_Brien_2022}

For an arbitrary operator $\hat O$ in addition to an LCU decomposition and 
separate treatment of each unitary components, 
one can do $\hat O$ block-encoding as a part of a 
larger unitary \cite{O_Brien_2022,molobs2023ft}
\bea
\hat U_L &=& \begin{pmatrix}
    \hat O / c & \hat U_{01} \\ 
    \hat U_{10} & \hat U_{11},
\end{pmatrix}
\eea
where $c$ is a constant to maintain the unitarity of $\hat U_L$ and 
$\hat U_{ij}$ are other blocks. Building the walker operator using 
$R_{\Psi} = 1- 2\ket{\Psi,0}\bra{\Psi,0}$ and $R_{U\Psi} = 1- 2\hat U_L\ket{\Psi,0}\bra{\Psi,0}\hat U_L^\dagger$ and estimating its phase can be 
used to obtain $|\bra{\Psi}\hat O\ket{\Psi}|$. The controlled version of 
$\hat U_L$ can deliver the sign of the expectation value as was discussed before. 

Yet another consideration of a general Hermitian operator $\hat O$ in Ref.~\cite{Knill:PRA2007} was to obtain its expectation value via measuring that of $\hat U(t) = e^{i\hat O t}$. Then 
for small $t$,  
using the $\langle \hat U(t)\rangle \approx 1+ it \langle \hat O\rangle$ approximation one can extract $\langle \hat O\rangle$. 

\subsection{Reducing Quantum Resources for QPE}

Reduction of the number of qubits and circuit depth of the QPE algorithm 
is an area of active recent research under the name of early fault-tolerant (EFT)
algorithms\cite{EFTrev}. The EFT methods are involving QPE framework for estimating ground state energy and properties. 
In our consideration, they are applied to a walker operator that is usually simpler than that obtained for 
the energy estimation (e.g., in the qubitization algorithm). Review \cite{EFTrev}
considers multiple EFT approaches, thus here we will not cover them in extensive details, instead we would like 
to provide basic ideas for common strategies of saving quantum resources. 

The common circuit for frugal QPE is the Hadamard test (see \fig{hadamard_circuit_fig}) where $\hat U$ is the walker operator. It was suggested as a circuit for iterative 
QPE by Kitaev \cite{Kitaev_book} requiring $\ket{\Psi}$ to be one of the 
eigenstates of the $\hat U = e^{i\hat H t}$ operator. 
If one considers the 
probability of getting $d=\pm 1$ in the measurement after 
applying $\hat U^k$ and the phase rotation $S(\phi)$, 
\bea
\Pr(d|k,\phi) = \frac{1}{2} [1+d \cos(k\theta+\phi)], 
\eea
it gives an illustration how deeper circuit (larger $k$) gives access 
to higher digits of $\theta$ and thus higher accuracy. On the other 
hand, considering usual limitations in the circuit depth one can 
obtain the accuracy by increasing the number of times $d$ is measured 
to have a better estimate of $\Pr(d|k,\phi)$ to deduce $\theta$ from lower 
$k$ circuits. 

In the usual scenarios when $\ket{\Psi}$ is not an eigenstate of the unitary operator,
$\text{Pr}(d|k,\phi)$ is comprised of signals from multiple eigenvalues. To extract 
eigenvalues in this case, one needs to perform more elaborate classical post processing. 
Multiple works were devoted to developing optimal post-processing techniques 
based on Bayesian analysis\cite{WiebeBPE,aQPE}, Fourier transform with various filters \cite{Kshirsagar_2024,LinPRXQ22,Wang_2023}, compressed sensing\cite{QPE_CS24}, 
and non-linear fitting\cite{LinPRXQ23}. 






\subsection{Robust Amplitude Estimation}

\begin{figure}
    \centering
    \includegraphics[width=0.9\columnwidth]{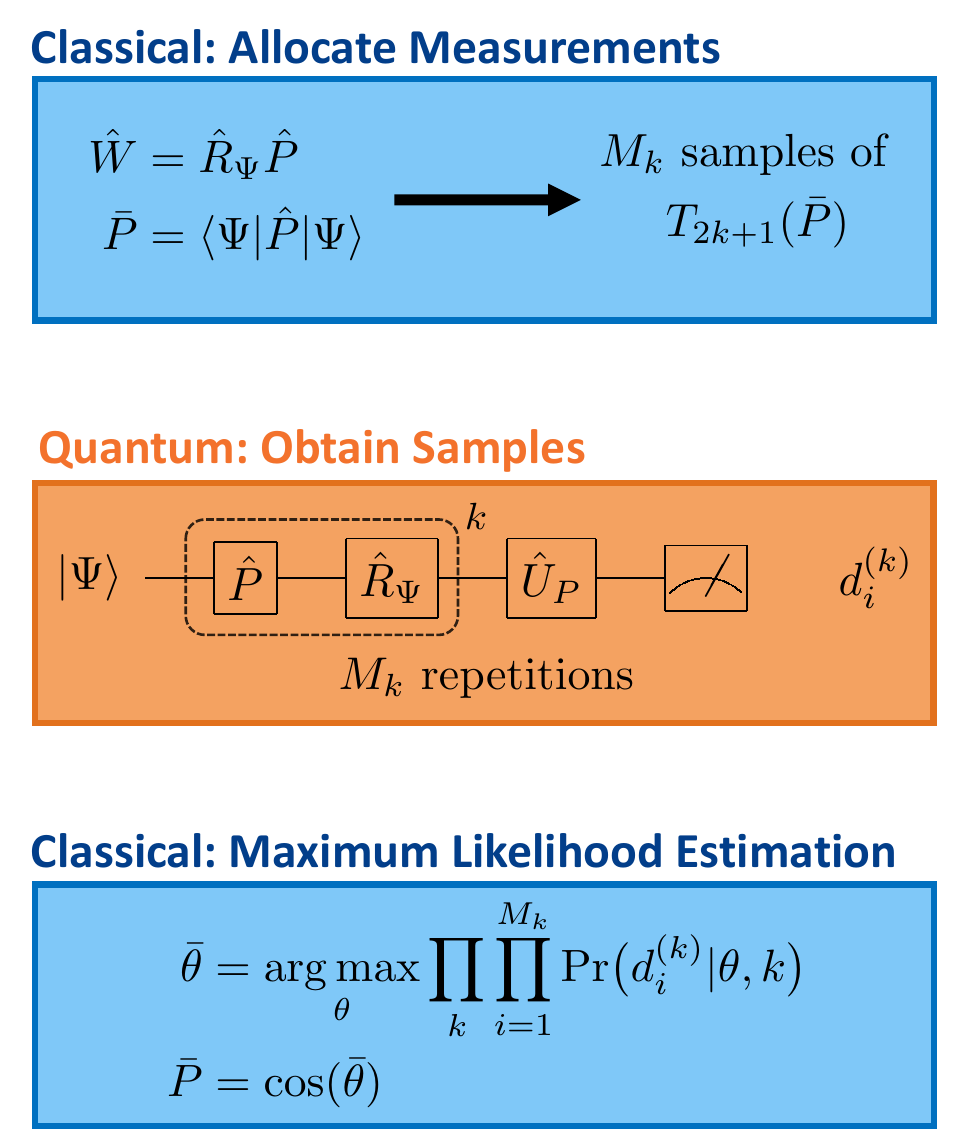}
    \caption{Diagrammatic representation of robust amplitude estimation. In the first step, $M_k$ measurements are allocated to estimate the Chebyshev polynomials $T_{2k+1}(\bar P) = \braket{\Psi|(\hat{W}^k)^\dagger \hat{P} \hat{W}^k|\Psi}$, where $\hat{W} = \hat{R}_{\Psi}\hat{P}$ is the walker operator. In the second step, the measurements are obtained. Here, $\hat{U}_P$ is a unitary that maps $\hat{P}$ to Ising form. The probability of observing $d_i^{(k)} \in \{-1,1\}$ is given by $\mathrm{Pr}(d_i^{(k)}|\theta, k)$ in Eq. \eqref{rae_prob}.  Lastly, maximum likelihood estimation is used to obtain a value for $\bar P$.}
    \label{fig:QPE_RAE}
\end{figure}

One of the successful implementations of ideas of flexible reduction
of the circuit depth from the full QPE size and still 
taking advantage of higher accuracy in estimation is the 
Robust Amplitude Estimation (RAE)\cite{RAE_PRX21,RAE22}. 
To reduce the circuit 
depth the full QPE is substituted with a few steps of a walker operator. 

To obtain $\bar{P} = \bra{\Psi} \hat P \ket{\Psi}$, the RAE scheme, summarized in \fig{fig:QPE_RAE}, applies to $\ket{\Psi}$ the walker operator, $\hat W = \hat R_{\Psi} \hat P$, where 
$\hat R_{\Psi} = 1-2\ket{\Psi}\bra{\Psi}$, several 
times and then measures $\hat P$. The expectation value of $\hat P$ for state $\ket{k} = \hat W^k \ket{\Psi}$ is related to that of the original state $\ket{\Psi}$:
\bea
\bra{k} \hat P \ket{k} &=& T_{2k+1}(\bar{P}),
\eea
where $T_{2k+1}$ is Chebyshev's polynomial of the first kind.
For example, 
\bea
\ket{k=1} &=& (\hat P - 2\bar{P})\ket{\Psi} \\ \notag
\bra{k=1} \hat P \ket{k=1} &=& 4 \bar{P}^3 - 3\bar{P} \\ 
&=& T_{3} (\bar{P}).
\eea
Here one measures $T_{3} (\bar{P})$ and extracts the value of 
$\bar{P}$ via classical post processing. 
To understand why RAE gives higher accuracy in the estimation of $\bar{P}$ via measuring $T_{2k+1} (\bar{P})$
it is instructive to do an error propagation analysis. If random 
variables $x$ and $y$ are related by $y = f(x)$, 
then measuring $y$ with corresponding variance $\mathrm{Var}(y)$ can be used to estimate $x$ with the variance given by 
$\mathrm{Var}(x) =\mathrm{Var}(y) [d f^{-1}(y)/dy]^2$.
Applying this formula to $x = \bar{P}$ and 
$y = T_{2k+1} (\bar{P}) = \cos[(2k+1)\arccos(\bar{P})]$ 
leads to the variance of the $k$-times estimator, 
$\mathrm{Var}_k$, to be 
\bea \label{eq:MSE_L}
\mathrm{Var}_k(\bar{P}) = \frac{\mathrm{Var}_0(\bar{P})}{(2k+1)^2},
\eea
where $\mathrm{Var}_0(\bar{P}) = 1-\bar{P}^2$.
Therefore, $k$ allows one to reduce the variance. 
The only issue with this consideration is that $\arccos$ function is multivalued, and thus the 
variance reduction can only be attained if one knows additionally which branch of the $\arccos$ function is relevant. The latter information 
can be obtained from a few preliminary regular ($k=0$) measurements of $\bar{P}$ (see Refs.~\cite{Suzuki_2020,kunitsa2024RAE} for details of ``layer scheduling"). In practice, the estimate is deduced by maximizing the likelihood of the observed measurements at various $k$.

In the original paper\cite{RAE_PRX21}, 
the analysis is done using Fisher information $I_F$ instead of variances,
\bea
I_F = \sum_{d} \frac{1}{\Pr(d|\bar{P})} \left( \frac{\partial}{\partial \bar{P}} \Pr(d|\bar{P})\right)^2, 
\eea
where $\Pr(d|\bar{P})$ is the likelihood function of getting measurement results $d=\pm 1$ assuming the mean of the distribution 
$\bar{P}$.
One advantage of considering $I_F$ is an ease of including various noise models.  A disadvantage of working with $I_F$ is that it gives a lower bound for estimator variance instead of true variance according to the Cramer-Rao bound $\mathrm{Var}(\bar{P}) \ge 1/I_F$. 
Defining as $\bar{P} = \cos(\theta)$ 
\bea
\Pr(d|\theta,k) = \frac{1}{2}(1+d\cos[(2k+1)\theta]).\label{rae_prob}
\eea
This leads to \eq{eq:MSE_L}, if we assume the saturation of the Cramer-Rao bound in the ideal noiseless case.

\section{Conclusions}

In this review, we have surveyed quantum measurement techniques for the quantum simulation of electronic structure. Central to this topic is the construction of additional measurement unitaries $\hat{U}$ enabling the extraction of information from the quantum computer, which is then processed classically to estimate the desired quantity. There are three aspects of the measurement process we explored in this review: (1) the classical algorithms for determining the unitaries $\hat{U}$; (2) the quantum circuits for implementing $\hat{U}$ on the quantum computer; and (3) the number of measurements required to achieve a desired accuracy. Selecting the optimal measurement algorithm for a given problem requires identifying and addressing the primary computational bottlenecks among these factors, while taking into account any potential overhead due to noise.

One can exploit algebraic properties of fermionic or qubit operators to partition the target Hamiltonian to measurable fragments. Originally, it was proposed to group together commuting, or qubit-wise commuting, Pauli products, which can be simultaneously measured. It was then realized that non-commuting fermionic or Pauli operators can also be measured together, provided that they can be easily transformed to a group of commuting Pauli operators by unitary transformations. The choice of fragment type influences the quantum circuit cost, and optimization of the fragments to minimize measurement cost can be done via greedy optimization algorithms, or using estimates of covariances between terms. The determination of the optimal fragments and corresponding measurement unitaries $\hat{U}$ is a challenging pre-processing task for a classical computer, and there are some open questions left. For example, in VQE, the quantum state is often expressed in terms of generators of unitary transformations and their amplitudes; one can ask whether this information can be used efficiently to optimize the Hamiltonian partitioning.

An alternative to Hamiltonian partitioning is classical shadow tomography. It can be interpreted as learning the expectation value of the Hamiltonian using various unitaries that transform different fragments of the Hamiltonian to Ising form. Classical shadow tomography is particularly useful in estimating the expectation values of multiple operators, and therefore has applications in estimating reduced density matrices, and in quantum subspace methods. For reduction of the measurement and circuit costs, one has the freedom to use different informationally-complete groups of unitaries, as well as the probability distribution to sample from. As shown in this review, when these optimization techniques are applied, classical shadow tomography becomes conceptually similar to grouping techniques. 

We also considered measurement techniques which enlarge the qubit space beyond what is required to encode the Hamiltonian and quantum state. One of the simplest approaches to estimate expectation values using an extended space is the Hadamard test. It can be used to estimate the expectation value of any unitary $\hat{U}$ whose quantum circuit decomposition is known, without having to diagonalize it. However, while the Hadamard test can be used in principle for estimating the expectation value of any operator, given a decomposition into a linear combination of unitaries, it suffers from large measurement and quantum circuit costs to implement in practice. However, using additional qubits naturally leads to the usage of informationally-complete POVMs, which allow for the measurement of the entire Hamiltonian in one shot using relatively shallow depth measurement circuits. Here, there is a classical overhead to determine the optimal measurement basis in the extended qubit space to minimize the measurement cost for the target Hamiltonian. POVMs are a very general framework for describing quantum measurements, and both Hamiltonian partitioning and classical shadow tomography can be described as special cases of expectation value estimation using POVMs. It is an interesting question whether this connection can lead to more efficient Hamiltonian partitionings.

The problem of measuring multiple operators and/or multiple states was also considered. Usually, the first step is to express all desired quantities in terms of expectation values, if that is not done already. For example, gradients of variational parameters in VQE can be written as the expectation value of a commutator, or as a linear combination of expectation values of the Hamiltonian via the parameter shift rule. Matrix elements can also be written in terms of expectation values of states in a larger Hilbert space. Measuring multiple operators leads to a more complex optimization problem, as one must optimize the measurement cost for all operators simultaneously. In cases where there is a large diversity of operators to be estimated, this can be computationally expensive. Nonetheless, classical shadow tomography can be straightforwardly applied in this case, and the various Hamiltonian partitioning techniques developed for estimating a single operator can be extended to the multiple operator case as well.

On current and near-term hardware, the presence of hardware noise introduces additional measurement overhead, while limiting the quantum circuit depths that are achievable in practice. Optimization of measurement circuits to minimize depth based on the target Hamiltonian and device specifications can reduce the effect of errors on the measurement process. Error mitigation techniques also can correct for the effect of noise, typically at the cost of additional measurements and classical post-processing.  The choice of measurement strategy also influences which error mitigation methods are viable. For example, when a measurable fragment shares symmetries with the quantum state, one can measure the fragment together with its symmetry operators and retain only the shots that correspond to the correct symmetry values. Nonetheless, current hardware constraints restrict practical applications predominantly to small molecular systems (e.g., H$_2$, LiH) in minimal basis sets. As such, error correction is needed so that measurement approaches based on additional unitary transformations can become scalable to larger systems of interest for industrial and fundamental research.  

Looking toward future error-corrected quantum computers, one can ask whether there will be a general switch to Heisenberg scaling methods from measurement methods which have a finite sampling error, scaling as $1/\epsilon^2$. The outlook seems positive, with the main issue being finding more efficient block-encodings of the measured operators. Despite this, one can imagine a somewhat early fault tolerant period \cite{EFTrev}, in which minimizing quantum circuit depth is more important than minimizing the measurement cost, in which $1/\epsilon^2$ methods may still be preferable. Another question is whether there will be a complete transition to using the QPE algorithm for electronic energy estimation. It has been pointed out that a weakness of the QPE algorithm is its reliance on being able to prepare an initial state with a good overlap with the target eigenstate. A problem for QPE when applied to quantum chemistry is that, due to the electronic Hamiltonian two-body nature, and for larger systems, it is easier to prepare a quantum state with better energy than that with a good overlap. Therefore, one can imagine new quantum algorithms coming to substitute QPE to avoid reliance on the overlap.  

\section*{Acknowledgements}
S.P. and A.F.I. acknowledge financial support from
the Natural Sciences and Engineering Council of Canada
(NSERC).

\appendix

\section{Group Theoretic Background} 
\label{app:group}

Here we provide quantum channel and group theory details that help to understand why the twirled channel $\mathcal M_{\mathcal G}$ is 
invertible with a suitable choice of group $\mathcal G$.  First, we introduce the Liouville super-operator formalism to represent quantum channels and simplify our expressions. Our notation and discussions will mainly follow those from \cite{Chen_robust_2021}.

\subsection{Liouville Representation of Quantum Channels}
Since quantum channels are linear on the space of density operators, we can re-express quantum channels as linear operators acting on the $d^2$ dimensional Hilbert space $\mathcal H_{d^2}$ of density operators, where, for $N$ qubits, $d = 2^N$. Denote a vectorized density $\hat \rho$ by $|\hat \rho\rangle\!\rangle$ and the super-operator corresponding to the channel $U$ as $\mathsf U$. 

\textit{Pauli product basis}: A suitable basis for $\mathcal H_{d^2}$ is the set of normalized, Hermitian Pauli operators $|\vec \sigma\rangle\!\rangle = |\hat \sigma_1\otimes \dots\otimes\hat \sigma_{N}\rangle\!\rangle, \hat \sigma_i \in \{\hat \sigma_0, \hat \sigma_x, \hat \sigma_y, \hat \sigma_z\}$ given by an $N$-fold tensor product of normalized Pauli matrices, $\hat \sigma_0 = \hat 1/\sqrt{2}, \hat \sigma_x = \hat x/\sqrt{2}, \hat \sigma_y = \hat y/\sqrt{2}, \hat \sigma_z = \hat z/\sqrt{2}$. Pauli product basis elements are orthonormal with respect to the Hilbert-Schmidt inner product, defined as $\langle \!\langle \vec\sigma_a | \vec\sigma_b\rangle\!\rangle = \Tr(\vec\sigma_a^\dagger\vec\sigma_b)$. The entries of a column vector representing the density is given by
\begin{equation}
[|\hat \rho\rangle\!\rangle]_a = \langle \!\langle \vec\sigma_a | \hat \rho\rangle\!\rangle =  \Tr(\vec\sigma_a^\dagger \hat \rho)
\end{equation}
and the entries of the matrix representation of the super-operator is given by
\begin{equation}
[\mathsf M]_{ab} = \langle \!\langle \vec\sigma_a | \mathcal M(\vec\sigma_b)\rangle\!\rangle =  \Tr(\vec\sigma_a^\dagger \mathcal M(\vec\sigma_b)).
\end{equation}
The matrix representation of the superoperator, in the basis of Pauli operators is also known as the Pauli transfer matrix (PTM).

\textit{Majorana product basis}: Product of Majorana operators, defined in Eq. \eqref{maj_from_ferm} form an alternative basis for $\mathcal H_{d^2}$, agnostic of fermion to qubit operator mapping. Define the k-degree Majorana product as a product of $k$ Majoranas, $\Gamma_{\vec \mu} = \prod_{i=1}^k \hat \gamma_{\mu_i}$. The set of vectorized Majorana products, $\{\ket{\Gamma_{\vec \mu}}\}$, forms a basis for $\mathcal H_{d^2}$. Since Majorana products are isomorphic to Pauli products, the matrix representation of a super-operator in the Majorana product basis is equivalent to that in the Pauli product basis, up to a row and column permutation.

\subsection{Quantum Channels and their Invertibility}
Quantum channels are linear maps between density matrices. Density matrices have a trace of $1$, and the eigenvalues of density matrices describe probabilities that are required to be non-negative. Thus, quantum channels are required to preserve trace and positivity of eigenvalues. Since a quantum channel $\phi$ may act on a subset of qubits of the target state, extended maps $I\otimes \phi$ with an identity map on the extended space are required to be positive. This property of quantum channels is known as complete positivity. Linear maps on densities that are completely positive and trace preserving (CPTP) are known as quantum channels, and are physically realizable by unitary evolutions and measurements.

A general quantum channel is not invertible. For invertibility, it is required that the superoperator representing the quantum channel to be non-singular. For instance, a non-invertible measurement channel has PTM $\mathsf M =  \mathrm{diag}[1, 0, 0, 1]^{\otimes N}$ which is singular. 
The inverse of a quantum channel need not necessarily be physical (CPTP) and can be implemented by stochastic methods\cite{temme2017znepec}. 
For example, the depolarization channel of depolarizing strength, $p$ has the form $\mathcal D(\hat \rho) = (1 - p)\hat \rho + p\hat \sigma_0^{\otimes N}$ with an inverse map $\mathcal D^{-1}(\hat \rho) = (1 - 2p)^{-1}\left[(1-p)\hat \rho - p\hat \sigma_0^{\otimes N}\right]$, which is not a positive map.

\subsection{Invertibility by Twirling}

Since $\ket{0}\bra{0} = (\hat 1 + \hat z)/\sqrt{2}$ and $\ket{1}\bra{1} = (\hat 1 - \hat z)/\sqrt{2}$, the super-operator representation of the measurement channel in the Pauli operator basis is
\begin{equation}
    \mathsf M  = (|\hat \sigma_0\rangle\!\rangle\langle\!\langle \hat \sigma_0| + |\hat \sigma_z\rangle\!\rangle\langle\!\langle \hat \sigma_z|)^{\otimes N}.
\end{equation}
Equation \eqref{eqn:twirl} is now re-expressed in the super-operator formalism as 
\begin{align}
\mathbb E_{U \in \mathcal G}\Big[\mathsf{U^\dagger MU}\Big]|\hat \rho\rangle\!\rangle\! = \mathsf{M}_{\mathcal G}|\hat \rho\rangle\!\rangle\!
\end{align}
where we denote the channel $\mathsf M$ twirled by the group $\mathcal G$ as $\mathsf M_{\mathcal G}$.
Since the unitary channels $\mathcal U\in \mathcal G$ form a group under composition, the super-operators $\mathsf{U}$ representing the unitary channel $\mathcal U$ form a multiplicative group. As the super-operators form a finite dimensional representation of the group $\mathcal G$, the Hilbert space $\mathcal H_{d^2}$ can be written as a direct sum of irreducible subspaces $\mathcal H_\lambda$. Irreducible subspaces are vector spaces that are closed under the action of elements of $\mathcal G$ and do not have non-trivial subspaces that are closed under the action of $\mathcal G$ elements. Likewise, the representations of the elements of the group are reducible to a direct sum of irreducible representations that act independently on irreducible subspaces $\mathcal H_\lambda$.

Schur's lemma states that the only element that commutes with the irreducible representation of all the elements of a group is a constant multiple of the representation of the identity operator on the irreducible subspace. By construction, $\mathsf M_{\mathcal G}$ commutes with all irreducible representations of all group elements in $\mathcal G$, and thus reduces to a linear sum of identity operators on the individual irreducible subspaces. Consequently, $\mathsf M_{\mathcal G}$ can be written as a linear sum of projectors onto the irreducible subspaces $\mathcal H_\lambda$ of the group $\mathcal G$ as
\begin{equation}
    \mathsf M_{\mathcal G} = \sum_\lambda\frac{\Tr(\mathsf M_{\mathcal G}\mathsf \Pi_{\mathcal G}^\lambda)}{d_\lambda} \mathsf \Pi_{\mathcal G}^\lambda,
\end{equation}
where $\mathsf \Pi_{\mathcal G}^\lambda$ is an orthogonal projector to subspace $\mathcal H_\lambda$, $\mathsf \Pi_{\mathcal G}^\lambda \mathsf \Pi_{\mathcal G}^{\lambda'} = \delta_{\lambda\lambda'} \mathsf \Pi_{\mathcal G}^{\lambda}$,  and $d_\lambda = \Tr(\mathsf \Pi_{\mathcal G}^\lambda)$ is the dimension of $\mathcal H_\lambda$. $\mathsf M_{\mathcal G}$ is the twirl of the operator $\mathsf M$ with the group $\mathcal G$ and the process is known as \textit{group twirling}. The group needs to satisfy the condition 
\begin{equation}\label{eqn:invertibility_cond}
\Tr(\mathsf M_{\mathcal G} \mathsf \Pi_{\mathcal G}^\lambda) \neq 0, ~\forall \lambda,
\end{equation}
for the composite channel $\mathsf{M}_{\mathcal G}$ to be invertible. Then 
the inverse is built as
\begin{equation}
    \mathsf M_{\mathcal G}^{-1}  = \sum_\lambda\frac{d_\lambda}{\Tr(\mathsf M_{\mathcal G}\mathsf \Pi_{\mathcal G}^\lambda)} \mathsf \Pi_{\mathcal G}^\lambda.
\end{equation}
By applying the channel inverse onto $\mathcal M_{\mathcal G}(\hat \rho)$, we can obtain the reconstructed state $\hat \rho$. 

Subsequently, the expectation value of the observable can be obtained as
\begin{align}
    \Tr(\hat O\hat \rho) \equiv& \langle\!\langle\hat O^\dagger|\hat \rho\rangle\!\rangle\\
    =& \mathbb E_{\mathcal U \in \mathcal G}\left[\langle\!\langle\hat O^\dagger| \mathsf M_{\mathcal G}^{-1}\mathsf U^\dagger \mathsf M\mathsf U |\hat \rho\rangle\!\rangle\right]\label{eqn:sup_exp_step}\\ \notag
    \equiv & \mathbb E_{\mathcal U \in \mathcal G}
    \Bigg{[}\sum_{\vec z \in \{1, -1\}^N} \Tr\left(\hat O \cdot \mathcal M_{\mathcal G}^{-1}\circ \mathcal U^\dagger (\ket{\vec z}\bra{\vec z})\right) \\
    \times & \langle \vec z |\mathcal U(\hat \rho)|\vec z\rangle\Bigg{]}
\end{align}
Here, $\Tr\left(\hat O \cdot  \mathcal M_{\mathcal G}^{-1}\circ \mathcal U^\dagger (\ket{\vec z}\bra{\vec z})\right)$ can be evaluated classically 
efficiently, while coefficients $\langle \vec z | \hat U\hat \rho\hat U^\dagger|\vec z\rangle$ are determined through repeated measurements.
The explicit form of $\Tr\left(\hat O \cdot  \mathcal M_{\mathcal G}^{-1}\circ \mathcal U^\dagger (\ket{\vec z}\bra{\vec z})\right)$ 
depends on the used group $\mathcal G$, thus we exemplify these 
expressions for a few commonly used groups of unitary transformations.

\subsection{Unitary Groups}

Finding groups $\mathcal G$ that satisfy the invertibility criterion in Eq. \eqref{eqn:invertibility_cond} for the measurement channel is non-trivial. We present a few discrete groups and discuss the invertibility of the corresponding twirled measurement channel.

\subsubsection{Global Clifford Unitaries}\label{app:globalcliffords}
The Clifford group of $N$ qubits, $\mathrm{Cl}(N)$ is defined as the group of all unitary evolutions that permute the set of non-identity $N$ qubit Pauli products amongst each other. \cite{Koenig_2014} Since the number of such valid permutations that preserve commutation relations between transformed Pauli products are finite, $\mathrm{Cl}(N)$ forms a discrete group. The twirled channel can be expressed as 
\begin{align}
    \mathsf M_{Cl(N)} =& \frac{1}{|\mathrm{Cl}(N)|} \sum_{U \in Cl(N)}\mathsf U^\dagger \mathsf M\mathsf U
    \end{align}
where $\mathsf U \in \mathrm{Cl}(N)$ are super-operator representations of the Clifford group that act on $\mathcal H_{d^2}$.
The span of identity, $\mathcal H :=\text{span}\{|\vec \sigma_0\rangle\!\rangle\}$ is trivially invariant and irreducible under unitary transformations. Since we can always find a Clifford unitary $U \in \mathrm{Cl}(N)$ that transforms a non-identity Pauli product to any other non-identity Pauli product, informally, this suggests that we cannot have any invariant subspaces of the complementary $d^2 - 1 $ dimensional space, $\mathcal H^{(1)} = \mathcal H_{d^2}\setminus \mathcal H^{(0)}$ spanned by any particular subset of Pauli products. Thus the super-operator representation of the Clifford group has two irreducible subspaces, $\mathcal H^{(0)}$, $\mathcal H^{(1)}$. Denote the projectors onto $\mathcal H^{(0)}$ and $\mathcal H^{(1)}$ by
\begin{equation}
    \mathsf \Pi_0 = |\vec\sigma_0\rangle\!\rangle\langle\!\langle \vec \sigma_0|, ~ \mathsf \Pi_1 = \mathsf I - \mathsf \Pi_0
\end{equation}
where $\mathsf I \in \mathcal L(\mathcal H_{d^2})$ is the super-operator representation of the identity channel $I(\hat \rho) = \hat \rho$. The twirled operator can be written as $\mathsf M_{\mathrm{Cl}(N)} = c_0 \mathsf \Pi_0 + c_1 \mathsf \Pi_1$, with constants evaluated as
\begin{align}
c_0 =& \frac{\Tr\left(\mathsf M \mathsf \Pi_0 \right)}{d_0} = 1\\
c_1 =& \frac{\Tr\left(\mathsf M \mathsf \Pi_1\right)}{d_1} = \frac{d-1}{d^2-1} = \frac{1}{d+1}
\end{align}
For the measurement channel $\mathsf M$, the Clifford group satisfies the invertibility criterion in Eq. \eqref{eqn:invertibility_cond}. The inverse of the twirled channel is
\begin{equation}
\mathsf M_{\mathrm{Cl}(N)}^{-1} = \mathsf \Pi_0 + (d+1)\mathsf \Pi_1 = (d+1)\mathsf I - d\mathsf \Pi_0
\end{equation}
The reconstructed density has the form
\begin{align}
\mathsf M_{\mathrm{Cl}(N)}^{-1} \mathbb E_{\mathcal  U }\left[\mathsf U^\dagger \mathsf M \mathsf U\right] |\hat \rho\rangle\!\rangle 
 =& (d+1)\mathbb E_{\mathcal U}\left[\mathsf U^\dagger \mathsf M\mathsf U \right]|\hat \rho\rangle\!\rangle \nonumber\\&- d\mathbb E_{\mathcal U}\left[ \mathsf \Pi_0\mathsf U^\dagger \mathsf M\mathsf U \right] |\hat \rho\rangle\!\rangle
\end{align}
The second term simplifies as
\begin{align}
    \mathsf \Pi_0 \mathsf U^\dagger \mathsf M\mathsf U |\hat \rho\rangle\!\rangle =&  |\vec\sigma_0\rangle\!\rangle\langle\!\langle \vec\sigma_0|\mathsf U^\dagger \mathsf M\mathsf U |\hat \rho\rangle\!\rangle\\
    \equiv& \frac{1}{d}\hat 1^{\otimes N} \Tr\left(\hat 1^{\otimes N}\cdot \mathcal U^\dagger \circ M\circ \mathcal U(\hat \rho)\right)\\
    = & \frac{1}{d}\hat 1^{\otimes N},
\end{align}
where the last step follows since unitary and measurement channels are trace preserving. Expanding the measurement channel, the reconstructed density has the form
\begin{align}
    \hat \rho =& (d+1)\mathbb E_{\mathcal U} \left[\sum_{\vec z \in \{1, -1\}^N}\mathcal U^\dagger(\ket{\vec z}\bra{\vec z})\bra{\vec z}\mathcal U(\hat \rho)\ket{\vec z}\right] - \hat 1^{\otimes N}.
\end{align}

Note that $\mathsf M_{\mathrm{Cl}(N)}$ has the form of the depolarizing channel with depolarizing strength $p=d/(d+1)$. In general, twirling a channel with $\mathrm{Cl}(N)$ creates a depolarizing channel with depolarizing strength dependent on the twirled channel.

\subsubsection{Local Clifford Unitaries}
Despite being finite, the cardinality of the Clifford group is combinatorially large \cite{Koenig_2014},
\begin{equation}
    |\mathrm{Cl}(N)| = 2^{N^2 + 2N}\prod_{j=1}^N (4^j - 1)
\end{equation}
and reconstructing the density to estimate observables will require a large sample size. Instead, we can consider a tensor product of smaller Clifford groups, defined on subsets of qubits. For instance, we consider the group consisting of tensor products of single qubit Clifford unitaries, $\mathrm{Cl}(1)^{\otimes N}$. As before, each single qubit Clifford unitary group $\mathrm{Cl}(1)$ divides the Hilbert space of the single qubit density into two irreducible subspaces, $\mathcal H^{(0)}$ and $\mathcal H^{(1)}$, spanned by the single qubit identity Pauli operator and the non-identity Pauli operators respectively. Denote the projectors onto these subspaces by $\mathsf \Pi^{(0)}$ and $\mathsf \Pi^{(1)}$ respectively. The tensor product group $\mathrm{Cl}(1)\otimes \mathrm{Cl}(1)$ has irreducible subspaces $\mathcal H^{(0)}\otimes \mathcal H^{(0)}$, $\mathcal H^{(0)}\otimes \mathcal H^{(1)}$, $\mathcal H^{(1)}\otimes \mathcal H^{(0)}$ and $\mathcal H^{(1)}\otimes \mathcal H^{(1)}$. As $\mathcal H^{(0)}$ is one dimensional, the first three subspaces are irreducible. The subspace $\mathcal H^{(1)}\otimes \mathcal H^{(1)}$ is irreducible as one can always find a unitary in $\mathrm{Cl}(1)\otimes \mathrm{Cl}(1)$ that transforms a two qubit non-identity Pauli product to any other two qubit non-identity Pauli product. This property can be seen as a consequence of the definition of Pauli products as a tensor product of Pauli operators. Extending this to $N$-qubits, $\mathrm{Cl}(1)^{\otimes N}$ has $2^{N}$ irreducible subspaces given by all possible $N$-tensor product combinations of $\mathcal H^{(0)}$ and $\mathcal H^{(1)}$. We introduce a vector with binary entries $\vec h = (h_1, \dots, h_{N}) \in \{0, 1\}^N$ to label the irreducible subspaces $\mathcal H^{(\vec h)} = \otimes_{i = 1}^{N} \mathcal H^{(h_i)}$,
where $\mathcal H^{(0)} = \mathrm{span}\{|\hat \sigma_0\rangle\!\rangle\}$ and $\mathcal H^{(1)} = \mathcal H_{2^2} \setminus \mathcal H^{(0)}$,
the projector onto these subspaces are $\mathsf \Pi^{(\vec h)} = \otimes_{i=1}^{N}\mathsf \Pi^{(h_i)}$. The inverse of the measurement channel twirled with $\mathrm{Cl}(1)^{\otimes N}$ is 
\begin{equation}
    \mathsf M_{\mathrm{Cl}(1)^{\otimes N}}^{-1} = \bigotimes_{j = 1}^{N}\left( \mathsf \Pi^{(0)} + 3\mathsf \Pi^{(1)}\right) =  \sum_{\vec h \in \{0, 1\}^N} 3^{|\vec h|} \mathsf \Pi^{(\vec h)}
\end{equation}
where $|\vec h|$ is the number of 1's in $\vec h$. 
The reconstructed density is given by
\begin{equation}
    \hat \rho = \mathbb E_{\mathcal U}\left[\sum_{\vec z\in \{1, -1\}^N} \bigotimes_{j=1}^{N}\left(3~ \mathcal U_j^\dagger\left(\ket{z_j}\bra{z_j}\right) - \hat 1\right)\bra{\vec z}\mathcal U(\hat \rho) \ket{\vec z}\right]
\end{equation}
where $\mathcal U = \bigotimes_{j=1}^{N} \mathcal  U_j$ is a tensor product of single qubit Clifford unitaries, $\mathcal  U_i\in \mathrm{Cl}(1)$ and $\ket{\vec z} = \ket{z_1z_2\dots z_N}$.

The estimate of Pauli product $\hat P = \otimes_{j=1}^N\hat \sigma_j$ on $\hat \rho$ is
\begin{align}
    \Tr(\hat \rho \hat P) = \mathbb E_{\mathcal U}
    \Bigg[&\sum_{z }\prod_{i=1}^N \bigg(3 \langle z_j|\mathcal U_j (\hat \sigma_j)|z_j\rangle \nonumber\\&- \Tr(\hat \sigma_j)\bigg)\langle z | \mathcal U(\hat \rho)|z\rangle\Bigg]
\end{align}
The factor $\left(3 \langle z_j|\mathcal U_j (\hat \sigma_j)|z_j\rangle - \Tr(\hat \sigma_j)\right)$ is equivalent to $f_j(\hat B_j, \hat \sigma_j)$ introduced in Eq. \eqref{cs_local_clif_estimator}, with the basis $\hat B_j$ defining the Clifford unitary $\mathcal U_j$. 

\subsubsection{Clifford-Fermionic Gaussian Unitaries}\label{app:cfgu}

Recall, the continuous group $\mathrm{FGU}(N)$ consists of unitaries in $\mathrm{Spin}(2N)$ of the form
\begin{equation}
    \hat U(e^A) = \exp\left(-\frac{1}{4}\sum_{\mu,\nu=0}^{2N-1} A_{\mu\nu}\hat \gamma_\mu \hat \gamma_\nu\right),
\end{equation}
where $A$ is a real antisymmetric matrix $A^T = -A$. 
The adjoint action of $\mathrm{FGU}(N)$ on Majorana operators is given by
\begin{equation}
    \hat U(e^A)^\dagger\left(\hat \gamma_{\mu}\right) = \sum_{\nu=0}^{2N - 1} \left(e^A\right)_{\mu\nu} \hat \gamma_\nu
\end{equation}
where $e^A \in \mathrm{SO}(2N)$ defines the transformation on the Majorana operators $\{\hat \gamma_\mu\}$. Informally, since this action conserves the degree when applied on a Majorana product, $\mathrm{FGU}(N)$ has $2N$ non-trivial irreducible representations that act on subspaces 
\begin{equation}
\mathcal H^{(k)} = \text{span}(\{\Gamma_{\vec \mu}: |\vec \mu| = k\}), ~~~k = 1, \dots, 2N.
\end{equation}
When $U(e^A)$ is restricted to a Clifford, $U(e^A) \in \mathrm{FGU}(N) \cap \mathrm{Cl}(N):= \mathrm{cFGU}(N)$, $e^A$ is simply a $(2N\times 2N)$ signed permutation matrix. The super-operator representation of the $\mathrm{cFGU}(N)$-twirled measurement channel, $\mathsf M_{\mathrm{cFGU}}$ is diagonal in the basis of Majorana operators and can be inverted. Since all operators of interest (RDMs, electronic Hamiltonians) consist of even powers of Majorana operators, it is sufficient to invert the twirled channel on $\mathcal H^{(\mathrm{even})} = \bigoplus_{k = 0}^{N} \mathcal H^{(2k)}$. Then the twirled channel has the form
\begin{equation}
\mathsf M_{\mathrm{cFGU}} = \sum_{k=0}^{N} \lambda_{2k} \Pi_{2k}, ~~~ \lambda_{2k} = {N\choose k}/{2N \choose 2k}
\end{equation}
and can be inverted on $\mathcal H^{(\mathrm{even})}$.

\subsubsection{Pauli Group}\label{app:pauli_group}

The Pauli group of $N$ qubits $\mathrm P(N)$ consists of Pauli products,
$\{\hat P_i\}$ with an additional phase $\{\pm 1, \pm i\}$. 
Elements of the Pauli group are unitary, and can be used to twirl quantum channels, but the measurement channel twirled with the Pauli group is not invertible. Since Pauli products either commute or anti-commute, the Pauli group divides $\mathcal H_{d^2}$ into $4^N$ one dimensional irreducible subspaces $\mathcal H^{(a)} = \mathrm{span}\{|\vec \sigma_{a}\rangle\!\rangle\}$
with corresponding projectors $\mathsf \Pi_{a} = |\vec \sigma_a\rangle\!\rangle\langle\!\langle \vec \sigma_a|$. $\Tr(\mathsf \Pi_a \mathsf M) = 0$ when $\vec \sigma_a$ consists $\hat \sigma_x$ or $\hat \sigma_y$ at any qubit position leading to non-invertibility of the twirled measurement channel.

\bibliography{ref}

\end{document}